\documentclass[english]{article}
\usepackage[T1]{fontenc}
\usepackage[latin9]{inputenc}
\usepackage{array}
\usepackage{float}
\usepackage{multirow}
\usepackage{amsmath}
\usepackage{amssymb}
\usepackage{graphicx}
\usepackage{setspace}

\makeatletter

\providecommand{\tabularnewline}{\\}

\newcommand{\lyxaddress}[1]{
\par {\raggedright #1
\vspace{1.4em}
\noindent\par}
}
\newenvironment{lyxcode}
{\par\begin{list}{}{
\setlength{\rightmargin}{\leftmargin}
\setlength{\listparindent}{0pt}% needed for AMS classes
\raggedright
\setlength{\itemsep}{0pt}
\setlength{\parsep}{0pt}
\normalfont\ttfamily}%
 \item[]}
{\end{list}}

\makeatother

\usepackage{babel}
\begin{document}

\title{Testing Selective Influence Directly Using Trackball Movement Tasks }

\author{Ru Zhang$^{1}$, Cheng-Ta Yang$^{2}$, and Janne V. Kujala$^{3}$ }
\maketitle

\lyxaddress{$^{1}$ Department of Psychology and Neuroscience, University of
Colorado, Boulder, Boulder, CO, USA }

\lyxaddress{$^{2}$ Department of Psychology, National Cheng Kung University,
No. 1, University Road, Tainan, Taiwan 701 }

\lyxaddress{$^{3}$ Department of Mathematical Information Technology, University
of Jyväskylä, FI-40014 Jyväskylä, Finland}
\begin{abstract}
Systems factorial technology (SFT; Townsend \& Nozawa, 1995) is regarded
as a useful tool to diagnose if features (or dimensions) of the investigated
stimulus are processed in a parallel or serial fashion. In order to
use SFT, one has to assume the speed to process each feature is influenced
by that feature only, termed as selective influence (Sternberg, 1969).
This assumption is usually untestable as the processing time for a
stimulus feature is not observable. Stochastic dominance is traditionally
used as an indirect evidence for selective influence (e.g., Townsend
\& Fifi\'{c}, 2004). However, one should keep in mind that selective
influence may be violated even when stochastic dominance holds. The
current study proposes a trackball movement paradigm for a direct
test of selective influence. The participants were shown a reference
stimulus and a test stimulus simultaneously on a computer screen.
They were asked to use the trackball to adjust the test stimulus until
it appeared to match the position or shape of the reference stimulus.
We recorded the reaction time, the parameters defined the reference
stimulus (denoted as $\alpha$ and $\beta$ ), and the parameters
defined the test stimulus (denoted as $A$ and $B$). It was expected
that the participants implemented the serial AND, parallel AND, or
coactive manner to adjust $A$ and $B$, and serial OR and parallel
OR strategies were prohibited. We tested selective influence of $\alpha$
and $\beta$ on the amount of time to adjust $A$ and $B$ through
testing selective influence of $\alpha$ and $\beta$ on the values
of $A$ and $B$ using the linear feasibility test (Dzhafarov \& Kujala,
2010). We found that when the test was passed and stochastic dominance
held, the inferred architecture was as expected, which was further
confirmed by the trajectory of $A$ and $B$ observed in each trial.
However, with stochastic dominance only SFT can suggest a prohibited
architecture. Our results indicate the proposed method is more reliable
for testing selective influence on the processing speed than examining
stochastic dominance only. 

\end{abstract}

\part*{Introduction}

\noindent A mental architecture is a hypothetical network of mental
processes when a cognitive task is being performed by a subject. Considering
a stimulus having only two features $\alpha$ and $\beta$. Let us
assume there is a channel to process the information of $\alpha$
and another channel to process the information of $\beta$. There
are numerous ways to arrange the two channels. One can inspect three
fundamental characteristics of the arrangements, that are architecture
(serial vs. parallel), stopping rule (OR vs. AND), and capacity (limited,
unlimited vs. super). A serial architecture processes one channel
after the preceding channel is completely executed. A parallel architecture
starts to process all the channels simultaneously but can terminate
them at different times. The OR rule means that the entire processing
can be completed as soon as any one of the channels is complete. If
all processes must be completely executed to ensure a response, then
it is the AND rule. Capacity measures the efficiency of information
processing as the workload, i.e., number of channels varies. When
executing a given channel is not affected by adding an additional
channel, the capacity is unlimited. Super capacity indicates that
processing efficiency of individual channels actually increases as
the workload is increased. Limited capacity indicates that the processing
efficiency decreases with the increased workload. Those properties
can be understood by inspecting the distributional behavior of response/reaction
time (RT) in a double-factorial paradigm (DFP) in the framework of
systems factorial technology (SFT; Townsend \& Nozawa, 1995; for further
development, see Schweickert, Giorgini, \& Dzhafarov, 2000; Dzhafarov,
Schweickert, \& Sung, 2004; Yang, Fifi\'{c}, \& Townsend, 2013; Zhang
\& Dzhafarov, 2015; Little, Altieri, Fifi\'{c}, \& Yang, 2017). There
are two critical types of manipulation that comprise the DFP: manipulation
of workload and manipulation of stimulus salience. This paradigm includes
a full factorial combination of the two types of manipulation, each
incorporating two levels for each stimulus feature. For instance,
in a visual detection task two dots (denoted as $\alpha$ and $\beta$)
are presented to the subjects, one on the left and the other on the
right. The researcher manipulates the stimulus salience by choosing
two different levels of brightness for each dot, level one for less
bright and level two for bright. So there are four stimuli for this
type of manipulation: $\alpha_{1}\beta_{1}$, $\alpha_{1}\beta_{2}$,
$\alpha_{2}\beta_{1}$, and $\alpha_{2}\beta_{2}$. The manipulation
of workload can be realized by tuning each dot on and off. So in some
trials only the left dot or the right dot is shown. This type of manipulation
introduces four additional stimuli to the experiment: $\alpha_{0}\beta_{1}$,
$\alpha_{0}\beta_{2}$, $\alpha_{1}\beta_{0}$, and $\alpha_{2}\beta_{0}$,
where the subscript 0 indicates the corresponding dot is off. The
trials with only one dot displays are named single-channel trials.
The trials with two dots display are named double-channel trials.
So far SFT has been widely used to investigate mental architectures
implemented in various cognitive tasks with short RT, such as the
simple detection task (Townsend \& Nozawa, 1995), Stroop task (Eidels,
Townsend, \& Algom, 2010), Gestalt principles (Eidels, Townsend, \&
Pomerantz, 2008), visual search (Fifi\'{c}, Townsend, \& Eidels, 2008),
short-term memory search (Townsend \& Fifi\'{c}, 2004), face perception
(Fifi\'{c} \& Townsend, 2010; Wenger \& Townsend, 2001; Yang, Fifi\'{c},
Chang, \& Little, 2018), attention (Yang, 2017), change detection
(Yang, 2011), audiovisual detection (Yang, Altieri, \& Little, 2018),
categorization (Fifi\'{c}, Little, \& Nosofsky, 2010), and even in
the clinical domain (Johnson, Blaha, Houpt, \& Townsend, 2010; Altieri
\& Yang, 2016). 

In order to implement SFT, one has to impose three assumptions on
the investigated system: ordering of the RT distributions (Townsend
\& Schweickert, 1989), selective influence (Sternberg, 1969), and
subject's adherence to a single type of mental architecture. Ordering
of the RT distributions is expressed as the ordering of survival functions:

\begin{align}
S_{\alpha_{1}}(t) & \geq S_{\alpha_{2}}(t),\nonumber \\
S_{\beta_{1}}(t) & \geq S_{\beta_{2}}(t).\label{ordering}
\end{align}
$S_{\alpha_{i}}(t)$ is the survival function for $T_{\alpha_{i}},i\in\left\{ 1,2\right\} $,
which is the duration for channel $\alpha$ at level $i$, and $S_{\beta_{j}}(t)$
is the survival function for $T_{\beta_{j}},j\in\left\{ 1,2\right\} $,
which is the duration for channel $\beta$ at level $j$. The ordering
assumption can be easily satisfied empirically as one can for instance
manipulate the brightness of a stimulus that the less bright stimulus
is processed slower than the bright one. 

SFT requires selective influence which can be understood as the duration
for channel $\alpha$ is affected by the changing of $\alpha$ but
not the changing of $\beta$, and the duration for channel $\beta$
is affected by the changing of $\beta$ but not the changing of $\alpha$.
It is written as

\begin{equation}
(T_{\alpha},T_{\beta})\looparrowleft(\alpha,\beta).\label{SI time}
\end{equation}
Dzhafarov and Kujala (2010) proposed a rigorous definition for it:
Random variables $T_{\alpha}$ and $T_{\beta}$ are selectively influenced
by $\alpha$ and $\beta$, if and only if there are measurable functions
$g_{\alpha}$ and $g_{\beta}$ and a random entity $C$ whose distribution
is independent of $\alpha$ and $\beta$, such that $(T_{\alpha},T_{\beta})\sim(g_{\alpha}(\alpha,C),g_{\beta}(\beta,C))$,
where $\sim$ stands for \textquotedbl{}is distributed as\textquotedbl{}.
The random entity $C$, in psychology, can be resolution of the monitor,
which is apparently independent of stimulus features $\alpha$ and
$\beta$, but influences $T_{\alpha}$ and $T_{\beta}$ simultaneously.
Please note Dzhafarov and Kujala's definition of selective influence
is not limited in the field of psychology: $\alpha$ and $\beta$
can be two external factors in any system and $T_{\alpha}$ and $T_{\beta}$
are the random variables in response to the external factors. The
readers should keep in mind in the framework of SFT the notion of
selective influence is confined on the relation between stimulus features
and the durations to process those features. 

Dzhafarov and Kujala (2010) also proposed an equivalent definition:
Selective influence (\ref{SI time}) holds if and only if there exists
a jointly distributed quadruple $(H_{\alpha_{1}},H_{\alpha_{2}},H_{\beta_{1}},H_{\beta_{2}})$,
such that 

\[
(H_{\alpha_{i}},H_{\beta_{j}})\sim(T_{\alpha_{i}},T_{\beta_{j}})\mid_{\alpha_{i}\beta_{j}},
\]
where $(T_{\alpha_{i}},T_{\beta_{j}})\mid_{\alpha_{i}\beta_{j}}$
represents the jointly distributed $T_{\alpha_{i}}$ and $T_{\beta_{j}}$
conditioned on the stimulus $\alpha_{i}\beta_{j}$. Dzhafarov and
Kujala (2010) developed the linear feasibility test (LFT) to establish
or falsify selective influence. Let us assume that $T_{\alpha}$ has
$m$ possible values: $\left\{ a_{1},a_{2},...,a_{m}\right\} $ and
$T_{\beta}$ has $n$ possible values: $\left\{ b_{1},b_{2},...,b_{n}\right\} $.
Let us write the joint probability for the vector $(H_{\alpha_{1}},H_{\alpha_{2}},H_{\beta_{1}},H_{\beta_{2}})$
as

\begin{doublespace}
\[
\Pr\left(H_{\alpha_{1}}=a_{\alpha_{1}},H_{\alpha_{2}}=a_{\alpha_{2}},H_{\beta_{1}}=b_{\beta_{1}},H_{\beta_{2}}=b_{\beta_{2}}\right)=Q_{a_{\alpha_{1}}a_{\alpha_{2}}b_{\beta_{1}}b_{\beta_{2}}},
\]
where $a_{\alpha_{1}},a_{\alpha_{2}}\in\left\{ a_{1},a_{2},...,a_{m}\right\} $
and $b_{\beta_{1}},b_{\beta_{2}}\in\left\{ b_{1},b_{2},...,b_{n}\right\} $
with constraints

\[
Q_{a_{\alpha_{1}}a_{\alpha_{2}}b_{\beta_{1}}b_{\beta_{2}}}\geq0,
\]
such that 
\end{doublespace}

\begin{flalign}
\Pr\left(H_{\alpha_{1}}=a_{\alpha_{1}},H_{\beta_{1}}=b_{\beta_{1}}\right) & =\sum_{a_{\alpha_{2}}b_{\beta_{2}}}Q_{a_{\alpha_{1}}a_{\alpha_{2}}b_{\beta_{1}}b_{\beta_{2}}}=\Pr\left(T_{\alpha_{1}}=a_{\alpha_{1}},T_{\beta_{1}}=b_{\beta_{1}}\right)\mid_{\alpha_{1}\beta_{1}},\nonumber \\
\Pr\left(H_{\alpha_{1}}=a_{\alpha_{1}},H_{\beta_{2}}=b_{\beta_{2}}\right) & =\sum_{a_{\alpha_{2}}b_{\beta_{1}}}Q_{a_{\alpha_{1}}a_{\alpha_{2}}b_{\beta_{1}}b_{\beta_{2}}}=\Pr\left(T_{\alpha_{1}}=a_{\alpha_{1}},T_{\beta_{2}}=b_{\beta_{2}}\right)\mid_{\alpha_{1}\beta_{2}},\nonumber \\
\Pr\left(H_{\alpha_{2}}=a_{\alpha_{2}},H_{\beta_{1}}=b_{\beta_{1}}\right) & =\sum_{a_{\alpha_{1}}b_{\beta_{2}}}Q_{a_{\alpha_{1}}a_{\alpha_{2}}b_{\beta_{1}}b_{\beta_{2}}}=\Pr\left(T_{\alpha_{2}}=a_{\alpha_{2}},T_{\beta_{1}}=b_{\beta_{1}}\right)\mid_{\alpha_{2}\beta_{1}},\nonumber \\
\Pr\left(H_{\alpha_{2}}=a_{\alpha_{2}},H_{\beta_{2}}=b_{\beta_{2}}\right) & =\sum_{a_{\alpha_{1}}b_{\beta_{1}}}Q_{a_{\alpha_{1}}a_{\alpha_{2}}b_{\beta_{1}}b_{\beta_{2}}}=\Pr\left(T_{\alpha_{2}}=a_{\alpha_{2}},T_{\beta_{2}}=b_{\beta_{2}}\right)\mid_{\alpha_{2}\beta_{2}}.\label{eq:JDC}
\end{flalign}
If the nonnegative solution for the $Q$ variables that satisfies
(\ref{eq:JDC}) exists, we say LFT is passed and selective influence
(\ref{SI time}) is established, otherwise selective influence is
falsified. 

Marginal selectivity is a necessary condition for selective influence.
It states that the marginal distribution of $T_{\alpha}$ does not
depend on $\beta$ and the marginal distribution of $T_{\beta}$ does
not depend on $\alpha$. It can be mathematically written as 

\begin{doublespace}
\begin{flalign}
\sum_{b_{\beta_{1}}}\Pr\left(T_{\alpha_{1}}=a_{\alpha_{1}},T_{\beta_{1}}=b_{\beta_{1}}\right)\mid_{\alpha_{1}\beta_{1}} & =\sum_{b_{\beta_{2}}}\Pr\left(T_{\alpha_{1}}=a_{\alpha_{1}},T_{\beta_{2}}=b_{\beta_{2}}\right)\mid_{\alpha_{1}\beta_{2}},\nonumber \\
\sum_{b_{\beta_{1}}}\Pr\left(T_{\alpha_{2}}=a_{\alpha_{2}},T_{\beta_{1}}=b_{\beta_{1}}\right)\mid_{\alpha_{2}\beta_{1}} & =\sum_{b_{\beta_{2}}}\Pr\left(T_{\alpha_{2}}=a_{\alpha_{2}},T_{\beta_{2}}=b_{\beta_{2}}\right)\mid_{\alpha_{2}\beta_{2}},\nonumber \\
\sum_{a_{\alpha_{1}}}\Pr\left(T_{\alpha_{1}}=a_{\alpha_{1}},T_{\beta_{1}}=b_{\beta_{1}}\right)\mid_{\alpha_{1}\beta_{1}} & =\sum_{a_{\alpha_{2}}}\Pr\left(T_{\alpha_{2}}=a_{\alpha_{2}},T_{\beta_{1}}=b_{\beta_{1}}\right)\mid_{\alpha_{2}\beta_{1}},\nonumber \\
\sum_{a_{\alpha_{1}}}\Pr\left(T_{\alpha_{1}}=a_{\alpha_{1}},T_{\beta_{1}}=b_{\beta_{2}}\right)\mid_{\alpha_{1}\beta_{2}} & =\sum_{a_{\alpha_{2}}}\Pr\left(T_{\alpha_{2}}=a_{\alpha_{2}},T_{\beta_{1}}=b_{\beta_{2}}\right)\mid_{\alpha_{2}\beta_{2}}.\label{marginal selectivity}
\end{flalign}
(\ref{eq:JDC}) implies marginal selectivity (\ref{marginal selectivity}).
If marginal selectivity is violated, the nonnegative solution for
(\ref{eq:JDC}) does not exist. 

Table \ref{tab:An-example-of} gives an example of joint probabilities
$\Pr\left(T_{\alpha_{i}}=a_{\alpha_{i}},T_{\beta_{j}}=b_{\beta_{j}}\right)\mid_{\alpha_{i}\beta_{j}}$,
where $i,j\in\left\{ 1,2\right\} $ and $m=n=2$. The numbers outside
the grids are marginal probabilities. 
\end{doublespace}

\begin{table}[H]
\begin{centering}
\caption{An example of joint probabilities of $(T_{\alpha_{i}},T_{\beta_{j}})$
given stimulus $\alpha_{i}\beta_{j}$, $i,j\in\left\{ 1,2\right\} $.\label{tab:An-example-of}}
\par\end{centering}
\begin{centering}
\begin{tabular}{|c|c|c|c}
\cline{1-3} 
$\alpha_{1}\beta_{1}$ & $T_{\beta_{1}}=b_{1}$ & $T_{\beta_{1}}=b_{2}$ & \tabularnewline
\cline{1-3} 
$T_{\alpha_{1}}=a_{1}$ & .2 & .2 & .4\tabularnewline
\cline{1-3} 
$T_{\alpha_{1}}=a_{2}$ & .1 & .5 & .6\tabularnewline
\cline{1-3} 
\multicolumn{1}{c}{} & \multicolumn{1}{c}{.3} & \multicolumn{1}{c}{.7} & \tabularnewline
\end{tabular}%
\begin{tabular}{|c|c|c|c}
\cline{1-3} 
$\alpha_{1}\beta_{2}$ & $T_{\beta_{2}}=b_{1}$ & $T_{\beta_{2}}=b_{2}$ & \tabularnewline
\cline{1-3} 
$T_{\alpha_{1}}=a_{1}$ & .3 & .1 & .4\tabularnewline
\cline{1-3} 
$T_{\alpha_{1}}=a_{2}$ & .4 & .2 & .6\tabularnewline
\cline{1-3} 
\multicolumn{1}{c}{} & \multicolumn{1}{c}{.7} & \multicolumn{1}{c}{.3} & \tabularnewline
\end{tabular}
\par\end{centering}
\centering{}%
\begin{tabular}{|c|c|c|c}
\cline{1-3} 
$\alpha_{2}\beta_{1}$ & $T_{\beta_{1}}=b_{1}$ & $T_{\beta_{1}}=b_{2}$ & \tabularnewline
\cline{1-3} 
$T_{\alpha_{2}}=a_{1}$ & .1 & .5 & .6\tabularnewline
\cline{1-3} 
$T_{\alpha_{2}}=a_{2}$ & .2 & .2 & .4\tabularnewline
\cline{1-3} 
\multicolumn{1}{c}{} & \multicolumn{1}{c}{.3} & \multicolumn{1}{c}{.7} & \tabularnewline
\end{tabular}%
\begin{tabular}{|c|c|c|c}
\cline{1-3} 
$\alpha_{2}\beta_{2}$ & $T_{\beta_{2}}=b_{1}$ & $T_{\beta_{2}}=b_{2}$ & \tabularnewline
\cline{1-3} 
$T_{\alpha_{2}}=a_{1}$ & .4 & .2 & .6\tabularnewline
\cline{1-3} 
$T_{\alpha_{2}}=a_{2}$ & .3 & .1 & .4\tabularnewline
\cline{1-3} 
\multicolumn{1}{c}{} & \multicolumn{1}{c}{.7} & \multicolumn{1}{c}{.3} & \tabularnewline
\end{tabular}
\end{table}
This example satisfies marginal selectivity since the conditions in
(\ref{marginal selectivity}) are met. Substituting the joint probabilities
in Table \ref{tab:An-example-of} into (\ref{eq:JDC}), 

\[
\left(\begin{array}{cccccccccccccccc}
1 & 1 & 0 & 0 & 1 & 1 & 0 & 0 & 0 & 0 & 0 & 0 & 0 & 0 & 0 & 0\\
0 & 0 & 1 & 1 & 0 & 0 & 1 & 1 & 0 & 0 & 0 & 0 & 0 & 0 & 0 & 0\\
0 & 0 & 0 & 0 & 0 & 0 & 0 & 0 & 1 & 1 & 0 & 0 & 1 & 1 & 0 & 0\\
0 & 0 & 0 & 0 & 0 & 0 & 0 & 0 & 0 & 0 & 1 & 1 & 0 & 0 & 1 & 1\\
1 & 0 & 1 & 0 & 1 & 0 & 1 & 0 & 0 & 0 & 0 & 0 & 0 & 0 & 0 & 0\\
0 & 1 & 0 & 1 & 0 & 1 & 0 & 1 & 0 & 0 & 0 & 0 & 0 & 0 & 0 & 0\\
0 & 0 & 0 & 0 & 0 & 0 & 0 & 0 & 1 & 0 & 1 & 0 & 1 & 0 & 1 & 0\\
0 & 0 & 0 & 0 & 0 & 0 & 0 & 0 & 0 & 1 & 0 & 1 & 0 & 1 & 0 & 1\\
1 & 1 & 0 & 0 & 0 & 0 & 0 & 0 & 1 & 1 & 0 & 0 & 0 & 0 & 0 & 0\\
0 & 0 & 1 & 1 & 0 & 0 & 0 & 0 & 0 & 0 & 1 & 1 & 0 & 0 & 0 & 0\\
0 & 0 & 0 & 0 & 1 & 1 & 0 & 0 & 0 & 0 & 0 & 0 & 1 & 1 & 0 & 0\\
0 & 0 & 0 & 0 & 0 & 0 & 1 & 1 & 0 & 0 & 0 & 0 & 0 & 0 & 1 & 1\\
1 & 0 & 1 & 0 & 0 & 0 & 0 & 0 & 1 & 0 & 1 & 0 & 0 & 0 & 0 & 0\\
0 & 1 & 0 & 1 & 0 & \text{0} & 0 & 0 & 0 & 1 & 0 & 1 & 0 & 0 & 0 & 0\\
0 & 0 & 0 & 0 & 1 & 0 & 1 & 0 & 0 & 0 & 0 & 0 & 1 & 0 & 1 & 0\\
0 & 0 & 0 & 0 & 0 & 1 & 0 & 1 & 0 & 0 & 0 & 0 & 0 & 1 & 0 & 1
\end{array}\right)\left(\begin{array}{c}
Q_{a_{1}a_{1}b_{1}b_{1}}\\
Q_{a_{1}a_{1}b_{1}b_{2}}\\
Q_{a_{1}a_{1}b_{2}b_{1}}\\
Q_{a_{1}a_{1}b_{2}b_{2}}\\
Q_{a_{1}a_{2}b_{1}b_{1}}\\
Q_{a_{1}a_{2}b_{1}b_{2}}\\
Q_{a_{1}a_{2}b_{2}b_{1}}\\
Q_{a_{1}a_{2}b_{2}b_{2}}\\
Q_{a_{2}a_{1}b_{1}b_{1}}\\
Q_{a_{2}a_{1}b_{1}b_{2}}\\
Q_{a_{2}a_{1}b_{2}b_{1}}\\
Q_{a_{2}a_{1}b_{2}b_{2}}\\
Q_{a_{2}a_{2}b_{1}b_{1}}\\
Q_{a_{2}a_{2}b_{1}b_{2}}\\
Q_{a_{2}a_{2}b_{2}b_{1}}\\
Q_{a_{2}a_{2}b_{2}b_{2}}
\end{array}\right)=\left(\begin{array}{c}
.2\\
.2\\
.1\\
.5\\
.3\\
.1\\
.4\\
.2\\
.1\\
.5\\
.2\\
.2\\
.4\\
.2\\
.3\\
.1
\end{array}\right),
\]
the nonnegative solution

\begin{doublespace}
\begin{align*}
(Q_{a_{1}a_{1}b_{1}b_{1}},\,Q_{a_{1}a_{1}b_{1}b_{2}},\,\ldots,\,Q_{a_{2}a_{2}b_{2}b_{2}})^{T}\\
=(0,0,0,0,.1,.1,.2,0,0,.1,.4,.1,0,0,0,0)^{T}
\end{align*}
establishes selective influence in this example.
\end{doublespace}

With the assumption of selective influence, as one manipulates the
features of the interested stimulus, the durations influenced by these
features vary and consequently the overall duration is changed as
well. With these factorial manipulations, each mental architecture
has a distributional pattern of RT that is different from other architectures.
$T_{\alpha}$ and $T_{\beta}$ are usually unobservable in empirical
studies, so selective influence of $\alpha$ and $\beta$ on $T_{\alpha}$
and $T_{\beta}$ cannot be tested directly. By imposing the assumption
of selective influence, ordering of the RT distributions is equivalent
to the four inequalities termed as stochastic dominance:

\begin{align}
S_{\alpha_{1}\beta_{1}}(t) & \geq S_{\alpha_{2}\beta_{1}}(t),\nonumber \\
S_{\alpha_{1}\beta_{2}}(t) & \geq S_{\alpha_{2}\beta_{2}}(t),\nonumber \\
S_{\alpha_{1}\beta_{1}}(t) & \geq S_{\alpha_{1}\beta_{2}}(t),\nonumber \\
S_{\alpha_{2}\beta_{1}}(t) & \geq S_{\alpha_{2}\beta_{2}}(t).\label{eq:sd}
\end{align}
$S_{\alpha_{i}\beta_{j}}(t)$ is the survival function for $T_{\alpha_{i}\beta_{j}}$,
which is the overall duration to process the stimulus $\alpha_{i}\beta_{j}$.
$T_{\alpha_{i}\beta_{j}}$ is usually observable, therefore stochastic
dominance is traditionally used to test selective influence (Townsend
\& Nozawa, 1995; Eidels, Townsend, \& Algom, 2010; Eidels, Townsend,
\& Pomerantz, 2008; Fifi\'{c}, Townsend, \& Eidels, 2008; Townsend
\& Fifi\'{c}, 2004; Fifi\'{c} \& Townsend, 2010; Wenger \& Townsend,
2001; Johnson, Blaha, Houpt, \& Townsend, 2010). However, one should
keep in mind that selective influence may be violated even when stochastic
dominance holds. Strictly speaking, stochastic dominance is only a
necessary condition for selective influence combined with ordering
of RT distributions. 

The third assumption states the subject maintains a single type of
mental architecture from trial to trial. This assumption could be
invalid as one may implement parallel AND for one trial and switch
to serial AND in another trial. In psychological research, it is usually
impossible to track the mental architecture in each trial. 

Three important properties were constructed in the framework of SFT:
\begin{itemize}
\item Mean Interaction Contrast (MIC)
\end{itemize}

\paragraph{
\[
\mathrm{MIC}(t)=\overline{T}_{\alpha_{1}\beta_{1}}-\overline{T}_{\alpha_{1}\beta_{2}}-\overline{T}_{\alpha_{2}\beta_{1}}+\overline{T}_{\alpha_{2}\beta_{2}},
\]
}

where $\overline{T}_{\alpha_{i}\beta_{j}}$ stands for the mean of
$T_{\alpha_{i}\beta_{j}}$.
\begin{itemize}
\item Survivor Interaction Contrast (SIC)
\end{itemize}

\paragraph{
\[
\mathrm{SIC}(t)=S_{\alpha_{1}\beta_{1}}(t)-S_{\alpha_{1}\beta_{2}}(t)-S_{\alpha_{2}\beta_{1}}(t)+S_{\alpha_{2}\beta_{2}}(t).
\]
}

\begin{figure}[H]
\centering{}\includegraphics[scale=0.36]{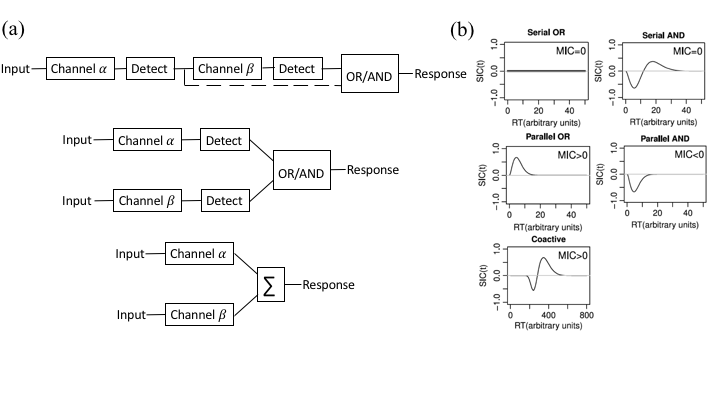}\caption{(a) Several classical mental architectures. (b) Characteristic behavior
of SIC and MIC for those architectures conditional on the three assumptions
(Yang, Fifi\'{c}, \& Townsend, 2013). \label{figure_architecture and SIC}}
\end{figure}

Four different combinations of architecture and stopping rule are
of the greatest traditional interest. They are serial OR, serial AND,
parallel OR, and parallel AND (Figure \ref{figure_architecture and SIC}(a)).
The overall duration for each model is a function of $T_{\alpha}$
and $T_{\beta}$. They are $T_{\alpha\beta}=T_{\alpha}$ or $T_{\beta}$,
$T_{\alpha\beta}=T_{\alpha}+T_{\beta}$, $T_{\alpha\beta}=\mathrm{min}(T_{\alpha},T_{\beta})$,
and $T_{\alpha\beta}=\mathrm{max}(T_{\alpha},T_{\beta})$, respectively.
In addition to the four models, the model with the information from
each parallel channel pooled toward a single decision is coactive.
Coactive processing is a special case of parallel models. It was previously
proved that for a serial OR model, both MIC = 0 and SIC = 0 across
time. For a serial AND model, MIC is zero and SIC fluctuates from
negative to positive and the sum of areas of its negative part and
positive part is zero. For a parallel OR model, both MIC and SIC are
positive. For a parallel AND model, both MIC and SIC are negative.
For a coactive model, MIC is positive and SIC fluctuates from negative
to positive and the area of the negative part is smaller than that
of the positive part. Having different characteristic patterns of
MIC and SIC, one can diagnose the mental architecture out of the five
candidate models (Figure \ref{figure_architecture and SIC}(b)). 

\begin{itemize}
\item Capacity (C)
\end{itemize}
In Townsend and Wenger (2004)'s paper, the capacity coefficients were
developed using temporal variables. For the AND stopping rule, 
\begin{lyxcode}
\begin{equation}
\mathrm{C}_{\mathrm{AND}}(t)=\frac{K_{\alpha}(t)+K_{\beta}(t)}{K_{\alpha\beta}(t)},\label{eq:capacity_AND}
\end{equation}
\end{lyxcode}
where $K(t)=\mathrm{ln}(F(t))=\mathrm{ln}(1-S(t))$. For the OR stopping
rule,
\begin{lyxcode}
\begin{equation}
\mathrm{C}_{\mathrm{OR}}(t)=\frac{H_{\alpha\beta}(t)}{H_{\alpha}(t)+H_{\beta}(t)},\label{eq:capacity_OR}
\end{equation}
\end{lyxcode}
where $H(t)=\int_{0}^{t}h(t)dt=\int_{0}^{t}\frac{f(t)}{S(t)}dt$,
where $f(t)$ is the density function. If $\mathrm{C_{AND/OR}}(t)>1$,
the capacity is super; if $\mathrm{C_{AND/OR}}(t)=1$, the capacity
is unlimited; if $\mathrm{C_{AND/OR}}(t)<1$, the capacity is limited. 

In this article we propose a trackball movement paradigm that can
test the assumptions that are usually untestable in other paradigms.
It includes two tasks: the dot position reproduction task (Experiments
1(a), 1(b), and 1(c)) and the floral shape reproduction task (Experiments
2(a), 2(b), and 2(c)). The participants were shown a reference stimulus
and a test stimulus simultaneously on a computer screen. They were
asked to use the trackball to adjust the test stimulus until it appeared
to match the position or shape of the reference stimulus. There were
two parameters defined the reference stimulus (denoted as $\alpha$
and $\beta$ ) and two parameters defined the test stimulus (denoted
as $A$ and $B$). So essentially the task goal was to match $\alpha$
and $\beta$ by adjusting $A$ and $B$. We tested selective influence
of $\alpha$ and $\beta$ on the amount of time to adjust $A$ and
$B$ through testing selective influence of $\alpha$ and $\beta$
on the values of $A$ and $B$ using LFT. We found that when the test
was passed and stochastic dominance held, the inferred architecture
about the adjustment of $A$ and $B$ was as expected (either parallel
AND or coactive), which was further confirmed by the trajectory of
$A$ and $B$ observed in each trial. The trajectory also confirmed
the assumption that the subjects maintained a stable stategy to respond
to the stimulus. However, with stochastic dominance only SFT can suggest
a prohibited architecture, e.g. parallel OR. Our results indicate
the proposed method is more reliable for testing the assumption of
selective influence on the processing speed than examining stochastic
dominance only. 

\part*{Method}

\section*{Participants }

Experiments 1(a), 2(a), and 2(b) were conducted at Purdue University
in USA. Experiments 1(b), 1(c), and 2(c) took place at National Cheng
Kung University (NCKU) in Taiwan (Table \ref{experiments and subjects}).
Three graduate students at Purdue University labeled as S1 to S3 participated
in Experiment 1(a) and Experiment 2(b). Three graduate students at
Purdue University labeled as S4 to S6 participated in Experiment 2(a).
Students at NCKU labeled as S7 to S11 participated in Experiment 1(b).
Students labeled as S12 to S16 participated in Experiment 1(c). Students
labeled as S17 to S21 participated in Experiment 2(c). The participants
at Purdue were aged 22-33 and the participants at NCKU were aged 19-30.

\begin{table}[H]
\caption{The experiments and the participants.}
\begin{centering}
\begin{tabular}{ccc}
\hline 
Experiment & Purdue & NCKU\tabularnewline
\hline 
1(a) & S1, S2, S3 & \tabularnewline
1(b) &  & S7, S8, S9, S10, S11\tabularnewline
1(c) &  & S12, S13, S14, S15, S16\tabularnewline
 2(a) & S4, S5, S6 & \tabularnewline
 2(b) & S1, S2, S3 & \tabularnewline
 2(c) &  & S17, S18, S19, S20, S21\tabularnewline
\hline 
\end{tabular}
\par\end{centering}
\label{experiments and subjects}
\end{table}

\section*{Stimuli and Procedure}

Visual stimuli consisting of dots and curves were presented on a flat-panel
monitor. They were grayish-white on a comfortably low intensity background.
The diameter of the dots and the width of the curves was 5 pixels
(px). The participants viewed the stimuli in darkness using a chin
rest with a forehead support at the distance of 90 cm from the monitor,
making 1 screen pixel approximately 62 sec arc. In each trial the
participants were asked to match a fixed reference stimulus by adjusting
a variable test stimulus by rotating a trackball using their dominant
hand. Once a response was made to the participant\textquoteright s
satisfaction, she or he clicked a button on the trackball device to
terminate this trial, and a new stimulus appeared half a second later.
There was no time pressure for each participant: They can make every
response with their own pace. Each experiment was consisted of several
sessions, each included hundreds of trials with a break in the middle.
Each such session was preceded by a practice series of 10 trials (which
were not analyzed). Each experiment took several days, one or two
sessions per day.

\subsection*{Experiment 1}

\begin{figure}
\begin{raggedright}
\includegraphics[bb=-100bp 0bp 616bp 465bp,scale=0.3]{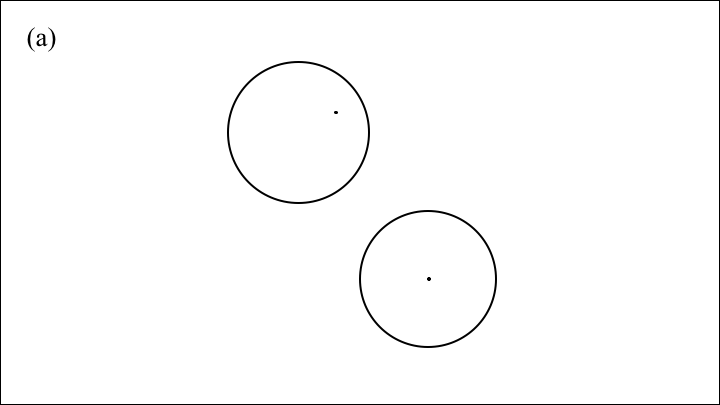}
\par\end{raggedright}
\begin{raggedright}
\includegraphics[bb=-100bp 0bp 616bp 465bp,scale=0.3]{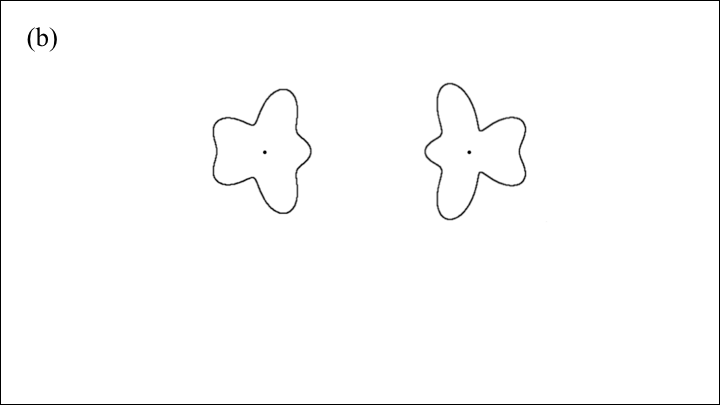}
\par\end{raggedright}
\caption{Stimuli used in (a) the dot position reproduction task and (b) the
shape reproduction task.}

\label{stimuli}
\end{figure}

Experiment 1 was a dot position reproduction task. In each trial the
participants were presented with two dots in two circles simultaneously
(Figure \ref{stimuli}(a)). The radii of the two circles were both
160 px. The upper left dot was the reference stimulus. It appeared
in the first quadrant of the circle and it was immovable. The dot
appeared in the center of its circle was the test stimulus. It was
movable. The participants were asked to move the movable test dot
until its location matched that of the fixed reference one. Once a
response was made, the program recorded the locations of the reference
dot and the test dot. The RT from the onset of the presentation of
stimuli to the button click in each trial was recorded. The program
also recorded the coordinates of the moving dot every 10 ms in each
trial. Experiment 1 contained three designs: 1(a), 1(b), and 1(c).
The significant differences among them were: Experiment 1(a) included
only double-channel trials, which did not allow computing the capacity
coefficient. In order to estimate the capacity, Experiments 1(b) and
1(c) had double-channel trials and single-channel trials that met
the requirement of DFP. The trials were presented in different ways
in the two designs: In Experiment 1(b), the two types of trials were
presented in a mixed way while in Experiment 1(c) the single-channel
trials were displayed separately from all the double-channel trials. 

\subsubsection*{Experiment 1(a)}

The horizontal coordinate of the reference dot with respect to the
center of its circle was randomly generated from the interval {[}20
px, 80 px{]} and the vertical coordinate was randomly generated from
{[}20 px, 80 px{]} too. We ran 1860 trials for each subject divided
equally in six sessions. 

\subsubsection*{Experiment 1(b)}

Experiment 1(b) had some trials in which either the horizontal coordinate
or the vertical coordinate of the reference dot was 0 px. In each
trial the two parameters of the reference dot were generated from
{[}20 px, 80 px{]} $\times$ {[}20 px, 80 px{]}, 0 px $\times$ {[}20
px, 80 px{]}, or {[}20 px, 80 px{]} $\times$ 0 px with probabilities
.5, .25, and .25 respectively. The trials generated from {[}20 px,
80 px{]} $\times$ {[}20 px, 80 px{]} were double-channel trials and
the trials generated from 0 px $\times$ {[}20 px, 80 px{]} and {[}20
px, 80 px{]} $\times$ 0 px were single-channel trials. There were
1680 trials in total divided equally in eight sessions. In all the
other aspects, Experiment 1(b) was identical to 1(a). 

\subsubsection*{Experiment 1(c)}

There were eight sessions, each including 210 trials. The first four
sessions contained double-channel trials only and the four sessions
ran later contained single-channel trials only. The two parameters
of reference stimulus in the double-channel sessions were randomly
generated from {[}20 px, 80 px{]} $\times$ {[}20 px, 80 px{]} in
each trial. The two parameters of reference stimulus in the single-channel
sessions were randomly generated from 0 px $\times$ {[}20 px, 80
px{]} or {[}20 px, 80 px{]} $\times$ 0 px. In all the other aspects,
Experiment 1(c) was identical to 1(b).

\subsection*{Experiment 2}

Experiment 2 was a floral shape reproduction task. Examples of two
floral shapes together with their centers are shown in Figure \ref{stimuli}(b).
Two such configurations were presented simultaneously in each trial.
The reference stimulus was on the left and it was fixed. The test
stimulus was on the right and it was modifiable. Both shapes were
generated from this function: 

\begin{align}
X & =\mathrm{cos}(.02\pi\triangle)[70+\alpha\mathrm{cos}(.06\pi\triangle)+\beta\mathrm{cos}(.1\pi\triangle)],\label{eq:floral shape-1}\\
Y & =\mathrm{sin}(.02\pi\triangle)[70+\alpha\mathrm{cos}(.06\pi\triangle)+\beta\mathrm{cos}(.1\pi\triangle)],\nonumber 
\end{align}
where $X$ and $Y$ are the horizontal coordinate (px) and vertical
coordinate (px) of the shape. $\triangle$ spans all the integers
from 0 to 99. $\alpha$ and $\beta$ in the function are amplitude
1 and amplitude 2 that determine the exact configuration of the floral
shape. 

We used $\alpha$ and $\beta$ to denote the amplitudes for the reference
shape. For the modifiable test shape, we replaced $\alpha$ with $A$
and $\beta$ with $B$. In other words, $A$ was represented as amplitude
1 and $B$ as amplitude 2 for the modifiable shape. The amplitudes
of the modifiable shape were initially selected from the interval
{[}-35 px, 35 px{]}. The participants were asked to match the reference
shape by modifying the test one. Since the computer program can only
read the horizontal move and the vertical move of the trackball, a
transformation function from the trackball move to the amplitude move
was imposed:

\begin{align}
A_{new} & =A+\frac{\mathrm{sign}(\triangle x)}{100}(70-A-|B|),\label{eq:transformation from px to amplitude-1}\\
B_{new} & =B+\frac{\mathrm{sign}(\triangle y)}{100}(70-|A|-B).\nonumber 
\end{align}
Here $\triangle x$ is the horizontal move of the trackball and $\triangle y$
is the vertical move of the trackball. $\triangle x$ and $\triangle y$
can be updated every 1 px or -1 px. In each trial, the program recorded
amplitude 1 and amplitude 2 of the reference shape and the finalized
test shape. The RT in each trial was recorded. 

Experiment 2 contained three designs: 2(a), 2(b), and 2(c). All the
trials in Experiment 2 were double-channel trials. The significant
differences among them were: Experiments 2(b) and 2(c) tracked the
change of amplitudes of the test shape every 10 ms within each trial
while Experiment 2(a) did not include that function. Experiment 2(c)
and Experiment 2(b) were almost identical. Experiment 2(c) was ran
to examine if the results obtained from Experiment 2(b) at Purdue
can be replicated at NCKU. 

\subsubsection*{Experiment 2(a)}

For each fixed reference shape, amplitude 1 ($\alpha$) was randomly
selected from an interval {[}-30 px, 30 px{]} and amplitude 2 ($\beta$)
was selected from the same interval. We ran 1890 trials for each participant
divided equally in nine sessions. 

\subsubsection*{Experiment 2(b)}

This experiment was identical to Experiment 2(a) except the program
recorded the amplitudes of the test shape that was being modified
every 10 ms in each trial. 

\subsubsection*{Experiment 2(c)}

This experiment contained four sessions, each containing 10 practice
trials and 200 main trials. In all the other aspects, it was identical
to Experiment 2(b). 

\part*{Results}

There were two parameters defined the reference stimulus (denoted
as $\alpha$ and $\beta$ ) and two parameters defined the test stimulus
(denoted as $A$ and $B$). Table \ref{tab: Factors--and} presents
what these parameters stand for. We expected in both experiments,
the subjects implemented the parallel AND, serial AND, or coactive
manner to adjust $A$ and $B$. The stopping rule OR should not be
used as in the experiments both features ($A$ and $B$) of the test
stimulus had to match the features ($\alpha$ and $\beta$ ) of the
reference stimulus. 

\begin{singlespace}
\begin{table}[H]
\caption{Parameters $(\alpha,\beta)$ of the reference stimuli and parameters
$(A,B)$ of the test stimuli.\label{tab: Factors--and}}
\centering{}%
\begin{tabular}{ccccc}
\hline 
Task & $\alpha$ & $\beta$ & $A$ & $B$\tabularnewline
\hline 
Dot position & Horizontal  & Vertical & Horizontal  & Vertical\tabularnewline
reproduction & coordinate & coordinate & coordinate of & coordinate of\tabularnewline
 & of the  & of the & the test & the test\tabularnewline
 & reference dot & reference dot & dot & dot\tabularnewline
 &  &  &  & \tabularnewline
Floral shape  & Amplitude 1 & Amplitude 2 & Amplitude 1  & Amplitude 2\tabularnewline
reproduction & of the & of the & of the  & of the\tabularnewline
 & reference shape & reference shape & test & test\tabularnewline
 &  &  & shape & shape\tabularnewline
\hline 
\end{tabular}
\end{table}

\end{singlespace}

\subsection*{Dot Position Reproduction Task}

\subsubsection*{Trackball Movements}

\begin{figure}[H]
\begin{centering}
\includegraphics[scale=0.5]{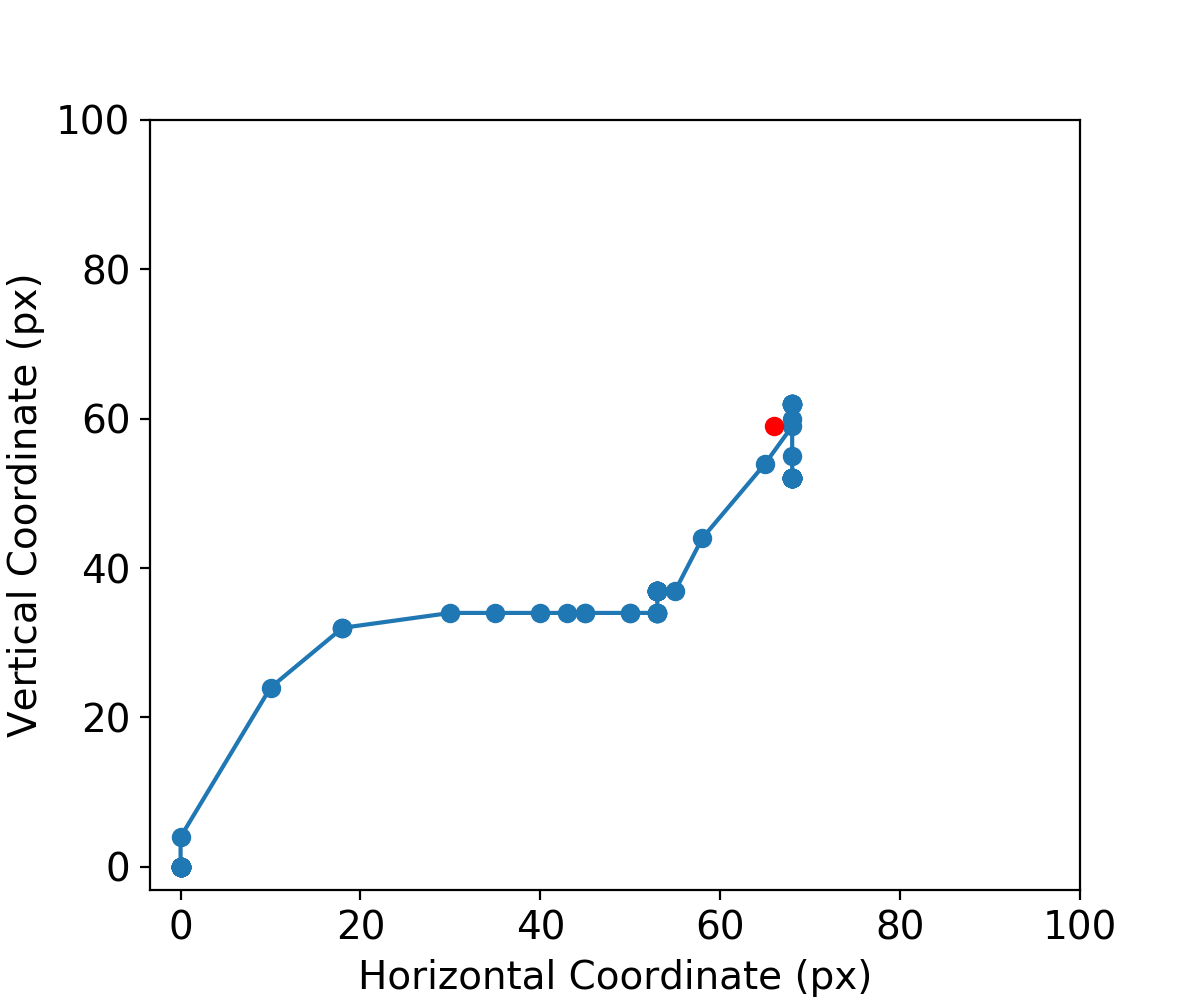}
\par\end{centering}
\caption{Trackball movements in a typical trial in the dot position reproduction
task (plotted every 50ms). }

\label{trackball move dot}
\end{figure}

Figure \ref{trackball move dot} shows the trackball movements in
a typical trial in the dot position reproduction task. The trajectory
of the trackball movements confirmed the third assumption of SFT that
the subject adheres a single type of mental architecture from trial
to trial. The red dot represents the location of the fixed reference
dot. The test dot (blue) started from (0 px, 0 px) and after a sequence
of movements for the horizontal coordinate and the vertical coordinate,
the final location was very close to the target location indicating
the coordinates were not adjusted in the parallel OR or serial OR
or serial AND manner: If the stopping rule OR was used, one should
expect the final location of the test dot aligned well with the target
either horizontally or vertically but not both. If serial AND was
used in the task, the two coordinates should not move simultaneously
as observed in Figure \ref{trackball move dot}. The trajectory implies
parallel AND or coactive were used by the subjects in the task. However
the trajectory is not able to differentiate parallel AND from coactive. 

\subsubsection*{Testing Selective Influence}

Selective influence of $\alpha$ and $\beta$ on $T_{\alpha}$ and
$T_{\beta}$ has ideally to be tested before SFT is implemented. In
our experimental paradigm, channel $\alpha$ and channel $\beta$
were characterized by two properties. One was the physical parameters
of the response, i.e., $A$ and $B$, and the other was the durations
for the channels, i.e., $T_{\alpha}$ and $T_{\beta}$. We speculated
that $(A,B)\looparrowleft(\alpha,\beta)$ is a sufficient (and perhaps
also necessary) condition for $(T_{\alpha},T_{\beta})\looparrowleft(\alpha,\beta)$.
If this speculation is accepted, we can test $(T_{\alpha},T_{\beta})\looparrowleft(\alpha,\beta)$
by inspecting whether $(A,B)\looparrowleft(\alpha,\beta)$. Please
be aware we used the finalized coordinates of $A$ and $B$ rather
than the intermediate coordinates while the dot was being moved for
the test of $(A,B)\looparrowleft(\alpha,\beta)$.

The outliers of $A$ and $B$ were handled in this way: We computed
$A-\alpha$ and $B-\beta$ for each trial. Any trial that was out
of 3 standard deviations of the set of $A-\alpha$ or $B-\beta$ was
considered as an outlier and was removed from further analysis. In
order to test selective influence and implement SFT, both $\alpha$
and $\beta$ should have discrete levels. A two by two factorial design
can be achieved in Experiment 1 if one splits $\alpha$ and $\beta$
with respect to 50 px. Of course other values can be chosen to make
the discretization. Here we use 50 px as an example. We label interval
{[}20 px, 50 px) one level and {[}50 px, 80 px{]} another. Table \ref{AB_dot}
presents the corresponding means and standard deviations of $A$ and
$B$ conditional on different combinations of $\alpha$ and $\beta$
for Experiment 1.

\begin{singlespace}
\begin{table}[H]
\caption{Means and standard deviations of the finalized horizontal coordinates
and the finalized vertical coordinates of the test dots for Experiment
1. }
\begin{centering}
\begin{tabular}{ccc}
\hline 
{\scriptsize{}Exp.} & {\scriptsize{}Subject} & {\tiny{}$([20\,\mathrm{px},50\,\mathrm{px}),[20\,\mathrm{px},50\,\mathrm{px}))$,
$([20\,\mathrm{px},50\,\mathrm{px}),[50\,\mathrm{px},80\,\mathrm{px}])$,
$([50\,\mathrm{px},80\,\mathrm{px}],[20\,\mathrm{px},50\,\mathrm{px}))$,
$([50\,\mathrm{px},80\,\mathrm{px}],[50\,\mathrm{px},80\,\mathrm{px}])$ }\tabularnewline
\hline 
{\scriptsize{}1(a)} & {\scriptsize{}S1} & {\tiny{}$(37.10\pm9.06,33.81\pm9.44)$, $(37.68\pm9.78,61.59\pm9.32)$,
$(66.57\pm9.78,33.88\pm9.32)$, $(66.56\pm8.05,63.03\pm9.35)$}\tabularnewline
{\scriptsize{}1(a)} & {\scriptsize{}S2} & {\tiny{}$(34.85\pm9.75,31.03\pm10.25)$, $(34.41\pm10.34,61.37\pm10.15)$,
$(62.84\pm10.68,33.17\pm10.94)$, $(62.79\pm10.20,64.16\pm10.55)$}\tabularnewline
{\scriptsize{}1(a)} & {\scriptsize{}S3} & {\tiny{}$(42.30\pm11.65,36.10\pm10.78)$, $(38.09\pm10.89,68.75\pm10.07)$,
$(75.01\pm9.13,33.84\pm8.55)$, $(72.41\pm10.21,63.07\pm11.24)$}\tabularnewline
{\scriptsize{}1(b)} & {\scriptsize{}S7} & {\tiny{}$(37.55\pm13.02,34.98\pm14.28)$, $(35.18\pm11.35,64.02\pm10.11)$,
$(66.02\pm10.43,34.95\pm11.74)$, $(66.85\pm10.79,61.89\pm10.07)$}\tabularnewline
{\scriptsize{}1(b)} & {\scriptsize{}S8} & {\tiny{}$(30.58\pm12.54,44.69\pm13.09)$, $(25.25\pm11.97,68.99\pm11.63)$,
$(59.54\pm11.24,38.93\pm14.41)$, $(52.03\pm12.16,69.30\pm12.30)$}\tabularnewline
{\scriptsize{}1(b)} & {\scriptsize{}S9} & {\tiny{}$(32.58\pm12.03,39.93\pm12.32)$, $(30.85\pm11.89,66.42\pm10.22)$,
$(61.21\pm11.53,40.29\pm11.78)$, $(56.99\pm11.44,64.92\pm11.81)$}\tabularnewline
{\scriptsize{}1(b)} & {\scriptsize{}S10} & {\tiny{}$(31.58\pm9.89,43.92\pm11.55)$, $(31.03\pm9.24,66.72\pm8.30)$,
$(57.93\pm9.17,42.23\pm11.41)$, $(58.41\pm9.87,67.01\pm9.38)$}\tabularnewline
{\scriptsize{}1(b)} & {\scriptsize{}S11} & {\tiny{}$(33.77\pm13.87,32.83\pm13.96)$, $(34.71\pm14.18,63.51\pm11.72)$,
$(66.49\pm10.62,26.71\pm13.05)$, $(67.32\pm10.76,60.70\pm11.16)$}\tabularnewline
{\scriptsize{}1(c)} & {\scriptsize{}S12} & {\tiny{}$(39.43\pm14.16,42.44\pm12.86)$, $(32.80\pm12.29,71.77\pm11.35)$,
$(68.43\pm9.90,35.12\pm13.31)$, $(60.75\pm11.11,67.63\pm12.69)$}\tabularnewline
{\scriptsize{}1(c)} & {\scriptsize{}S13} & {\tiny{}$(37.07\pm7.79,39.52\pm8.96)$, $(35.59\pm9.67,62.94\pm10.86)$,
$(58.81\pm8.06,39.13\pm10.27)$, $(61.04\pm9.04,65.60\pm10.75)$}\tabularnewline
{\scriptsize{}1(c)} & {\scriptsize{}S14} & {\tiny{}$(40.20\pm10.03,36.52\pm11.65)$, $(38.82\pm10.43,61.83\pm10.94)$,
$(64.83\pm9.37,35.00\pm10.45)$, $(66.54\pm9.77,61.38\pm8.17)$}\tabularnewline
{\scriptsize{}1(c)} & {\scriptsize{}S15} & {\tiny{}$(42.43\pm10.26,46.76\pm9.87)$, $(38.66\pm10.08,73.33\pm11.93)$,
$(69.78\pm10.47,44.11\pm8.95)$, $(67.53\pm9.37,69.77\pm9.94)$}\tabularnewline
{\scriptsize{}1(c)} & {\scriptsize{}S16} & {\tiny{}$(42.76\pm8.99,43.38\pm10.01)$, $(39.62\pm10.17,68.13\pm9.98)$,
$(65.96\pm8.71,42.22\pm10.37)$, $(64.06\pm9.46,67.65\pm9.95)$}\tabularnewline
\hline 
\end{tabular}
\par\end{centering}
\label{AB_dot}
\end{table}

\end{singlespace}

We then conducted four two sample Kolmogorov\textendash Smirnov(KS)
tests for each subject to examine marginal selectivity (\ref{marginal selectivity}):
$T_{\alpha_{i}}$ and $T_{\beta_{j}}$ were replaced with $A_{i}$
and $B_{j}$ that stand for the coordinates of the test dot conditional
on the reference dot $\alpha_{i}\beta_{j},i,j\in\left\{ 1,2\right\} $.
Table 5 presents the statistics for the tests. Each column of numbers
represents a particular paired comparison for the subjects. For instance,
$([20\,\mathrm{px},50\,\mathrm{px}),)$ compared the $A\mathrm{s}$
across different levels of $\beta$ but fixed $\alpha=[20\,\mathrm{px},50\,\mathrm{px})$.
$(,[50\,\mathrm{px},80\,\mathrm{px}])$ compared the $B\mathrm{s}$
across different levels of $\alpha$ but fixed $\beta=[50\,\mathrm{px},80\,\mathrm{px}]$.
We conclude that marginal selectivity was confirmed for S1, S7, S10,
S13, and S14 ($\mathrm{alpha}=.05/4$ for the Bonferroni adjustment). 

\begin{table}[H]
\begin{raggedright}
\caption{Two sample KS tests for marginal selectivity for Experiment 1. \label{test marginal selecitivity_dot}}
\par\end{raggedright}
\begin{centering}
\begin{tabular}{ccccccc}
\hline 
\multirow{2}{*}{{\scriptsize{}Exp.}} & \multirow{2}{*}{{\scriptsize{}Subject}} & \multirow{2}{*}{{\scriptsize{}$([20\,\mathrm{px},50\,\mathrm{px}),)$}} & \multirow{2}{*}{{\scriptsize{}$([50\,\mathrm{px},80\,\mathrm{px}],)$}} & \multirow{2}{*}{{\scriptsize{}$(,[20\,\mathrm{px},50\,\mathrm{px}))$}} & \multirow{2}{*}{{\scriptsize{}$(,[50\,\mathrm{px},80\,\mathrm{px}])$}} & {\scriptsize{}Marginal }\tabularnewline
 &  &  &  &  &  & {\scriptsize{}selectivity?}\tabularnewline
\hline 
1(a) & S1 & .046(.697) & .061(.397) & .048(.672) & .066(.272) & Yes\tabularnewline
1(a) & S2  & .050(.616) & .055(.514) & .102(.017) & .107(.011) & No\tabularnewline
1(a) & S3 & .156(.000) & .122(.003) & .125(.002) & .275(.000) & No\tabularnewline
1(b) & S7 & .138(.049) & .067(.756) & .070(.696) & .106(.230) & Yes\tabularnewline
1(b) & S8 & .189(.001) & .306(.000) & .180(.002) & .058(.927) & No\tabularnewline
1(b) & S9 & .095(.304) & .196(.001) & .078(.546) & .145(.037) & No\tabularnewline
1(b) & S10 & .081(.467) & .051(.980) & .099(.310) & .057(.917) & Yes\tabularnewline
1(b) & S11 & .196(.001) & .099(.274) & .076(.608) & .058(.892) & No\tabularnewline
1(c) & S12 & .252(.000) & .326(.000) & .215(.000) & .200(.001) & No\tabularnewline
1(c) & S13 & .140(.037) & .140(.038) & .077(.550) & .143(.039) & Yes\tabularnewline
1(c) & S14 & .100(.299) & .139(.038) & .090(.361) & .103(.276) & Yes\tabularnewline
1(c) & S15 & .182(.002) & .131(.073) & .171(.007) & .199(.001) & No\tabularnewline
1(c) & S16 & .187(.002) & .110(.168) & .079(.524) & .063(.840) & No\tabularnewline
\hline 
\end{tabular}
\par\end{centering}
Note: Each number outside of the brackets is the KS statistic value
and each number in the brackets is the $p$ value.
\end{table}

For those who passed the test of marginal selectivity, we investigated
if selective influence secured by conducting LFT. LFT is applicable
only if $A$ and $B$ are discrete. One can choose any value to create
two levels for $A$ and $B$ or discretize $A$ and $B$ into multiple
levels. For example, we created two levels for $A$: \{smaller than
or equal to 50 px, larger than 50 px\}, labeled as $\left\{ a_{1},a_{2}\right\} $,
and two levels for $B$: \{smaller than or equal to 50 px, larger
than 50 px\}, labeled as $\left\{ b_{1},b_{2}\right\} $. The numbers
in the cells of Table \ref{joint prob_dot} are the joint probabilities
for the discretized $(A_{ij},\:B_{ij}),i,j\in\left\{ 1,2\right\} $
and the numbers outside are the marginal probabilities. 

\begin{table}[H]
\caption{Joint distributions for the discretized $(A_{ij},\:B_{ij}),i,j\in\left\{ 1,2\right\} $
for S1, S7, S10, S13, and S14. S1 participated in Experiment 1(a).
S7 and S10 participated in Experiment 1(b). S13 and S14 participated
in Experiment 1(c).}
\begin{centering}
\begin{tabular}{|c|c|c|c|c|c|c|c|}
\hline 
\multicolumn{8}{|c|}{S1}\tabularnewline
\hline 
$([20\,\mathrm{px},50\,\mathrm{px}),$ & \multirow{2}{*}{$B=b_{1}$} & \multirow{2}{*}{$B=b_{2}$} &  & $([20\,\mathrm{px},50\,\mathrm{px}),$ & \multirow{2}{*}{$B=b_{1}$} & \multirow{2}{*}{$B=b_{2}$} & \multirow{2}{*}{}\tabularnewline
$[20\,\mathrm{px},50\,\mathrm{px}))$ &  &  &  & $[50\,\mathrm{px},80\,\mathrm{px}])$ &  &  & \tabularnewline
\cline{1-3} \cline{5-7} 
$A=a_{1}$ & .8811 & .0419 & .9230 & $A=a_{1}$ & .1361 & .7711 & .9072\tabularnewline
\cline{1-3} \cline{5-7} 
$A=a_{2}$ & .0749 & .0022 & .0771 & $A=a_{2}$ & .0173 & .0756 & .0928\tabularnewline
\cline{1-3} \cline{5-7} 
\multicolumn{1}{|c}{} & \multicolumn{1}{c}{.9560} & \multicolumn{1}{c}{.0441} & \multicolumn{1}{c}{} & \multicolumn{1}{c}{} & \multicolumn{1}{c}{.1534} & \multicolumn{1}{c}{.8467} & \tabularnewline
\cline{1-3} \cline{5-7} 
$([50\,\mathrm{px},80\,\mathrm{px}],$ & \multirow{2}{*}{$B=b_{1}$} & \multirow{2}{*}{$B=b_{2}$} &  & $([50\,\mathrm{px},80\,\mathrm{px}],$ & \multirow{2}{*}{$B=b_{1}$} & \multirow{2}{*}{$B=b_{2}$} & \tabularnewline
$[20\,\mathrm{px},50\,\mathrm{px}))$ &  &  &  & $[50\,\mathrm{px},80\,\mathrm{px}])$ &  &  & \tabularnewline
\cline{1-3} \cline{5-7} 
$A=a_{1}$ & .0590 & .0024 & .0614 & $A=a_{1}$ & .0069 & .0161 & .0230\tabularnewline
\cline{1-3} \cline{5-7} 
$A=a_{2}$ & .9127 & .0259 & .9386 & $A=a_{2}$ & .1034 & .8736 & .9770\tabularnewline
\cline{1-3} \cline{5-7} 
\multicolumn{1}{|c}{} & \multicolumn{1}{c}{.9717} & \multicolumn{1}{c}{.0283} & \multicolumn{1}{c}{} & \multicolumn{1}{c}{} & \multicolumn{1}{c}{.1103} & \multicolumn{1}{c}{.8897} & \tabularnewline
\hline 
\end{tabular}
\par\end{centering}
\begin{centering}
\begin{tabular}{|c|c|c|c|c|c|c|c|}
\hline 
\multicolumn{8}{|c|}{S7}\tabularnewline
\hline 
$([20\,\mathrm{px},50\,\mathrm{px}),$ & \multirow{2}{*}{$B=b_{1}$} & \multirow{2}{*}{$B=b_{2}$} &  & $([20\,\mathrm{px},50\,\mathrm{px}),$ & \multirow{2}{*}{$B=b_{1}$} & \multirow{2}{*}{$B=b_{2}$} & \multirow{2}{*}{}\tabularnewline
$[20\,\mathrm{px},50\,\mathrm{px}))$ &  &  &  & $[50\,\mathrm{px},80\,\mathrm{px}])$ &  &  & \tabularnewline
\cline{1-3} \cline{5-7} 
$A=a_{1}$ & .7122 & .1220 & .8342 & $A=a_{1}$ & .0904 & .8079 & .8983\tabularnewline
\cline{1-3} \cline{5-7} 
$A=a_{2}$ & .1610 & .0049 & .1659 & $A=a_{2}$ & .0113 & .0904 & .1017\tabularnewline
\cline{1-3} \cline{5-7} 
\multicolumn{1}{|c}{} & \multicolumn{1}{c}{.8732} & \multicolumn{1}{c}{.1269} & \multicolumn{1}{c}{} & \multicolumn{1}{c}{} & \multicolumn{1}{c}{.1017} & \multicolumn{1}{c}{.8983} & \tabularnewline
\cline{1-3} \cline{5-7} 
$([50\,\mathrm{px},80\,\mathrm{px}],$ & \multirow{2}{*}{$B=b_{1}$} & \multirow{2}{*}{$B=b_{2}$} &  & $([50\,\mathrm{px},80\,\mathrm{px}],$ & \multirow{2}{*}{$B=b_{1}$} & \multirow{2}{*}{$B=b_{2}$} & \tabularnewline
$[20\,\mathrm{px},50\,\mathrm{px}))$ &  &  &  & $[50\,\mathrm{px},80\,\mathrm{px}])$ &  &  & \tabularnewline
\cline{1-3} \cline{5-7} 
$A=a_{1}$ & .0769 & .0103 & .0872 & $A=a_{1}$ & .015 & .06 & .075\tabularnewline
\cline{1-3} \cline{5-7} 
$A=a_{2}$ & .8308 & .0820 & .9128 & $A=a_{2}$ & .135 & .79 & .927\tabularnewline
\cline{1-3} \cline{5-7} 
\multicolumn{1}{|c}{} & \multicolumn{1}{c}{.9077} & \multicolumn{1}{c}{.0923} & \multicolumn{1}{c}{} & \multicolumn{1}{c}{} & \multicolumn{1}{c}{.150} & \multicolumn{1}{c}{.85} & \tabularnewline
\hline 
\end{tabular}
\par\end{centering}
\begin{centering}
\begin{tabular}{|c|c|c|c|c|c|c|c|}
\hline 
\multicolumn{8}{|c|}{S10}\tabularnewline
\hline 
$([20\,\mathrm{px},50\,\mathrm{px}),$ & \multirow{2}{*}{$B=b_{1}$} & \multirow{2}{*}{$B=b_{2}$} &  & $([20\,\mathrm{px},50\,\mathrm{px}),$ & \multirow{2}{*}{$B=b_{1}$} & \multirow{2}{*}{$B=b_{2}$} & \multirow{2}{*}{}\tabularnewline
$[20\,\mathrm{px},50\,\mathrm{px}))$ &  &  &  & $[50\,\mathrm{px},80\,\mathrm{px}])$ &  &  & \tabularnewline
\cline{1-3} \cline{5-7} 
$A=a_{1}$ & .6866 & .2935 & .9801 & $A=a_{1}$ & .0223 & .9688 & .9911\tabularnewline
\cline{1-3} \cline{5-7} 
$A=a_{2}$ & .0199 & 0 & .0199 & $A=a_{2}$ & 0 & .0089 & .0089\tabularnewline
\cline{1-3} \cline{5-7} 
\multicolumn{1}{|c}{} & \multicolumn{1}{c}{.7065} & \multicolumn{1}{c}{.2935} & \multicolumn{1}{c}{} & \multicolumn{1}{c}{} & \multicolumn{1}{c}{.0223} & \multicolumn{1}{c}{.9777} & \tabularnewline
\cline{1-3} \cline{5-7} 
$([50\,\mathrm{px},80\,\mathrm{px}],$ & \multirow{2}{*}{$B=b_{1}$} & \multirow{2}{*}{$B=b_{2}$} &  & $([50\,\mathrm{px},80\,\mathrm{px}],$ & \multirow{2}{*}{$B=b_{1}$} & \multirow{2}{*}{$B=b_{2}$} & \tabularnewline
$[20\,\mathrm{px},50\,\mathrm{px}))$ &  &  &  & $[50\,\mathrm{px},80\,\mathrm{px}])$ &  &  & \tabularnewline
\cline{1-3} \cline{5-7} 
$A=a_{1}$ & .1618 & .0636 & .2254 & $A=a_{1}$ & .0126 & .2075 & .2201\tabularnewline
\cline{1-3} \cline{5-7} 
$A=a_{2}$ & .5780 & .1965 & .7745 & $A=a_{2}$ & .0503 & .7296 & .7799\tabularnewline
\cline{1-3} \cline{5-7} 
\multicolumn{1}{|c}{} & \multicolumn{1}{c}{.7398} & \multicolumn{1}{c}{.2601} & \multicolumn{1}{c}{} & \multicolumn{1}{c}{} & \multicolumn{1}{c}{.0629} & \multicolumn{1}{c}{.9371} & \tabularnewline
\hline 
\end{tabular}
\par\end{centering}
\begin{centering}
\begin{tabular}{|c|c|c|c|c|c|c|c|}
\hline 
\multicolumn{8}{|c|}{S13}\tabularnewline
\hline 
$([20\,\mathrm{px},50\,\mathrm{px}),$ & \multirow{2}{*}{$B=b_{1}$} & \multirow{2}{*}{$B=b_{2}$} &  & $([20\,\mathrm{px},50\,\mathrm{px}),$ & \multirow{2}{*}{$B=b_{1}$} & \multirow{2}{*}{$B=b_{2}$} & \multirow{2}{*}{}\tabularnewline
$[20\,\mathrm{px},50\,\mathrm{px}))$ &  &  &  & $[50\,\mathrm{px},80\,\mathrm{px}])$ &  &  & \tabularnewline
\cline{1-3} \cline{5-7} 
$A=a_{1}$ & .8357 & .1063 & .9420 & $A=a_{1}$ & .1105 & .8368 & .9473\tabularnewline
\cline{1-3} \cline{5-7} 
$A=a_{2}$ & .0531 & .0048 & .0579 & $A=a_{2}$ & .0211 & .0316 & .0527\tabularnewline
\cline{1-3} \cline{5-7} 
\multicolumn{1}{|c}{} & \multicolumn{1}{c}{.8888} & \multicolumn{1}{c}{.1111} & \multicolumn{1}{c}{} & \multicolumn{1}{c}{} & \multicolumn{1}{c}{.1316} & \multicolumn{1}{c}{.8684} & \tabularnewline
\cline{1-3} \cline{5-7} 
$([50\,\mathrm{px},80\,\mathrm{px}],$ & \multirow{2}{*}{$B=b_{1}$} & \multirow{2}{*}{$B=b_{2}$} &  & $([50\,\mathrm{px},80\,\mathrm{px}],$ & \multirow{2}{*}{$B=b_{1}$} & \multirow{2}{*}{$B=b_{2}$} & \tabularnewline
$[20\,\mathrm{px},50\,\mathrm{px}))$ &  &  &  & $[50\,\mathrm{px},80\,\mathrm{px}])$ &  &  & \tabularnewline
\cline{1-3} \cline{5-7} 
$A=a_{1}$ & .1014 & .0242 & .1256 & $A=a_{1}$ & .0107 & .1123 & .1230\tabularnewline
\cline{1-3} \cline{5-7} 
$A=a_{2}$ & .7488 & .1256 & .8744 & $A=a_{2}$ & .0963 & .7807 & .8770\tabularnewline
\cline{1-3} \cline{5-7} 
\multicolumn{1}{|c}{} & \multicolumn{1}{c}{.8502} & \multicolumn{1}{c}{.1498} & \multicolumn{1}{c}{} & \multicolumn{1}{c}{} & \multicolumn{1}{c}{.1070} & \multicolumn{1}{c}{.8930} & \tabularnewline
\hline 
\end{tabular}
\par\end{centering}
\label{joint prob_dot}
\end{table}

\begin{table}[H]
\begin{raggedright}
Table 6: Joint distributions for the discretized $(A_{ij},\:B_{ij}),i,j\in\left\{ 1,2\right\} $
for S1, S7, S10, S13, and S14 (continued). S1 participated in Experiment
1(a). S7 and S10 participated in Experiment 1(b). S13 and S14 participated
in Experiment 1(c).
\par\end{raggedright}
\begin{centering}
\begin{tabular}{|c|c|c|c|c|c|c|c|}
\hline 
\multicolumn{8}{|c|}{S14}\tabularnewline
\hline 
$([20\,\mathrm{px},50\,\mathrm{px}),$ & \multirow{2}{*}{$B=b_{1}$} & \multirow{2}{*}{$B=b_{2}$} &  & $([20\,\mathrm{px},50\,\mathrm{px}),$ & \multirow{2}{*}{$B=b_{1}$} & \multirow{2}{*}{$B=b_{2}$} & \multirow{2}{*}{}\tabularnewline
$[20\,\mathrm{px},50\,\mathrm{px}))$ &  &  &  & $[50\,\mathrm{px},80\,\mathrm{px}])$ &  &  & \tabularnewline
\cline{1-3} \cline{5-7} 
$A=a_{1}$ & .7104 & .1257 & .8361 & $A=a_{1}$ & .1064 & .7394 & .8458\tabularnewline
\cline{1-3} \cline{5-7} 
$A=a_{2}$ & .1530 & .0109 & .1639 & $A=a_{2}$ & .0426 & .1117 & .1543\tabularnewline
\cline{1-3} \cline{5-7} 
\multicolumn{1}{|c}{} & \multicolumn{1}{c}{.8634} & \multicolumn{1}{c}{.1366} & \multicolumn{1}{c}{} & \multicolumn{1}{c}{} & \multicolumn{1}{c}{.1490} & \multicolumn{1}{c}{.8511} & \tabularnewline
\cline{1-3} \cline{5-7} 
$([50\,\mathrm{px},80\,\mathrm{px}],$ & \multirow{2}{*}{$B=b_{1}$} & \multirow{2}{*}{$B=b_{2}$} &  & $([50\,\mathrm{px},80\,\mathrm{px}],$ & \multirow{2}{*}{$B=b_{1}$} & \multirow{2}{*}{$B=b_{2}$} & \tabularnewline
$[20\,\mathrm{px},50\,\mathrm{px}))$ &  &  &  & $[50\,\mathrm{px},80\,\mathrm{px}])$ &  &  & \tabularnewline
\cline{1-3} \cline{5-7} 
$A=a_{1}$ & .0593 & .0085 & .0678 & $A=a_{1}$ & 0 & .0511 & .0511\tabularnewline
\cline{1-3} \cline{5-7} 
$A=a_{2}$ & .8517 & .0805 & .9322 & $A=a_{2}$ & .1080 & .8409 & .9489\tabularnewline
\cline{1-3} \cline{5-7} 
\multicolumn{1}{|c}{} & \multicolumn{1}{c}{.9110} & \multicolumn{1}{c}{.0890} & \multicolumn{1}{c}{} & \multicolumn{1}{c}{} & \multicolumn{1}{c}{.1080} & \multicolumn{1}{c}{.8920} & \tabularnewline
\hline 
\end{tabular}
\par\end{centering}
\label{joint prob_dot-1}
\end{table}

We observe the marginal probabilities not exactly equal across conditions,
for instance {\scriptsize{}
\begin{alignat*}{1}
\mathrm{Pr}(A & =a_{1})\mid_{([20\,\mathrm{px},50\,\mathrm{px}),[20\,\mathrm{px},50\,\mathrm{px}))}=.9230\neq\mathrm{Pr}(A=a_{1})\mid_{([20\,\mathrm{px},50\,\mathrm{px}),[50\,\mathrm{px},80\,\mathrm{px}])}=.9072,\\
\mathrm{Pr}(A & =a_{1})\mid_{([50\,\mathrm{px},80\,\mathrm{px}],[20\,\mathrm{px},50\,\mathrm{px}))}=.0614\neq\mathrm{Pr}(A=a_{1})\mid_{([50\,\mathrm{px},80\,\mathrm{px}],[50\,\mathrm{px},80\,\mathrm{px}])}=.0230,\\
\mathrm{Pr}(B & =b_{1})\mid_{([20\,\mathrm{px},50\,\mathrm{px}),[20\,\mathrm{px},50\,\mathrm{px}))}=.9560\neq\mathrm{Pr}(B=b_{1})\mid_{([50\,\mathrm{px},80\,\mathrm{px}],[20\,\mathrm{px},50\,\mathrm{px}))}=.9717,\\
\mathrm{Pr}(B & =b_{1})\mid_{([20\,\mathrm{px},50\,\mathrm{px}),[50\,\mathrm{px},80\,\mathrm{px}])}=.1534\neq\mathrm{Pr}(B=b_{1})\mid_{([50\,\mathrm{px},80\,\mathrm{px}],[50\,\mathrm{px},80\,\mathrm{px}])}=.1103,
\end{alignat*}
}for S1. We considered .9230 and .9072, .0614 and .0230, .9560 and
.9717, and .1534 and .1103 statistically equal as marginal selectivity
was established statistically for that subject (Table \ref{test marginal selecitivity_dot}). 

In order to implement LFT, one requirement is (\ref{marginal selectivity})
has to be strictly hold. In order to fulfill this requirement, we
forced {\scriptsize{}
\begin{align*}
\mathrm{Pr}(A & =a_{1})\mid_{([20\,\mathrm{px},50\,\mathrm{px}),[20\,\mathrm{px},50\,\mathrm{px}))}=\mathrm{Pr}(A=a_{1})\mid_{([20\,\mathrm{px},50\,\mathrm{px}),[50\,\mathrm{px},80\,\mathrm{px}])}=(.9230+.9072)/2=.9151,\\
\mathrm{Pr}(A & =a_{2})\mid_{([20\,\mathrm{px},50\,\mathrm{px}),[20\,\mathrm{px},50\,\mathrm{px}))}=\mathrm{Pr}(A=a_{2})\mid_{([20\,\mathrm{px},50\,\mathrm{px}),[50\,\mathrm{px},80\,\mathrm{px}])}=(.0771+.0928)/2=.0849,\\
\mathrm{Pr}(B & =b_{1})\mid_{([20\,\mathrm{px},50\,\mathrm{px}),[20\,\mathrm{px},50\,\mathrm{px}))}=\mathrm{Pr}(B=b_{1})\mid_{([50\,\mathrm{px},80\,\mathrm{px}],[20\,\mathrm{px},50\,\mathrm{px}))}=(.9560+.9717)/2=.9638,\\
\mathrm{Pr}(B & =b_{2})\mid_{([20\,\mathrm{px},50\,\mathrm{px}),[20\,\mathrm{px},50\,\mathrm{px}))}=\mathrm{Pr}(B=b_{2})\mid_{([50\,\mathrm{px},80\,\mathrm{px}],[20\,\mathrm{px},50\,\mathrm{px}))}=(.0441+.0283)/2=.0362,
\end{align*}
}and averaged the other marginal probability pairs in the same way.
The joint probabilities in the cells of Table \ref{joint prob_dot}
were modified by keeping the value of $\mathrm{Pr}(A=a_{1},B=b_{1})$
for each stimulus and changing values for $P(A=a_{1},B=b_{2})$, $P(A=a_{2},B=b_{1})$,
and $P(A=a_{2},B=b_{2})$ according to the change of the corresponding
marginal probabilities. For instance, for S1 for stimulus $([20\,\mathrm{px},50\,\mathrm{px}),[20\,\mathrm{px},50\,\mathrm{px}))$,
the modified joint probabilities are

\begin{flalign*}
\mathrm{Pr}(A & =a_{1},B=b_{1})=.8811,\\
\mathrm{Pr}(A & =a_{1},B=b_{2})=.9151-.8811=.0340,\\
\mathrm{Pr}(A & =a_{2},B=b_{1})=.9638-.8811=.0827,\\
\mathrm{Pr}(A & =a_{2},B=b_{2})=.0849-.0827=.0022.
\end{flalign*}

LFT was passed as we were able to find nonnegative solutions for LFT
(Table \ref{LFT_dot}). It indicates selective influence of $\alpha$
and $\beta$ on $A$ and $B$ was established for these subjects.
We found LFT passed for all the other ways of discretization that
we tried for $A$ and $B$. We then considered $(T_{\alpha},T_{\beta})\looparrowleft(\alpha,\beta)$
was successfully established for S1, S7, S10, S13, and S14.

\begin{table}[h]
\caption{Solutions for LFT for S1, S7, S10, S13, and S14 in Experiment 1. S1
participated in Experiment 1(a). S7 and S10 participated in Experiment
1(b). S13 and S14 participated in Experiment 1(c).}
\begin{centering}
\begin{tabular}{cc}
\hline 
Subject & $(Q_{a_{1}a_{1}b_{1}b_{1}},\,Q_{a_{1}a_{1}b_{1}b_{2}},\,\ldots,\,Q_{a_{2}a_{2}b_{2}b_{2}})^{T}$\tabularnewline
\hline 
S1 & $(.0069,0,0,0,.0909,.7833,.034,0,0,.0353,0,0,0,.0474,0,.0022)^{T}$\tabularnewline
S7 & $(0,0,0,.0041,0,.7616,.0904,.0101,.015,.0619,0,0,.0204,.0365,0,0)^{T}$\tabularnewline
S10 & $(.0126,.1348,0,.0609,0,.5614,.03,.1859,0,.0144,0,0,0,0,0,0)^{T}$\tabularnewline
S13 & $(.0019,.0657,0,.0013,.001,.7671,.1076,0,0,.0338,.0088,.0128,0,0,0,0)^{T}$\tabularnewline
\multirow{1}{*}{S14} & $(0,0,0,.0001,0,.7281,.1064,.0063,0,.0593,0,0,.0221,.0777,0,0)^{T}$\tabularnewline
\hline 
\end{tabular}
\par\end{centering}
\label{LFT_dot}
\end{table}

\subsubsection*{Testing Stochastic Dominance}

For each participant, we considered the trial with RT outside of 5
standard deviations of the entire set of RTs an outlier and it was
not included in the further analysis. In the earlier part of this
article, we created two levels for $\alpha$ and $\beta$, {[}20 px,
50 px) one level and {[}50 px, 80 px{]} another. In order to test
stochastic dominance, what exactly level 1 and level 2 $\alpha$ and
$\beta$ stand for should be identified. We observed some subjects
spent more time when the target location was far from the center of
the circle and others spent more time to move to a location closer
to the center as they had to be more careful with their action. The
exact assignment of level 1 and level 2 for each subject can be found
in Table 8. The assignment was based on the observation of the dataset.
The interval range that was processed more slowly was labeled level
one and the other range was labeled level two.

\begin{table}[H]
\begin{raggedright}
\caption{The assignment of level 1 and level 2 for Experiment 1.}
\par\end{raggedright}
\begin{centering}
\begin{tabular}{ccc}
\hline 
\multirow{2}{*}{Experiment} & Level 1: $[20\,\mathrm{px},\,50\,\mathrm{px})$ & Level 1: $[50\,\mathrm{px},\,80\,\mathrm{px}]$\tabularnewline
 & Level 2: $[50\,\mathrm{px},\,80\,\mathrm{px}]$ & Level 2: $[20\,\mathrm{px},\,50\,\mathrm{px})$\tabularnewline
\hline 
1(a) & S1, S2, S3 & \tabularnewline
1(b) &  & S7, S8, S9, S10, S11\tabularnewline
1(c) & S12, S14, S15, S16 & S13\tabularnewline
\hline 
\end{tabular}
\par\end{centering}
\label{level 1=0000262}
\end{table}

The left column of Figure \ref{Survival_1_dot} presents the survival
functions of RT for the subjects who passed the test of selective
influence. In order to test if those survival functions satisfy stochastic
dominance (\ref{eq:sd}), two one tail KS tests were performed on
each of the four paired variables. For instance, in order to test
the first inequality in (\ref{eq:sd}), we required the maximum of
$S_{\alpha_{1}\beta_{1}}(t)-S_{\alpha_{1}\beta_{2}}(t)$ larger than
or equal to 0 and the maximum of $S_{\alpha_{1}\beta_{2}}(t)-S_{\alpha_{1}\beta_{1}}(t)$
equal to zero. The statistical results (Table \ref{test stochastic dominance_dot_1})
support the assumption of stochastic dominance for subject S1, S7,
S10, S13, and S14 with the assignment presented in Table 8 as for
each subject the $p$ values in the bottom row were larger than the
critical value $\mathrm{alpha}=.05/4$. Note for S13 and S14, the
four survival functions of RT were not statistically different from
each other. 

The left column of Figure \ref{SIC_MIC_failed subjects} presents
the survival functions of RT for those subjects who did not pass the
test of selective influence. They also passed the test of stochastic
dominance ($\mathrm{alpha}=.05/4$, Table \ref{test stochastic dominance_failed subjects}).
Note that for subject S8, S15, and S16, the four survival functions
of RT were not statistically different from each other. 

\begin{table}[H]
\caption{$p$ values of the one tail KS tests for stochastic dominance for
S1, S7, S10, S13, and S14 in Experiment 1. S1 participated in Experiment
1(a). S7 and S10 participated in Experiment 1(b). S13 and S14 participated
in Experiment 1(c).}
\begin{centering}
\begin{tabular}{cccc}
\hline 
\multicolumn{4}{c}{S1}\tabularnewline
\hline 
$S_{\alpha_{1}\beta_{1}}>S_{\alpha_{1}\beta_{2}}$ & $S_{\alpha_{1}\beta_{1}}>S_{\alpha_{2}\beta_{1}}$ & $S_{\alpha_{1}\beta_{2}}>S_{\alpha_{2}\beta_{2}}$ & $S_{\alpha_{2}\beta_{1}}>S_{\alpha_{2}\beta_{2}}$\tabularnewline
.075(.077) & .047(.379) & .098(.014) & .113(.004)\tabularnewline
$S_{\alpha_{1}\beta_{1}}<S_{\alpha_{1}\beta_{2}}$ & $S_{\alpha_{1}\beta_{1}}<S_{\alpha_{2}\beta_{1}}$ & $S_{\alpha_{1}\beta_{2}}<S_{\alpha_{2}\beta_{2}}$ & $S_{\alpha_{2}\beta_{1}}<S_{\alpha_{2}\beta_{2}}$\tabularnewline
.048(.344) & .038(.531) & .007(.977) & .005(.990)\tabularnewline
\hline 
\multicolumn{4}{c}{S7}\tabularnewline
\hline 
$S_{\alpha_{1}\beta_{1}}>S_{\alpha_{1}\beta_{2}}$ & $S_{\alpha_{1}\beta_{1}}>S_{\alpha_{2}\beta_{1}}$ & $S_{\alpha_{1}\beta_{2}}>S_{\alpha_{2}\beta_{2}}$ & $S_{\alpha_{2}\beta_{1}}>S_{\alpha_{2}\beta_{2}}$\tabularnewline
.114(.076) & .132(.037) & .302(.000) & .289(.000)\tabularnewline
$S_{\alpha_{1}\beta_{1}}<S_{\alpha_{1}\beta_{2}}$ & $S_{\alpha_{1}\beta_{1}}<S_{\alpha_{2}\beta_{1}}$ & $S_{\alpha_{1}\beta_{2}}<S_{\alpha_{2}\beta_{2}}$ & $S_{\alpha_{2}\beta_{1}}<S_{\alpha_{2}\beta_{2}}$\tabularnewline
.007(.989) & .006(.993) & .015(.958) & .015(.960)\tabularnewline
\hline 
\multicolumn{4}{c}{S10}\tabularnewline
\hline 
$S_{\alpha_{1}\beta_{1}}>S_{\alpha_{1}\beta_{2}}$ & $S_{\alpha_{1}\beta_{1}}>S_{\alpha_{2}\beta_{1}}$ & $S_{\alpha_{1}\beta_{2}}>S_{\alpha_{2}\beta_{2}}$ & $S_{\alpha_{2}\beta_{1}}>S_{\alpha_{2}\beta_{2}}$\tabularnewline
.086(.299) & .143(.022) & .181(.002) & .099(.128)\tabularnewline
$S_{\alpha_{1}\beta_{1}}<S_{\alpha_{1}\beta_{2}}$ & $S_{\alpha_{1}\beta_{1}}<S_{\alpha_{2}\beta_{1}}$ & $S_{\alpha_{1}\beta_{2}}<S_{\alpha_{2}\beta_{2}}$ & $S_{\alpha_{2}\beta_{1}}<S_{\alpha_{2}\beta_{2}}$\tabularnewline
.024(.907) & .004(.996) & .015(.958) & .019(.927)\tabularnewline
\hline 
\multicolumn{4}{c}{S13}\tabularnewline
\hline 
$S_{\alpha_{1}\beta_{1}}>S_{\alpha_{1}\beta_{2}}$ & $S_{\alpha_{1}\beta_{1}}>S_{\alpha_{2}\beta_{1}}$ & $S_{\alpha_{1}\beta_{2}}>S_{\alpha_{2}\beta_{2}}$ & $S_{\alpha_{2}\beta_{1}}>S_{\alpha_{2}\beta_{2}}$\tabularnewline
.146(.015) & .062(.481) & .107(.094) & .139(.022)\tabularnewline
$S_{\alpha_{1}\beta_{1}}<S_{\alpha_{1}\beta_{2}}$ & $S_{\alpha_{1}\beta_{1}}<S_{\alpha_{2}\beta_{1}}$ & $S_{\alpha_{1}\beta_{2}}<S_{\alpha_{2}\beta_{2}}$ & $S_{\alpha_{2}\beta_{1}}<S_{\alpha_{2}\beta_{2}}$\tabularnewline
.016(.951) & .036(.783) & .073(.334) & .004(.996)\tabularnewline
\hline 
\multicolumn{4}{c}{S14}\tabularnewline
\hline 
$S_{\alpha_{1}\beta_{1}}>S_{\alpha_{1}\beta_{2}}$ & $S_{\alpha_{1}\beta_{1}}>S_{\alpha_{2}\beta_{1}}$ & $S_{\alpha_{1}\beta_{2}}>S_{\alpha_{2}\beta_{2}}$ & $S_{\alpha_{2}\beta_{1}}>S_{\alpha_{2}\beta_{2}}$\tabularnewline
.041(.736) & .064(.430) & .105(.135) & .039(.732)\tabularnewline
$S_{\alpha_{1}\beta_{1}}<S_{\alpha_{1}\beta_{2}}$ & $S_{\alpha_{1}\beta_{1}}<S_{\alpha_{2}\beta_{1}}$ & $S_{\alpha_{1}\beta_{2}}<S_{\alpha_{2}\beta_{2}}$ & $S_{\alpha_{2}\beta_{1}}<S_{\alpha_{2}\beta_{2}}$\tabularnewline
.095(.192) & .034(.795) & .012(.972) & .064(.439)\tabularnewline
\hline 
\end{tabular}
\par\end{centering}
Note: Each number outside of the brackets is the KS statistic value
and each number in the brackets is the $p$ value.

\label{test stochastic dominance_dot_1}
\end{table}

\begin{table}[H]
\caption{One tail KS tests for stochastic dominance for S2, S3, S8, S9, S11,
S12, S15, and S16 in Experiment 1. S2 and S3 participated in Experiment
1(a). S8, S9, and S11 participated in Experiment 1(b). S12, S15, and
S16 participated in Experiment 1(c).}
\begin{centering}
\begin{tabular}{cccc}
\hline 
\multicolumn{4}{c}{S2}\tabularnewline
\hline 
$S_{\alpha_{1}\beta_{1}}>S_{\alpha_{1}\beta_{2}}$ & $S_{\alpha_{1}\beta_{1}}>S_{\alpha_{2}\beta_{1}}$ & $S_{\alpha_{1}\beta_{2}}>S_{\alpha_{2}\beta_{2}}$ & $S_{\alpha_{2}\beta_{1}}>S_{\alpha_{2}\beta_{2}}$\tabularnewline
.132(.000) & .065(.153) & .040(.493) & .086(.040)\tabularnewline
$S_{\alpha_{1}\beta_{1}}<S_{\alpha_{1}\beta_{2}}$ & $S_{\alpha_{1}\beta_{1}}<S_{\alpha_{2}\beta_{1}}$ & $S_{\alpha_{1}\beta_{2}}<S_{\alpha_{2}\beta_{2}}$ & $S_{\alpha_{2}\beta_{1}}<S_{\alpha_{2}\beta_{2}}$\tabularnewline
.012(.936) & .026(.742) & .039(.513) & .023(.799)\tabularnewline
\hline 
\multicolumn{4}{c}{S3}\tabularnewline
\hline 
$S_{\alpha_{1}\beta_{1}}>S_{\alpha_{1}\beta_{2}}$ & $S_{\alpha_{1}\beta_{1}}>S_{\alpha_{2}\beta_{1}}$ & $S_{\alpha_{1}\beta_{2}}>S_{\alpha_{2}\beta_{2}}$ & $S_{\alpha_{2}\beta_{1}}>S_{\alpha_{2}\beta_{2}}$\tabularnewline
.160(.000) & .179(.000) & .052(.309) & .035(.595)\tabularnewline
$S_{\alpha_{1}\beta_{1}}<S_{\alpha_{1}\beta_{2}}$ & $S_{\alpha_{1}\beta_{1}}<S_{\alpha_{2}\beta_{1}}$ & $S_{\alpha_{1}\beta_{2}}<S_{\alpha_{2}\beta_{2}}$ & $S_{\alpha_{2}\beta_{1}}<S_{\alpha_{2}\beta_{2}}$\tabularnewline
.000(1.000) & .000(1.000) & .038(.543) & .062(.190)\tabularnewline
\hline 
\multicolumn{4}{c}{S8}\tabularnewline
\hline 
$S_{\alpha_{1}\beta_{1}}>S_{\alpha_{1}\beta_{2}}$ & $S_{\alpha_{1}\beta_{1}}>S_{\alpha_{2}\beta_{1}}$ & $S_{\alpha_{1}\beta_{2}}>S_{\alpha_{2}\beta_{2}}$ & $S_{\alpha_{2}\beta_{1}}>S_{\alpha_{2}\beta_{2}}$\tabularnewline
.058(.561) & .082(.312) & .099(.137) & .087(.218)\tabularnewline
$S_{\alpha_{1}\beta_{1}}<S_{\alpha_{1}\beta_{2}}$ & $S_{\alpha_{1}\beta_{1}}<S_{\alpha_{2}\beta_{1}}$ & $S_{\alpha_{1}\beta_{2}}<S_{\alpha_{2}\beta_{2}}$ & $S_{\alpha_{2}\beta_{1}}<S_{\alpha_{2}\beta_{2}}$\tabularnewline
.038(.786) & .062(.516) & .048(.628) & .024(.887)\tabularnewline
\hline 
\multicolumn{4}{c}{S9}\tabularnewline
\hline 
$S_{\alpha_{1}\beta_{1}}>S_{\alpha_{1}\beta_{2}}$ & $S_{\alpha_{1}\beta_{1}}>S_{\alpha_{2}\beta_{1}}$ & $S_{\alpha_{1}\beta_{2}}>S_{\alpha_{2}\beta_{2}}$ & $S_{\alpha_{2}\beta_{1}}>S_{\alpha_{2}\beta_{2}}$\tabularnewline
.041(.728) & .079(.315) & .163(.005) & .121(.053)\tabularnewline
$S_{\alpha_{1}\beta_{1}}<S_{\alpha_{1}\beta_{2}}$ & $S_{\alpha_{1}\beta_{1}}<S_{\alpha_{2}\beta_{1}}$ & $S_{\alpha_{1}\beta_{2}}<S_{\alpha_{2}\beta_{2}}$ & $S_{\alpha_{2}\beta_{1}}<S_{\alpha_{2}\beta_{2}}$\tabularnewline
.091(.209) & .029(.858) & .015(.957) & .040(.728)\tabularnewline
\hline 
\multicolumn{4}{c}{S11}\tabularnewline
\hline 
$S_{\alpha_{1}\beta_{1}}>S_{\alpha_{1}\beta_{2}}$ & $S_{\alpha_{1}\beta_{1}}>S_{\alpha_{2}\beta_{1}}$ & $S_{\alpha_{1}\beta_{2}}>S_{\alpha_{2}\beta_{2}}$ & $S_{\alpha_{2}\beta_{1}}>S_{\alpha_{2}\beta_{2}}$\tabularnewline
.169(.004) & .128(.041) & .071(.374) & .142(.019)\tabularnewline
$S_{\alpha_{1}\beta_{1}}<S_{\alpha_{1}\beta_{2}}$ & $S_{\alpha_{1}\beta_{1}}<S_{\alpha_{2}\beta_{1}}$ & $S_{\alpha_{1}\beta_{2}}<S_{\alpha_{2}\beta_{2}}$ & $S_{\alpha_{2}\beta_{1}}<S_{\alpha_{2}\beta_{2}}$\tabularnewline
.010(.980) & .021(.919) & .053(.580) & .028(.854)\tabularnewline
\hline 
\multicolumn{4}{c}{S12}\tabularnewline
\hline 
$S_{\alpha_{1}\beta_{1}}>S_{\alpha_{1}\beta_{2}}$ & $S_{\alpha_{1}\beta_{1}}>S_{\alpha_{2}\beta_{1}}$ & $S_{\alpha_{1}\beta_{2}}>S_{\alpha_{2}\beta_{2}}$ & $S_{\alpha_{2}\beta_{1}}>S_{\alpha_{2}\beta_{2}}$\tabularnewline
.148(.011) & .089(.191) & .080(.296) & .138(.025)\tabularnewline
$S_{\alpha_{1}\beta_{1}}<S_{\alpha_{1}\beta_{2}}$ & $S_{\alpha_{1}\beta_{1}}<S_{\alpha_{2}\beta_{1}}$ & $S_{\alpha_{1}\beta_{2}}<S_{\alpha_{2}\beta_{2}}$ & $S_{\alpha_{2}\beta_{1}}<S_{\alpha_{2}\beta_{2}}$\tabularnewline
.025(.884) & .056(.521) & .051(.613) & .023(.899)\tabularnewline
\hline 
\multicolumn{4}{c}{S15}\tabularnewline
\hline 
$S_{\alpha_{1}\beta_{1}}>S_{\alpha_{1}\beta_{2}}$ & $S_{\alpha_{1}\beta_{1}}>S_{\alpha_{2}\beta_{1}}$ & $S_{\alpha_{1}\beta_{2}}>S_{\alpha_{2}\beta_{2}}$ & $S_{\alpha_{2}\beta_{1}}>S_{\alpha_{2}\beta_{2}}$\tabularnewline
.041(.715) & .062(.493) & .049(.621) & .050(.627)\tabularnewline
$S_{\alpha_{1}\beta_{1}}<S_{\alpha_{1}\beta_{2}}$ & $S_{\alpha_{1}\beta_{1}}<S_{\alpha_{2}\beta_{1}}$ & $S_{\alpha_{1}\beta_{2}}<S_{\alpha_{2}\beta_{2}}$ & $S_{\alpha_{2}\beta_{1}}<S_{\alpha_{2}\beta_{2}}$\tabularnewline
.104(.120) & .097(.174) & .032(.816) & .044(.699)\tabularnewline
\hline 
\multicolumn{4}{c}{S16}\tabularnewline
\hline 
$S_{\alpha_{1}\beta_{1}}>S_{\alpha_{1}\beta_{2}}$ & $S_{\alpha_{1}\beta_{1}}>S_{\alpha_{2}\beta_{1}}$ & $S_{\alpha_{1}\beta_{2}}>S_{\alpha_{2}\beta_{2}}$ & $S_{\alpha_{2}\beta_{1}}>S_{\alpha_{2}\beta_{2}}$\tabularnewline
.124(.048) & .131(.028) & .136(.030) & .139(.022)\tabularnewline
$S_{\alpha_{1}\beta_{1}}<S_{\alpha_{1}\beta_{2}}$ & $S_{\alpha_{1}\beta_{1}}<S_{\alpha_{2}\beta_{1}}$ & $S_{\alpha_{1}\beta_{2}}<S_{\alpha_{2}\beta_{2}}$ & $S_{\alpha_{2}\beta_{1}}<S_{\alpha_{2}\beta_{2}}$\tabularnewline
.001(1.000) & .000(1.000) & .005(.995) & .034(.800)\tabularnewline
\hline 
\end{tabular}
\par\end{centering}
Note: Each number outside of the brackets is the KS statistic value
and each number in the brackets is the $p$ value.

\label{test stochastic dominance_failed subjects}
\end{table}

\subsubsection*{Diagnosing Architectures According to SIC and MIC in the Presence
of $(A,B)\looparrowleft(\alpha,\beta)$ and Stochastic Dominance}

\begin{figure}[H]
\includegraphics[scale=0.4]{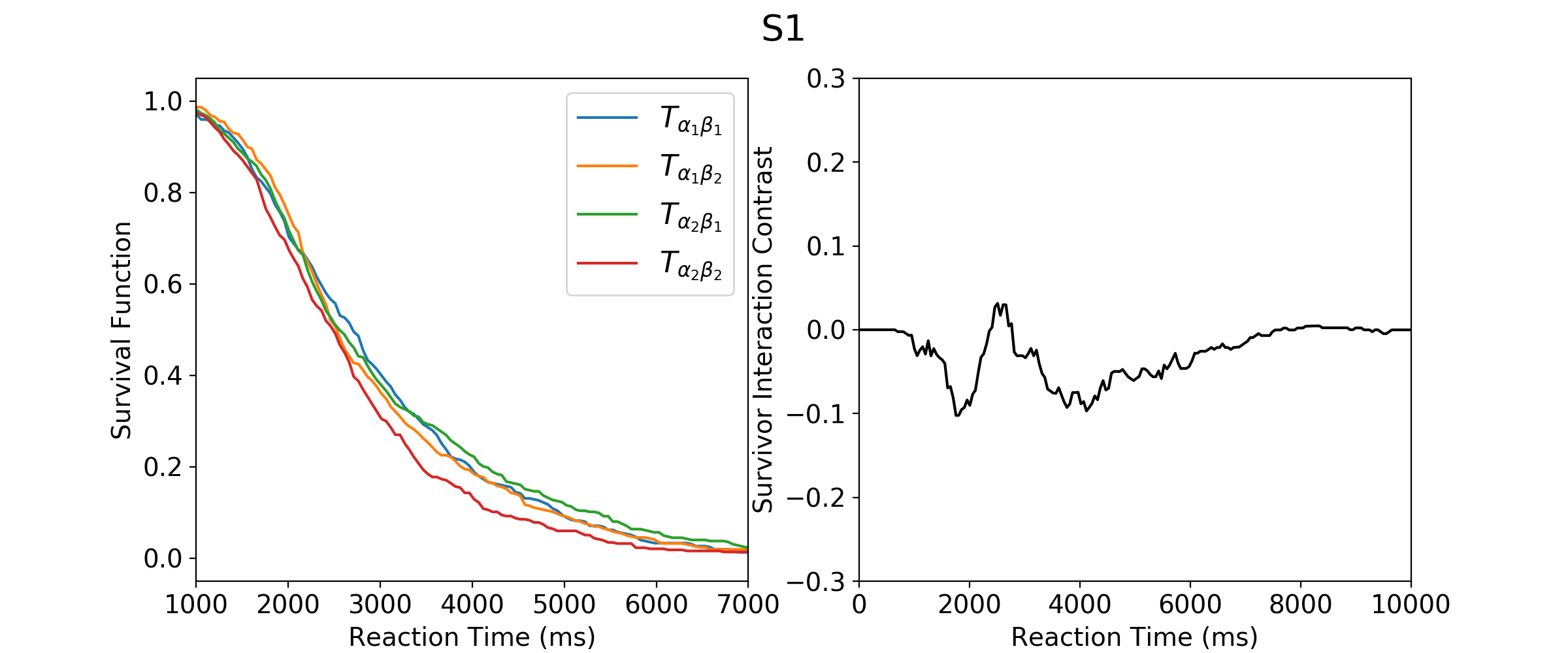}

\includegraphics[scale=0.4]{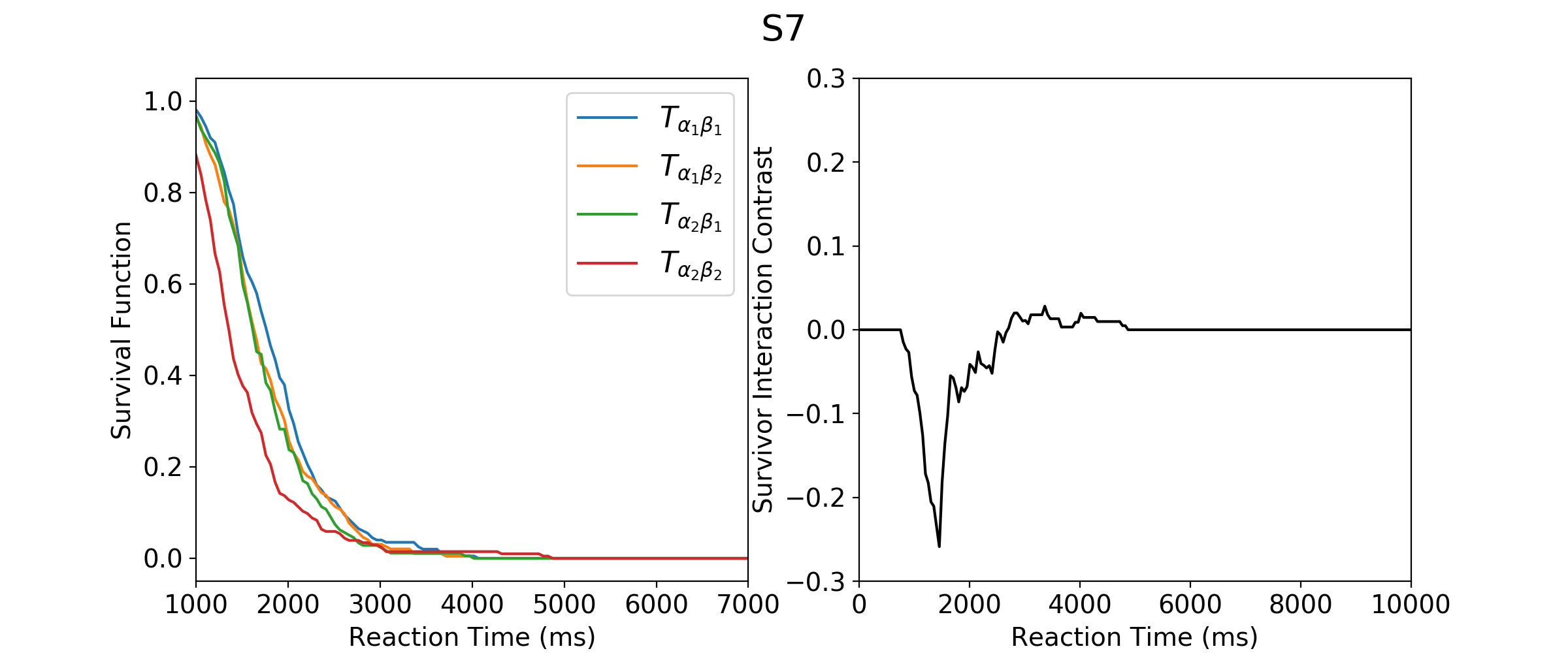}

\includegraphics[scale=0.4]{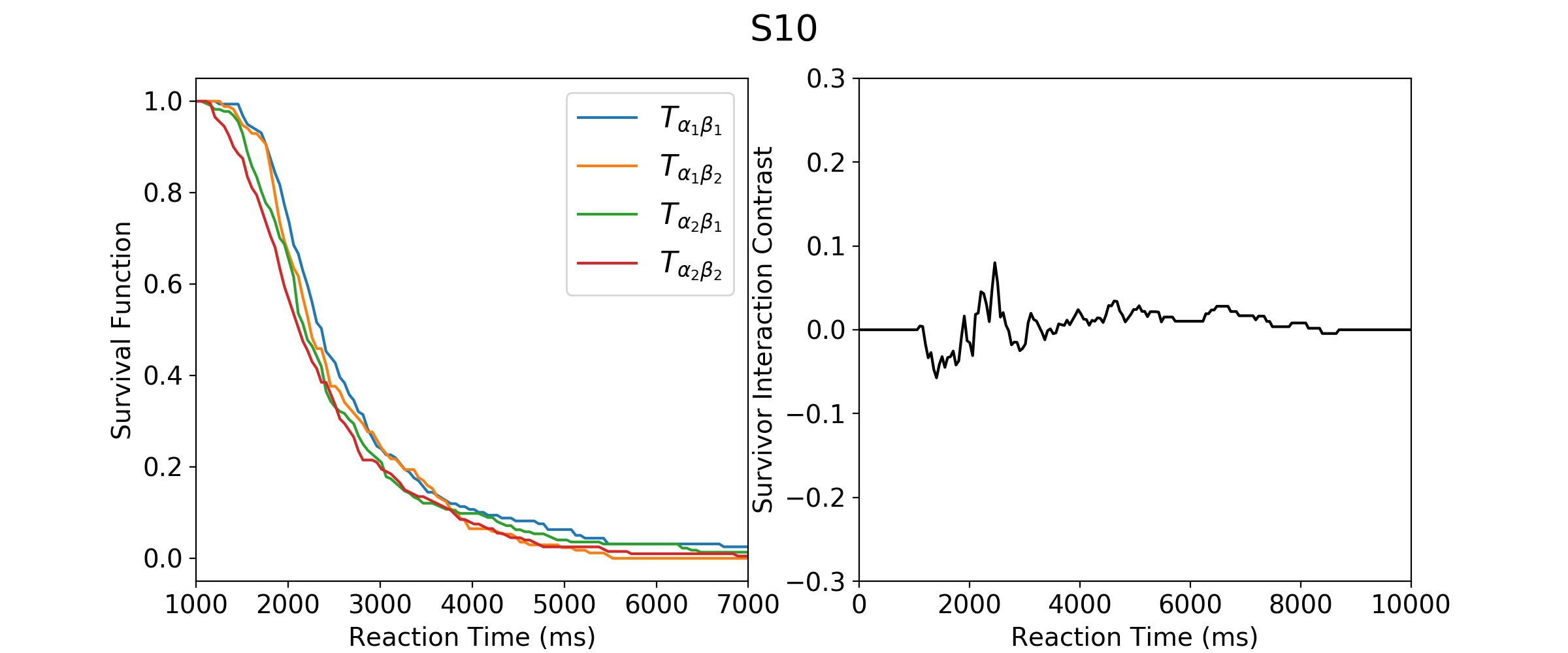}

\includegraphics[scale=0.4]{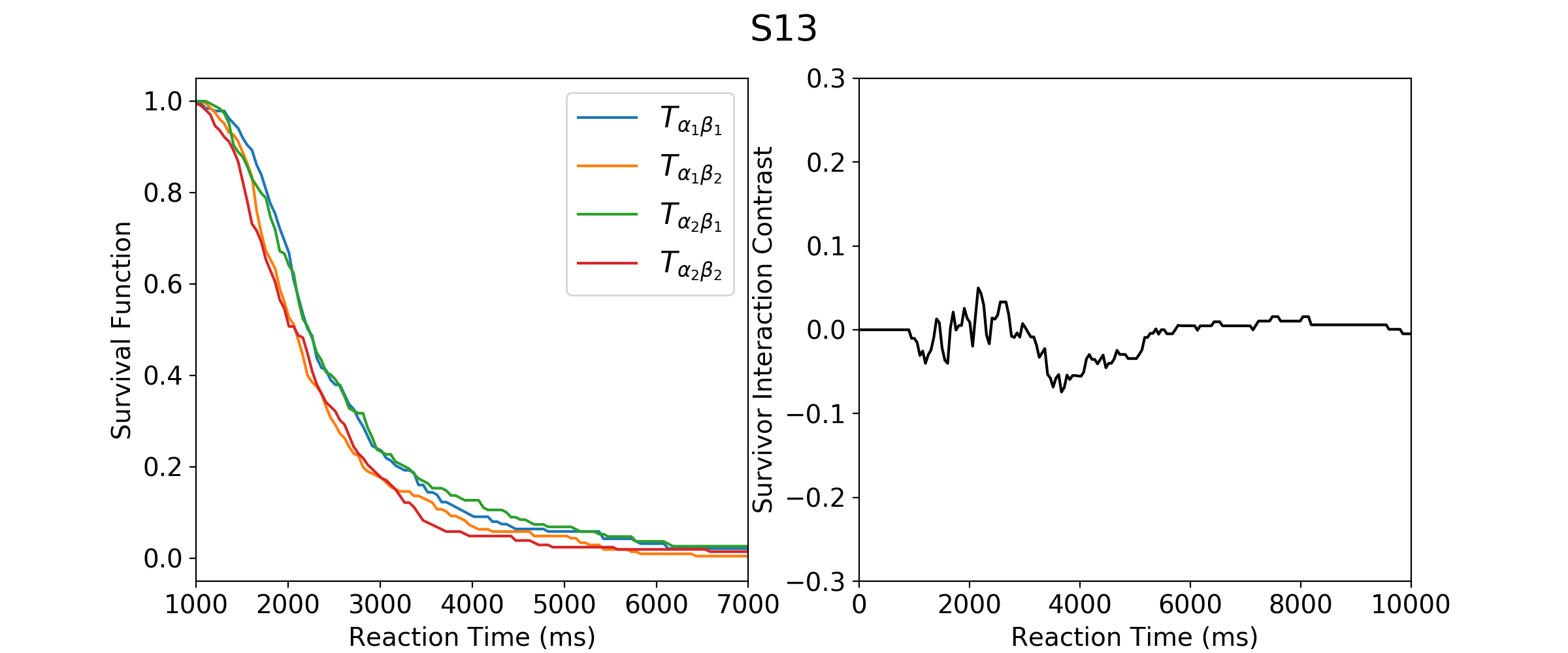}

\caption{Survival functions of RT and SIC for S1, S7, S10, S13, and S14 in
Experiment 1. S1 participated in Experiment 1(a). S7 and S10 participated
in Experiment 1(b). S13 and S14 participated in Experiment 1(c).}

\label{Survival_1_dot}
\end{figure}

With the confirmation of ordering of the RT distributions, selective
influence, and subject's adherence to a single type of mental architecture,
we then diagnosed how the horizontal and vertical coordinates of the
test dot were adjusted by investigating the behavior of SIC and MIC
for subject S1, S7, S10, S13, and S14. The SIC curves for these subjects
are displayed in the right column of Figure \ref{Survival_1_dot}.
We implemented the R package developed by Houpt, Blaha, McIntire,
Havig, and Townsend (2014) to inspect the statistical significance
of SIC and MIC. Table \ref{statistics of SIC and MIC} includes the
statistics for SIC and MIC and the inferred architectures from SFT.
$D^{+}$ is the most positive point of SIC and $D^{-}$ is the most
negative point of SIC. We chose $\mathrm{alpha}=.33$ not the conventional
critical value .05 here (Fox \& Houpt, 2016) as the null hypothesis
is SIC = 0 for all values of RT and MIC = 0 and hence conservative
alpha levels bias the tests toward indicating a serial OR signature
(flat SIC and zero MIC).

\begin{figure}[H]
\includegraphics[scale=0.4]{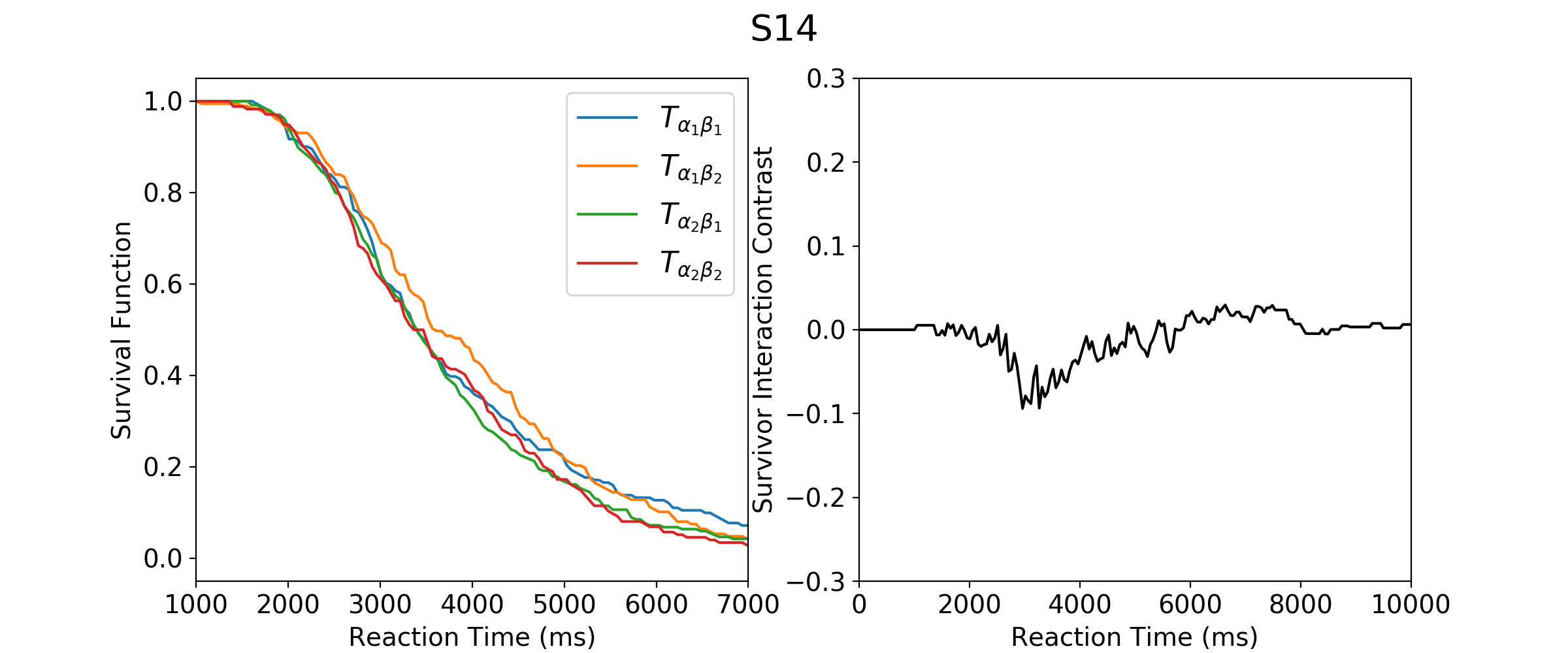}
\begin{raggedright}
Figure 4: Survival functions of RT and SIC for S1, S7, S10, S13, and
S14 in Experiment 1 (continued). S1 participated in Experiment 1(a).
S7 and S10 participated in Experiment 1(b). S13 and S14 participated
in Experiment 1(c). 
\par\end{raggedright}
\label{Survival_1_dot-1}
\end{figure}

S1 and S7 agreed with the characteristic properties for parallel AND
as $D^{-}$ was significant, $D^{+}$ was not significant, and MIC
was significantly smaller than zero. $D^{+}$, $D^{-}$, and MIC for
S10 and S13 were not significantly different from zero, indicating
a lack of statistical power to draw a conclusion. For S14, he/she
had a significant $D^{-}$ and insignificant $D^{+}$, which favored
the parallel AND model. However this subject did not have a significant
MIC, which was not aligned with parallel AND. To be cautious, we conclude
that S14's strategy was uncertain. In the very beginning we expected
that all the subjects in this experiment adjusted the coordinates
of the test dots in the parallel AND or serial AND or coactive manner.
The trajectory of the trackball movements excluded serial AND. The
architectures inferred for S1, S7, S10, S13, and S14 were either parallel
AND or uncertain, which was consistent with the earlier expectation
and the trackball move.

\begin{table}[h]
\caption{The statistics of SIC and MIC and the inferred architectures for S1,
S7, S10, S13, and S14 in Experiment 1. S1 participated in Experiment
1(a). S7 and S10 participated in Experiment 1(b). S13 and S14 participated
in Experiment 1(c).}
\begin{centering}
\begin{tabular}{ccccc}
\hline 
Subject & $D^{+}$($p$ value) & $D^{-}$($p$ value) & MIC($p$ value) & Architecture\tabularnewline
\hline 
S1 & .038(.733) & .105(.096) & -271.47(.144) & Parallel AND\tabularnewline
S7 & .030(.921) & .269(.001) & -145.48(.014) & Parallel AND\tabularnewline
\textit{\emph{S10}} & \textit{\emph{.085(.522)}} & \textit{\emph{.083(.538)}} & 54.74(.895) & \textit{\emph{-}}\tabularnewline
S13 & .085(.510) & .076(.582) & -65.234(.999) & -\tabularnewline
S14 & .050(.795) & .111(.321) & 103.86(.592) & Uncertain\tabularnewline
\hline 
\end{tabular}
\par\end{centering}
\label{statistics of SIC and MIC}
\end{table}

\subsubsection*{Estimating Capacity}

As mentioned in the beginning of this article, DFP includes two types
of manipulation. In Experiments 1(b) and 1(c), we manipulated the
workload by adding some reference dots with either the horizontal
coordinate or the vertical coordinate equal to 0 px. So sometimes
one channel was loaded to the subjects' action as they had to adjust
only one coordinate of the test dot and in the other trials two channels
were loaded. The manipulation of stimulus salience was realized by
assigning the horizontal coordinate of the reference dot to level
one ($\alpha_{1}$) or level two ($\alpha_{2}$) and the vertical
coordinate to level one ($\beta_{1}$) or level two ($\beta_{2}$),
so the processing speed of level 1 was slower than the speed of level
2. In the DFP, there are eight types of stimuli. The trials display
stimuli $\alpha_{0}\beta_{1}$, $\alpha_{0}\beta_{2}$, $\alpha_{1}\beta_{0}$,
and $\alpha_{2}\beta_{0}$ are single-channel trials. The trials display
stimuli $\alpha_{1}\beta_{1}$, $\alpha_{1}\beta_{2}$, $\alpha_{2}\beta_{1}$,
and $\alpha_{2}\beta_{2}$ are double-channel trials. In Experiment
1(b), the double-channel trials and the single-channel trials were
presented to the subjects in an intermixed way: In each trial, each
of the eight stimuli had the same chance to be shown. In Experiment
1(b), the double-channel trials and the single-channel trials were
presented in the separate experimental sessions. 

We anticipated the subjects in Experiment 1 used the AND stopping
rule to make responses as both the horizontal coordinate and the vertical
coordinate of the test dot had to match those of the reference dot.
The experimental design of Experiment 1(b) and 1(c) allowed the computation
of $K_{\alpha\beta}(t)$, $K_{\beta}(t)$, and $K_{\alpha}(t)$ in
(\ref{eq:capacity_AND}): $K_{\alpha\beta}(t)$ was computed from
the double-channel trials, $K_{\beta}(t)$ was from the single-channel
trials with stimuli $\alpha_{0}\beta_{1}$ and $\alpha_{0}\beta_{2}$,
and $K_{\alpha}(t)$ was from the stimuli $\alpha_{1}\beta_{0}$ and
$\alpha_{2}\beta_{0}$.

\begin{figure}[H]
\begin{centering}
\includegraphics[scale=0.4]{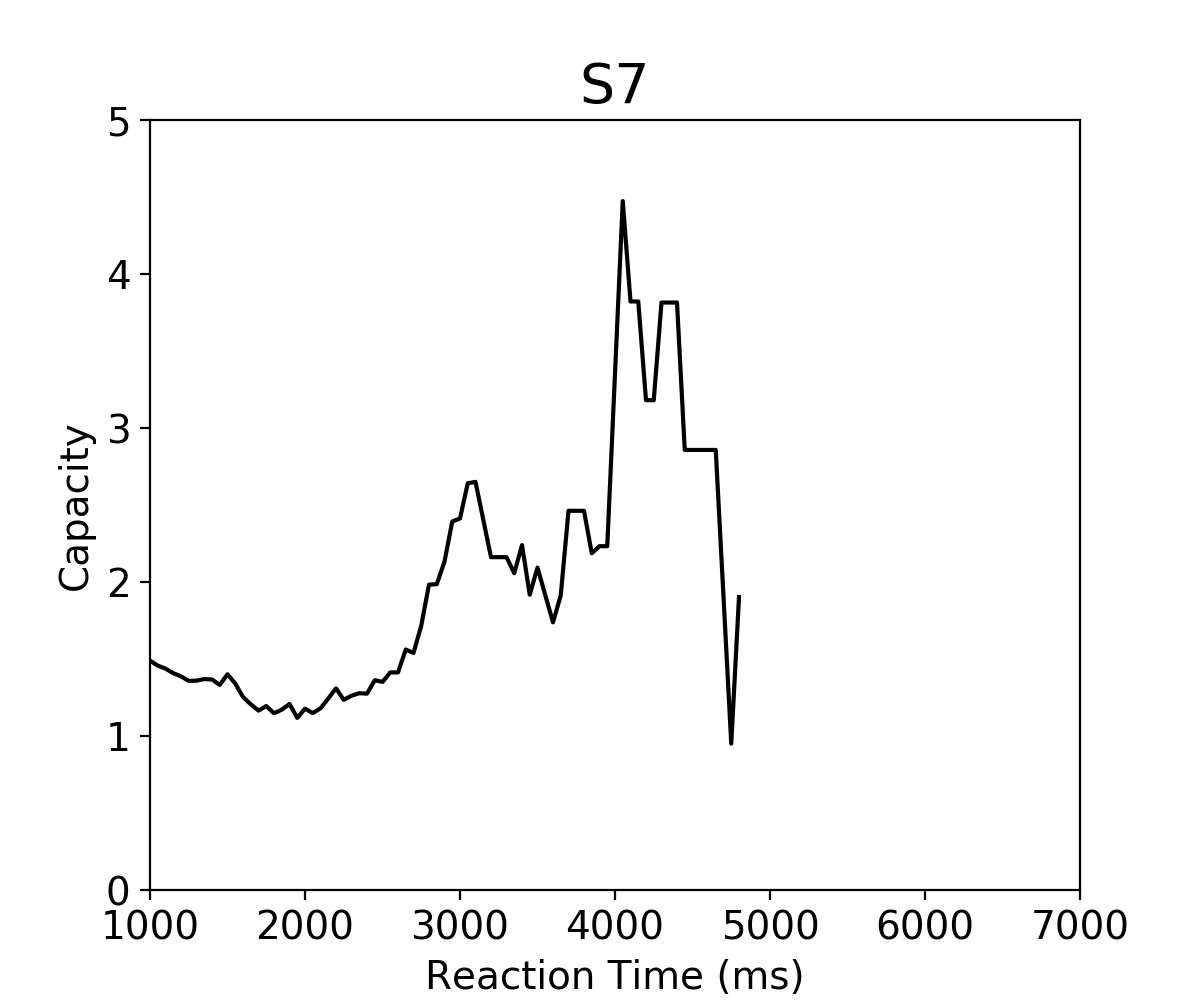}\includegraphics[scale=0.4]{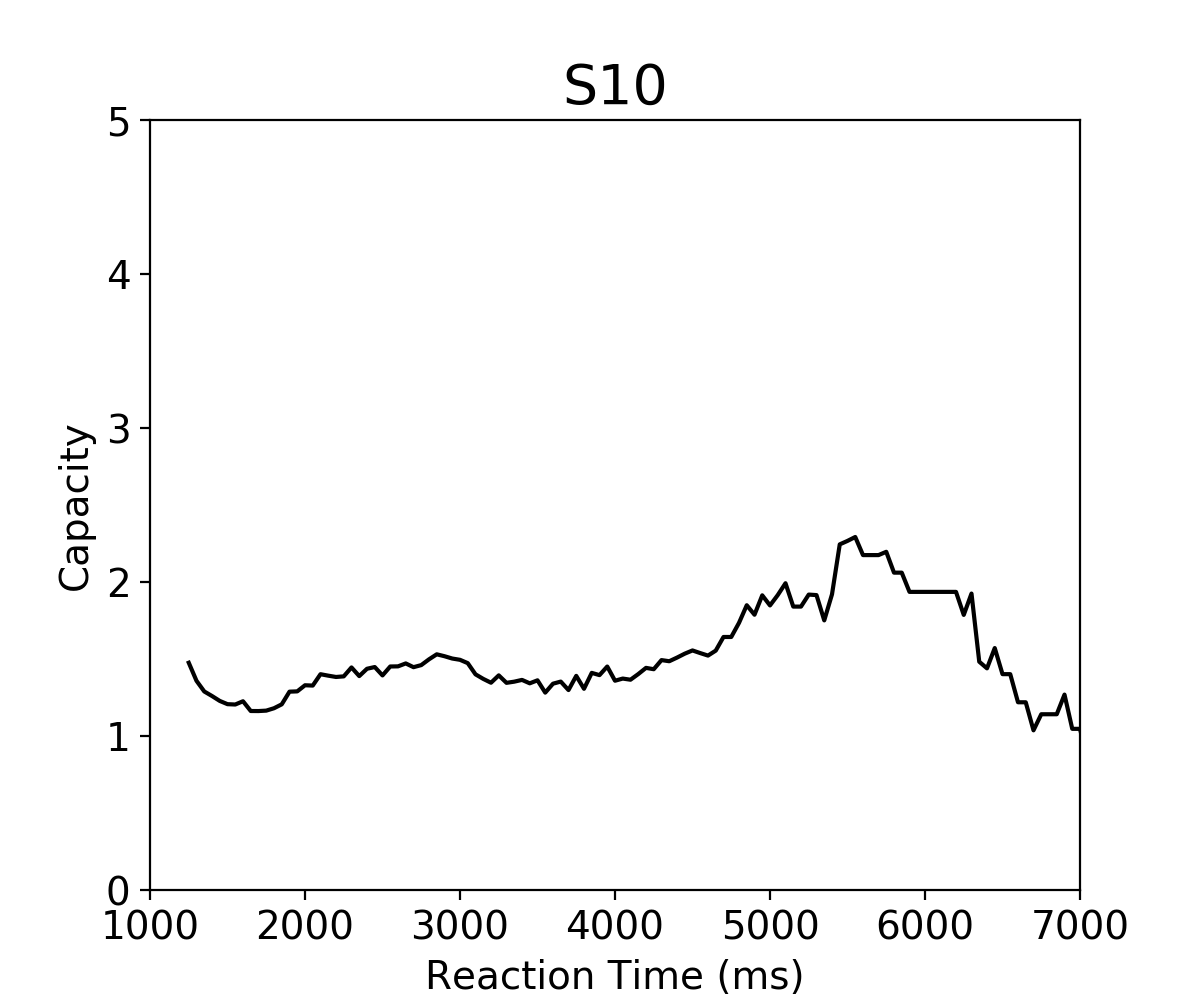}
\par\end{centering}
\begin{centering}
\includegraphics[scale=0.4]{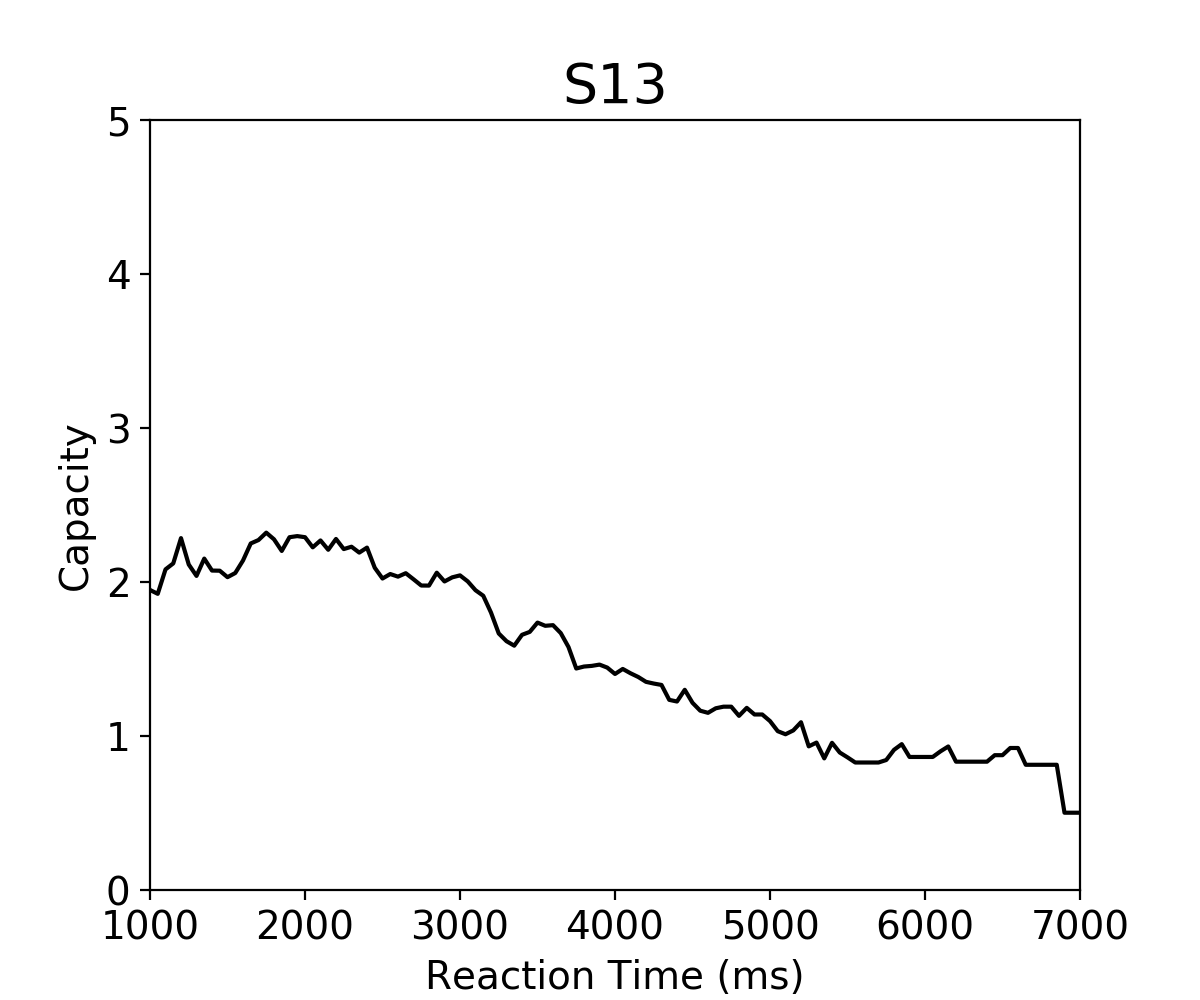}\includegraphics[scale=0.4]{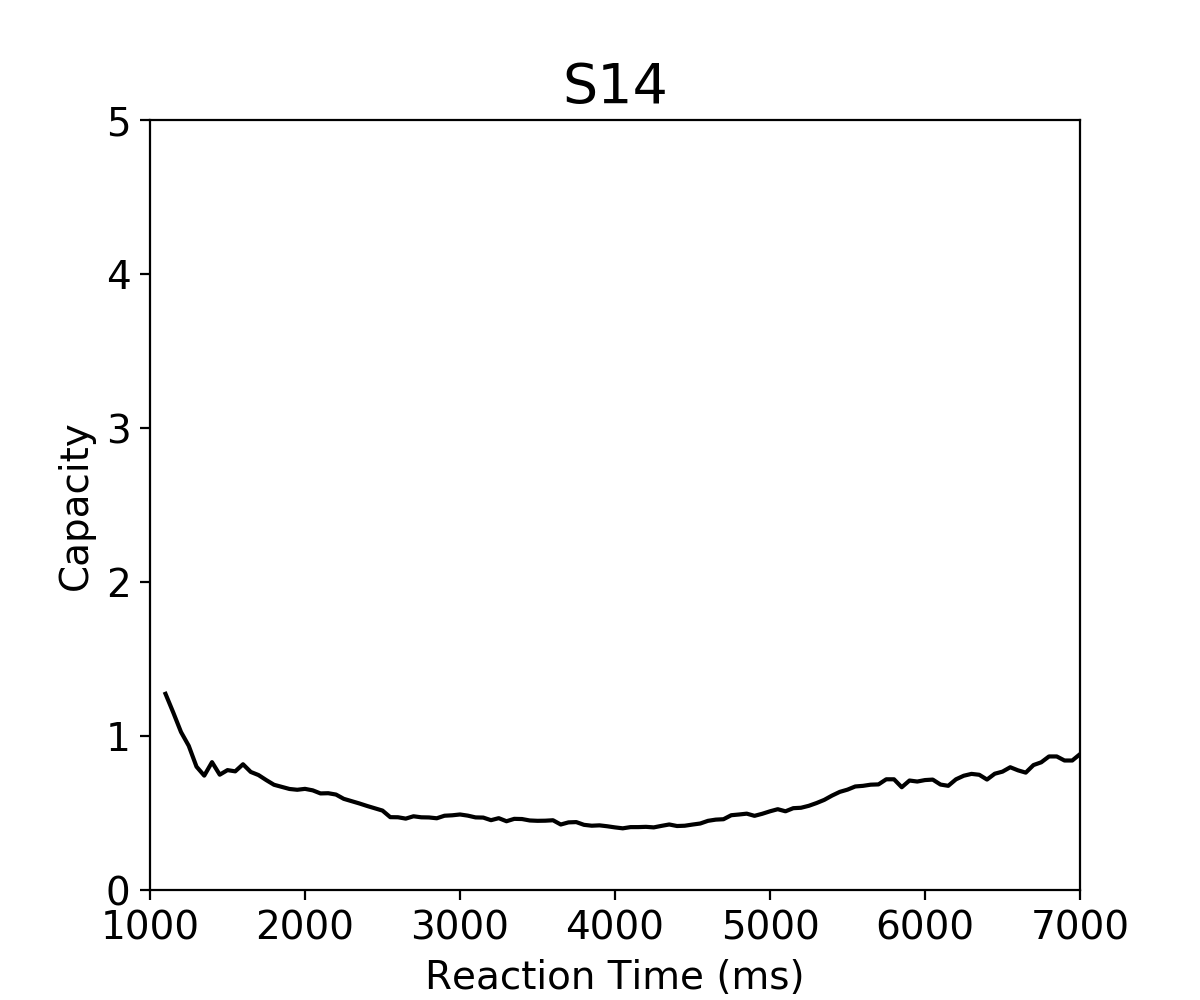}
\par\end{centering}
\caption{The capacity coefficients for S7, S10, S13, and S14 in Experiment
1. S7 and S10 participated in Experiment 1(b). S13 and S14 participated
in Experiment 1(c).}

\label{capacity_dot}
\end{figure}

The statistics of the capacity coefficient was computed using the
R package developed by Houpt, Blaha, McIntire, Havig, and Townsend
(2014). The capacity for S7, S10, and S13 (Figure \ref{capacity_dot})
was super ($p<.001$) indicating adding one channel speeded up the
processing of the other channel and for S14 it was limited indicating
adding one channel slowed down the other channel.

\subsubsection*{The Consequence of Absence of $(A,B)\looparrowleft(\alpha,\beta)$}

Subject S2, S3, S8, S9, S11, S12, S15, and S16 did not pass the test
for $(A,B)\looparrowleft(\alpha,\beta)$. We considered it a failure
for the establishment of $(T_{\alpha},T_{\beta})\looparrowleft(\alpha,\beta)$.
Those subjects all passed the test of stochastic dominance (Table
\ref{test stochastic dominance_failed subjects}). We investigated
their architectures without the security of selective influence. 

\begin{figure}[H]
\includegraphics[scale=0.4]{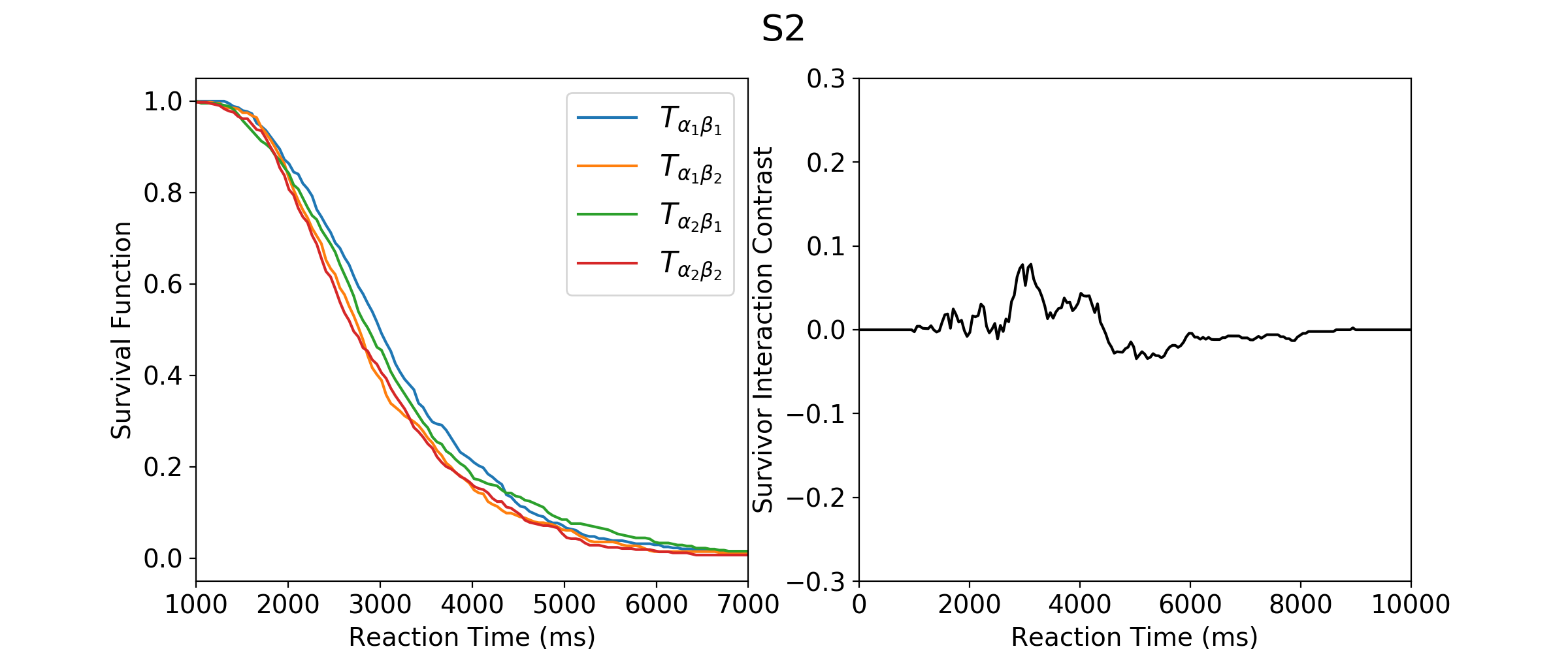}

\includegraphics[scale=0.4]{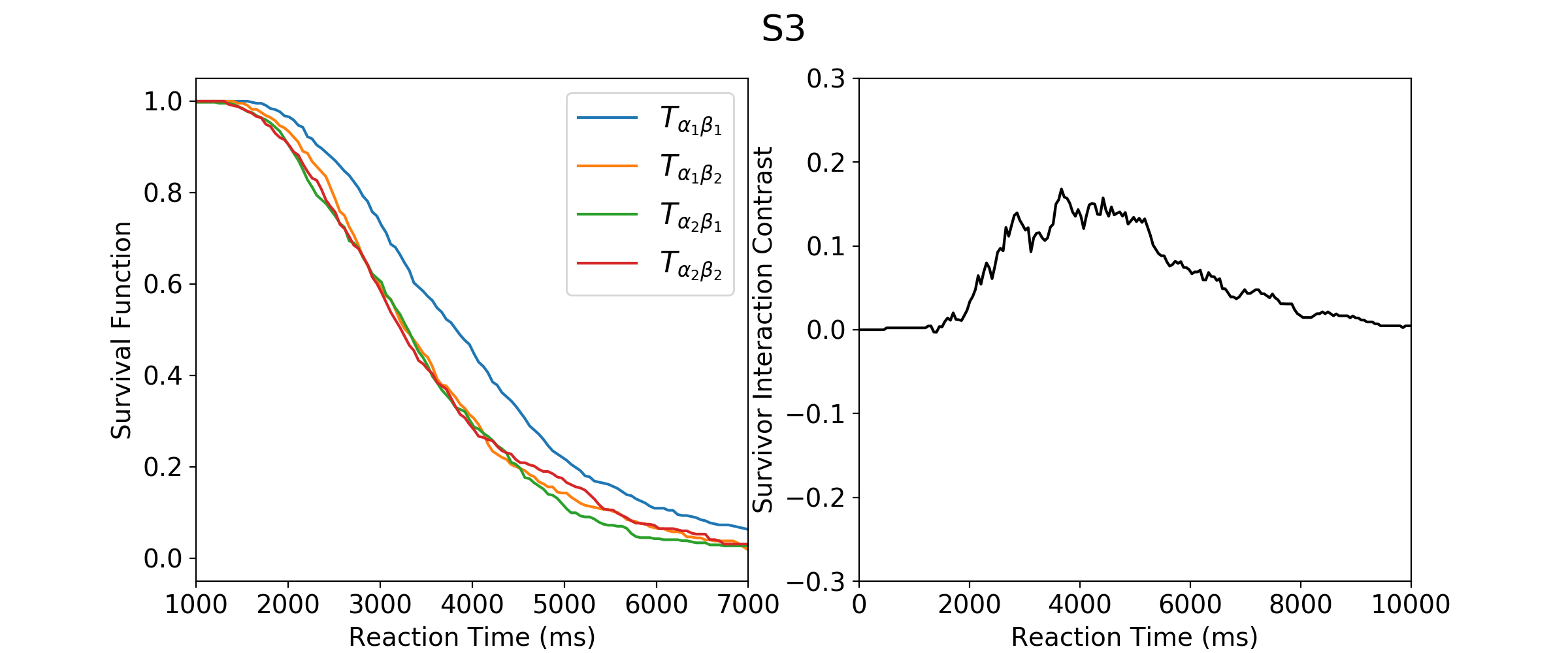}

\includegraphics[scale=0.4]{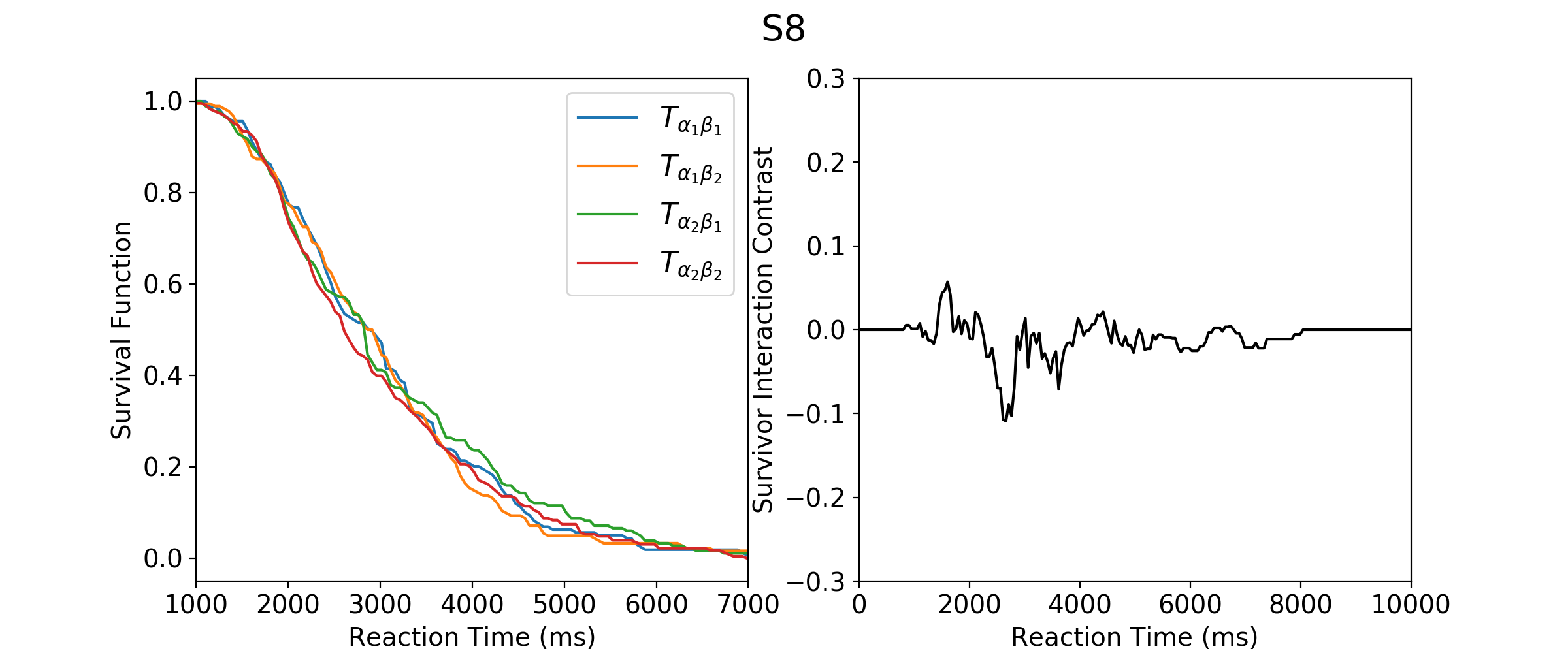}

\includegraphics[scale=0.4]{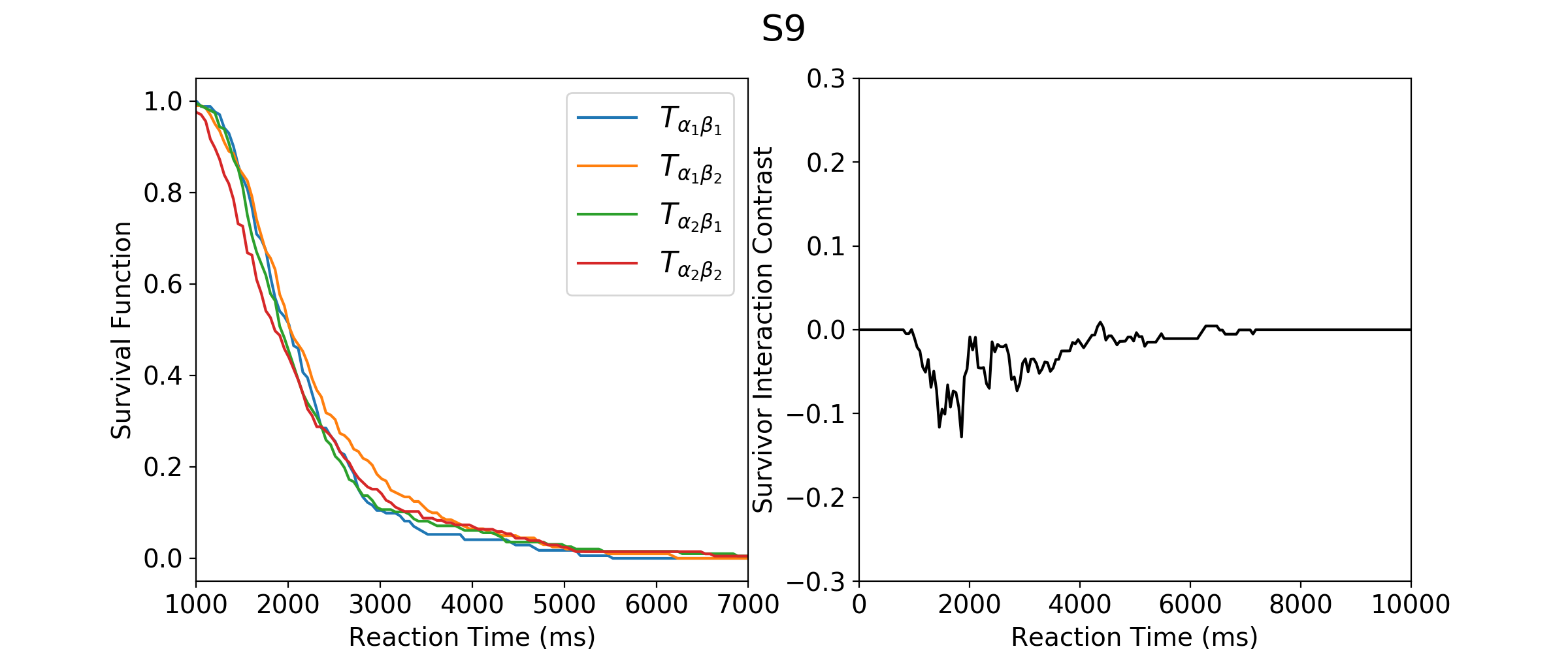}

\caption{Survival functions of RT and SIC for S2, S3, S8, S9, S11, S12, S15,
and S16 in Experiment 1. S2 and S3 participated in Experiment 1(a).
S8, S9, and S11 participated in Experiment 1(b). S12, S15, and S16
participated in Experiment 1(c).}

\label{SIC_MIC_failed subjects}
\end{figure}

The right column of Figure \ref{SIC_MIC_failed subjects} displays
the SIC curves for S2, S3, S8, S9, S11, S12, S15, and S16. The statistics
for SIC and MIC and the inferred architectures for these subjects
can be found in Table \ref{statistics of SIC and MIC_dot_1}. $D^{+}$
was significant and $D^{-}$ was not significant ($\mathrm{alpha}=.33$)
for S2, favoring the parallel OR model. However MIC for this subject
was not significantly greater than zero, which was not consistent
with the property for parallel OR. Therefore we considered the architecture
for this subject uncertain as the conclusions from SIC and MIC did
not converge. The referred architecture indicates S3 implemented the
parallel OR manner to make responses as $D^{+}$ was significant,
$D^{-}$ was not significant, and MIC was significantly greater than
zero. Subject S8, S11, S12, and S16's SIC and MIC were not significant,
then we considered their architectures were not diagnostic. For subject
S9, $D^{+}$ was not significant, $D^{-}$ was significant, and MIC
was significantly smaller than zero, supporting the parallel AND signature.
For S15, $D^{+}$ was not significant and $D^{-}$ was significant,
favoring the parallel AND model. However MIC was not significantly
different than zero, which did not align with parallel AND. Therefore
we considered the architecture for this person uncertain. For the
subjects examined in this section, we observed a prohibited signature:
parallel OR for S3. The diagnosis for S3 seemed questionable.

\begin{figure}[H]
\includegraphics[scale=0.4]{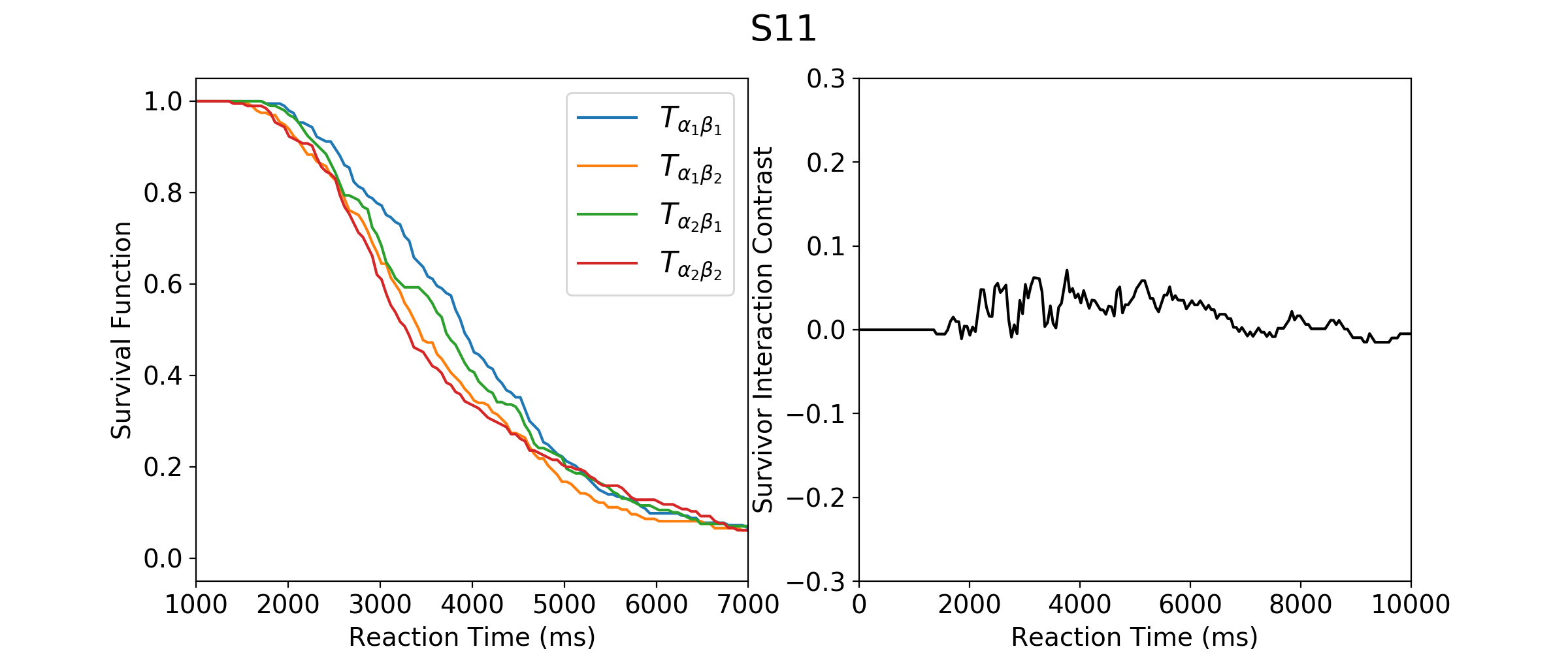}

\includegraphics[scale=0.4]{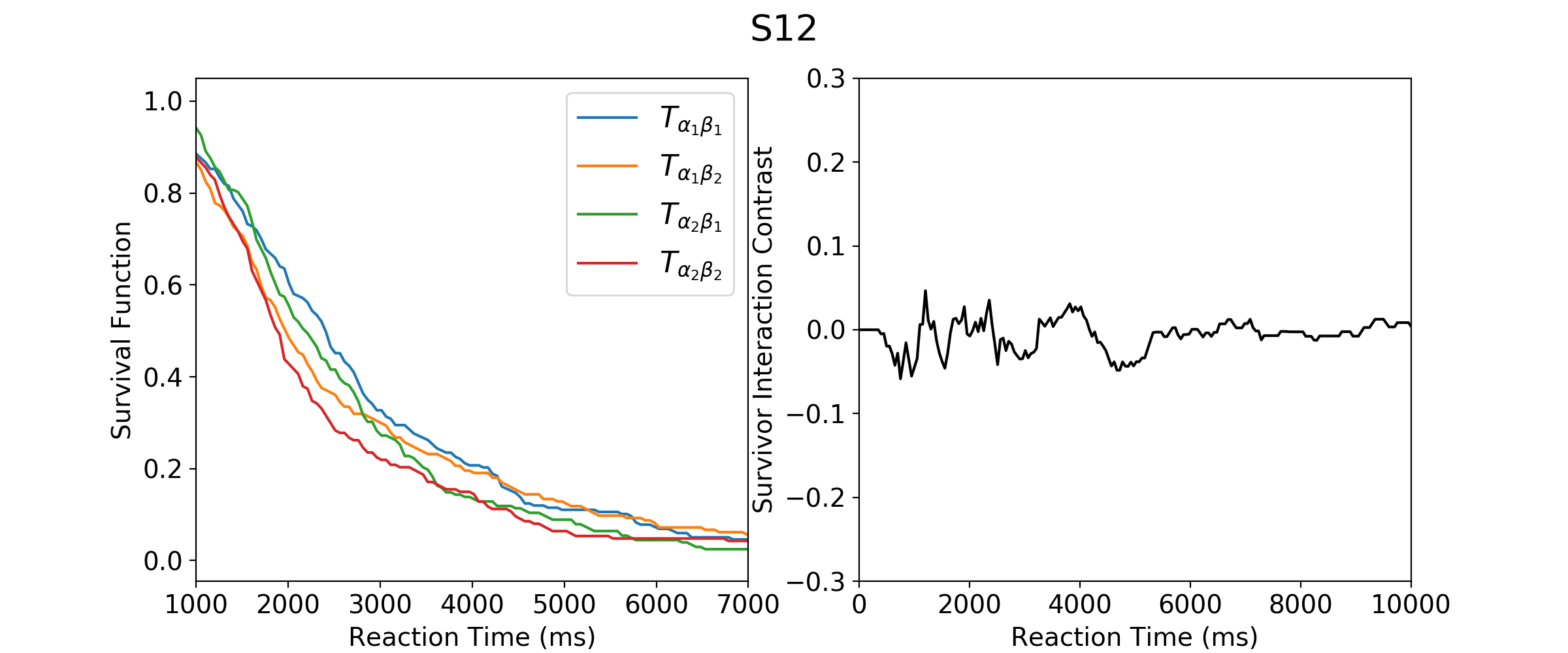}

\includegraphics[scale=0.4]{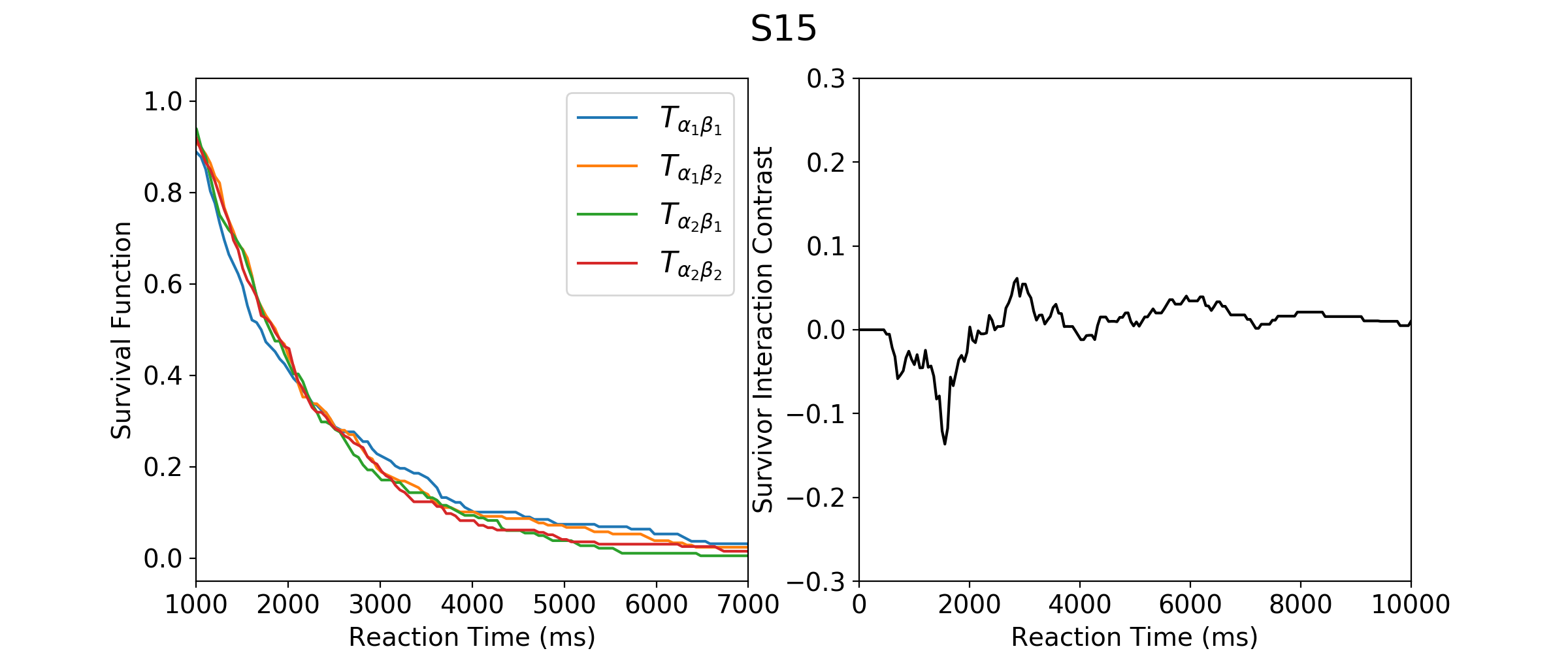}

\includegraphics[scale=0.4]{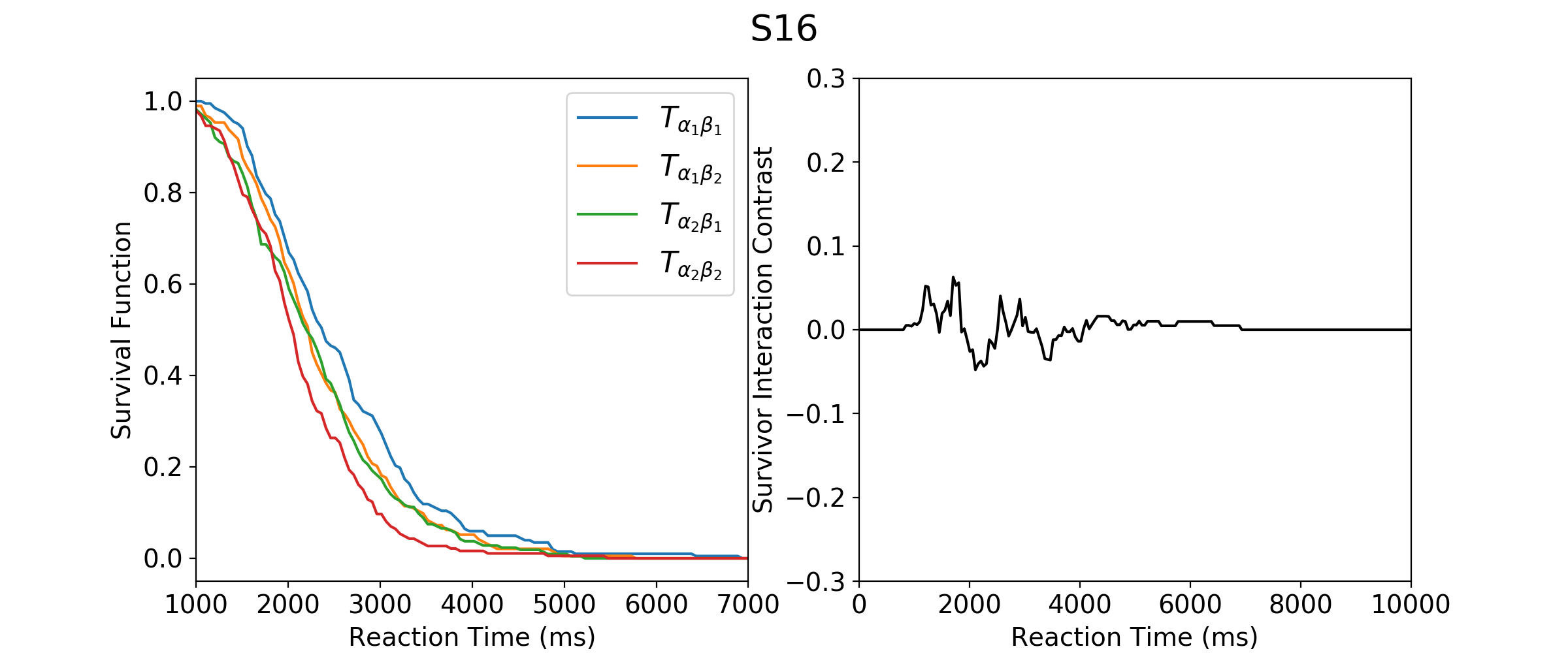}

Figure 6: Survival functions of RT and SIC for S2, S3, S8, S9, S11,
S12, S15, and S16 in Experiment 1 (continued). S2 and S3 participated
in Experiment 1(a). S8, S9, and S11 participated in Experiment 1(b).
S12, S15, and S16 participated in Experiment 1(c).

\label{SIC_MIC_failed subjects-1}
\end{figure}

\begin{table}[h]
\caption{The statistics of SIC and MIC and the inferred architectures for S2,
S3, S8, S9, S11, S12, S15, and S16 in Experiment 1. S2 and S3 participated
in Experiment 1(a). S8, S9, and S11 participated in Experiment 1(b).
S12, S15, and S16 participated in Experiment 1(c).}
\begin{centering}
\begin{tabular}{ccccc}
\hline 
Subject & $D^{+}$($p$ value) & $D^{-}$($p$ value) & MIC($p$ value) & Architecture\tabularnewline
\hline 
S2 & .085(.210) & .036(.761) & 8.65(.544) & Uncertain\tabularnewline
S3 & .159(.005) & .005(.995) & 563.25(.000) & Parallel OR\tabularnewline
\textit{\emph{S8}} & \textit{\emph{.069(.649)}} & \textit{\emph{.088(.495)}} & \textit{\emph{-23.637(.951)}} & \textit{\emph{-}}\tabularnewline
\textit{\emph{S9}} & \textit{\emph{.009(.993)}} & \textit{\emph{.138(.173)}} & \textit{\emph{-169.08(.199)}} & \textit{\emph{Parallel AND}}\tabularnewline
S11 & .074(.594) & .015(.980) & 122.69(.485) & -\tabularnewline
S12 & .067(.656) & .068(.646) & -81.801(.835) & -\tabularnewline
S15 & .077(.574) & .133(.188) & 67.986(.560) & Uncertain\tabularnewline
\multirow{1}{*}{S16} & .059(.719) & .079(.553) & -39.131(.825) & -\tabularnewline
\hline 
\end{tabular}
\par\end{centering}
\label{statistics of SIC and MIC_dot_1}
\end{table}

\subsubsection*{Discussions}

In this experiment, regardless of the test for $(A,B)\looparrowleft(\alpha,\beta)$
passed or not, stochastic dominance was not violated. However, given
$(A,B)\looparrowleft(\alpha,\beta)$ established, the architectures
diagnosed from SFT were not out of expectation. If $(A,B)\looparrowleft(\alpha,\beta)$
was absent, SFT led to an architecture that was indeed prohibited
for one subject out of eight, which indicated an absence of $(T_{\alpha},T_{\beta})\looparrowleft(\alpha,\beta)$
at least for that particular person. Therefore using stochastic dominance
only as an evidence for selective influence of $\alpha$ and $\beta$
on $T_{\alpha}$ and $T_{\beta}$ is risky as it may result in an
incorrect diagnosis about the architecture. Combining stochastic dominance
and $(A,B)\looparrowleft(\alpha,\beta)$ results in a more trustable
inference.

No single-channel trials were used in Experiment 1(a). In this experiment,
we found that it took more time to move a test dot to the target location
when the target was closer to the center of the circle. In Experiment
1(b), the participants had 50\% chance to view a double-channel stimulus
and 50\% chance to view a single-channel stimulus in any trial. We
found all the subjects spent more time to make a response when the
reference dot was further to the center of the circle. In Experiment
1(c), the single-channel stimuli and the double-channel stimuli did
not display to the participants in the mixed way. Rather the single-channel
trials were presented only when all the double-channel trials were
shown. We observed in Experiment 1(c) the RT was ordered in the same
way as in Experiment 1(a) for four participants out of five. It indicates
by mixing the single-channel trials with the double-channel trials
in the experimental design, the ordering of RT for those double-channel
trials was reversed. We name it context effect. 

\subsection*{Floral Shape Reproduction Task}

\subsubsection*{Trackball Movements}

\begin{figure}[H]
\begin{centering}
\includegraphics[scale=0.5]{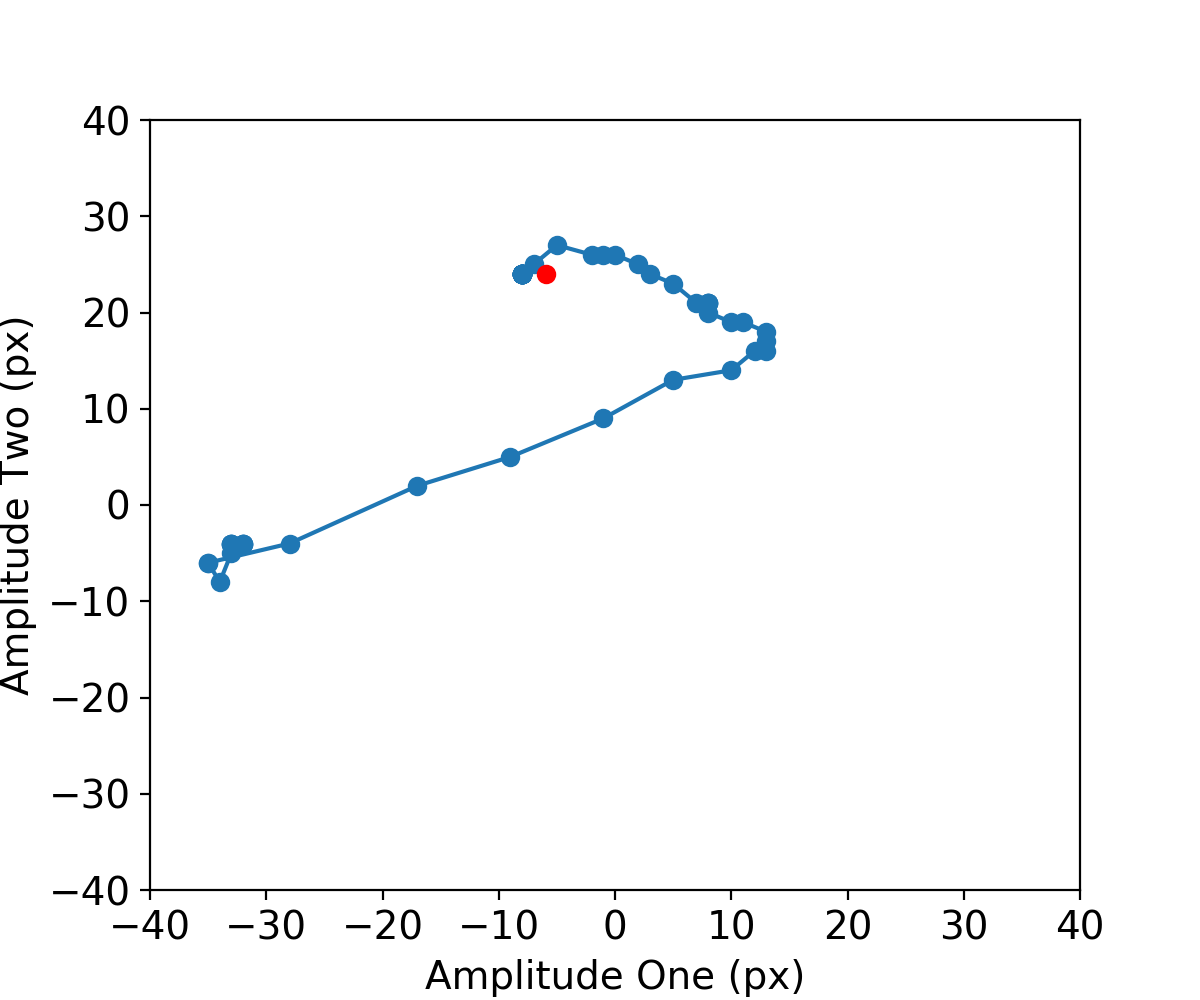}
\par\end{centering}
\caption{Typical trackball movements in a trial in the floral shape reproduction
task (plotted every 50 ms).}

\label{trackball move_dot}
\end{figure}

Figure \ref{trackball move_dot} shows the trackball movements in
a typical trial in the floral shape reproduction task. The trajectory
of the trackball movements confirmed the assumption of SFT that the
subject adheres to a single type of mental architecture from trial
to trial. The red dot represents the amplitudes of the fixed reference
shape. The test shape (blue) started from amplitudes (-31.828 px,
-4.468 px) and after a sequence of adjustments for amplitude one and
amplitude two, the finalized shape was very close to the target shape
indicating the amplitudes were not adjusted in the parallel OR or
serial OR or serial AND manner: If the stopping rule OR was used,
one should expect the finalized test shape matched well with the reference
shape either in amplitude one or amplitude two but not both. If serial
AND was used, one should expect the trajectory moved only along the
direction of amplitude one or along the direction of amplitude two
in each step but not diagonally as observed in the plot. The trajectory
implies parallel AND or coactive were used by the subjects in the
task. However the trajectory is not able to differentiate parallel
AND from coactive. 

\subsubsection*{Testing Selective Influence}

In the dot position reproduction task, the dot move on the screen
was directly reflected by the trackball move in the hand: If one moved
the trackball to the right, the test dot also moved to the right.
In the shape reproduction task, there was no apparent correspondence
between the trackball move and the change of the shape: Each trackball
move was transformed to the change of the amplitudes, defined by function
(\ref{eq:transformation from px to amplitude-1}). The horizontal
and vertical coordinates of each shape was defined by the transformation
from amplitudes (\ref{eq:floral shape-1}). The transformation functions
were designed for reasons. With these functions the floral shape reproduction
task was not as straightforward as the dot position reproduction task
to the participants, increasing the chance to have $A$ and $B$ selectively
influenced by $\alpha$ and $\beta$. In addition as we can see from
function (\ref{eq:transformation from px to amplitude-1}), when $A$
or $B$ is negative, it is updated more sensitively to the move of
the trackball than positive $A$ or $B$. So it was expected that
RT was short when the reference shape had negative amplitudes and
long with positive amplitudes. 

In order to test selective influence of $\alpha$ and $\beta$ on
the finalized $A$ and $B$, $\alpha$ and $\beta$ had to be discretized
for instance:

\begin{align}
\alpha_{1} & =(0\,\mathrm{px},30\,\mathrm{px}],\alpha_{2}=[-30\,\mathrm{px},0\,\mathrm{px}],\label{level_shape}\\
\beta_{1} & =(0\,\mathrm{px},30\,\mathrm{px}],\beta_{2}=[-30\,\mathrm{px},0\,\mathrm{px}].\nonumber 
\end{align}

The outliers of $A$ and $B$ were handled in this way: We computed
$A-\alpha$ and $B-\beta$ for each trial. Any trial that was out
of 3 standard deviations of the set of $A-\alpha$ or $B-\beta$ was
considered as an outlier and was removed from further analysis. Table
\ref{AB_shape} presents the corresponding means and standard deviations
of $A$ and $B$ for Experiment 2.

\begin{singlespace}
\begin{table}[H]
\caption{Means and standard deviations of the finalized amplitudes of the test
shapes for Experiment 2. }
\begin{centering}
\begin{tabular}{ccc}
\hline 
{\scriptsize{}Exp.} & {\scriptsize{}Subject} & {\tiny{}$((0\,\mathrm{px},30\,\mathrm{px}],(0\,\mathrm{px},30\,\mathrm{px}])$,
$((0\,\mathrm{px},30\,\mathrm{px}],[-30\,\mathrm{px},0\,\mathrm{px}])$,
$([-30\,\mathrm{px},0\,\mathrm{px}],(0\,\mathrm{px},30\,\mathrm{px}])$,
$([-30\,\mathrm{px},0\,\mathrm{px}],[-30\,\mathrm{px},0\,\mathrm{px}])$ }\tabularnewline
\hline 
{\scriptsize{}2(a)} & {\scriptsize{}S4} & {\tiny{}$(15.69\pm8.93,15.86\pm8.93)$, $(14.39\pm8.98,-15.74\pm9.23)$,
$(-14.66\pm8.97,15.05\pm9.53)$, $(-14.73\pm9.47,-15.25\pm9.02)$}\tabularnewline
{\scriptsize{}2(a)} & {\scriptsize{}S5} & {\tiny{}$(14.91\pm8.05,15.48\pm8.71)$, $(14.43\pm8.41,-14.97\pm8.58)$,
$(-14.33\pm8.78,15.21\pm8.44)$, $(-14.55\pm8.52,-15.03\pm8.09)$}\tabularnewline
{\scriptsize{}2(a)} & {\scriptsize{}S6} & {\tiny{}$(14.36\pm9.34,13.74\pm8.70)$, $(14.67\pm9.00,-15.26\pm8.48)$,
$(-14.95\pm9.25,14.14\pm8.26)$, $(-15.94\pm8.77,-15.52\pm8.71)$}\tabularnewline
{\scriptsize{}2(b)} & {\scriptsize{}S1} & {\tiny{}$(15.95\pm8.63,15.00\pm8.21)$, $(15.25\pm8.46,-14.62\pm8.51)$,
$(-14.65\pm8.21,14.83\pm8.34)$, $(-13.60\pm8.68,-15.05\pm8.67)$}\tabularnewline
{\scriptsize{}2(b)} & {\scriptsize{}S2} & {\tiny{}$(15.84\pm8.75,14.50\pm8.65)$, $(14.40\pm9.17,-15.71\pm8.73)$,
$(-15.77\pm8.54,14.99\pm8.94)$, $(-14.33\pm9.07,-15.23\pm8.90)$}\tabularnewline
{\scriptsize{}2(b)} & {\scriptsize{}S3} & {\tiny{}$(13.81\pm8.78,14.95\pm8.29)$, $(13.38\pm9.12,-14.85\pm9.41)$,
$(-13.65\pm8.62,14.60\pm8.81)$, $(-15.46\pm8.57,-15.40\pm8.41)$}\tabularnewline
{\scriptsize{}2(c)} & {\scriptsize{}S17} & {\tiny{}$(14.05\pm12.64,13.20\pm9.57)$, $(9.56\pm12.34,-13.08\pm10.76)$,
$(-7.22\pm12.94,14.61\pm9.62)$, $(-13.64\pm11.83,-14.60\pm9.38)$}\tabularnewline
{\scriptsize{}2(c)} & {\scriptsize{}S18} & {\tiny{}$(15.22\pm8.73,14.79\pm8.68)$, $(12.92\pm9.35,-15.20\pm9.01)$,
$(-13.75\pm8.36,14.57\pm8.15)$, $(-15.04\pm9.05,-16.12\pm8.14)$}\tabularnewline
{\scriptsize{}2(c)} & {\scriptsize{}S19} & {\tiny{}$(15.65\pm8.79,14.93\pm8.58)$, $(14.28\pm8.82,-14.34\pm7.67)$,
$(-15.03\pm8.50,13.74\pm8.30)$, $(-14.21\pm8.36,-14.06\pm8.28)$}\tabularnewline
{\scriptsize{}2(c)} & {\scriptsize{}S20} & {\tiny{}$(15.82\pm9.00,15.18\pm7.93)$, $(14.22\pm9.22,-15.54\pm8.38)$,
$(-14.67\pm8.77,14.65\pm8.57)$, $(-17.91\pm9.89,-15.26\pm8.82)$}\tabularnewline
{\scriptsize{}2(c)} & {\scriptsize{}S21} & {\tiny{}$(13.88\pm9.72,15.10\pm7.90)$, $(11.04\pm10.16,-15.23\pm9.15)$,
$(-12.40\pm10.11,15.82\pm9.43)$, $(-16.88\pm9.53,-14.42\pm8.91)$}\tabularnewline
\hline 
\end{tabular}
\par\end{centering}
\label{AB_shape}
\end{table}

\end{singlespace}

We then conducted four two sample KS tests for each subject to examine
marginal selectivity (\ref{marginal selectivity}): $T_{\alpha_{i}}$
and $T_{\beta_{j}}$ were replaced with $A_{i}$ and $B_{j}$ that
stand for the amplitudes of the test shape conditional on the reference
shape $\alpha_{i}\beta_{j},i,j\in\left\{ 1,2\right\} $. Table 14
presents the statistics for the tests. Each column of numbers represents
a particular paired comparison for the subjects. For instance, $((0\,\mathrm{px},30\,\mathrm{px}],)$
compared the $A\mathrm{s}$ across different levels of $\beta$ but
fixed $\alpha=(0\,\mathrm{px},30\,\mathrm{px}]$. $(,[-30\,\mathrm{px},0\,\mathrm{px}])$
compared the $B\mathrm{s}$ across different levels of $\alpha$ but
fixed $\beta=[-30\,\mathrm{px},0\,\mathrm{px}]$. We conclude that
marginal selectivity was confirmed for S4, S5, S6, S1, S18, and S19
($\mathrm{alpha}=.05/4$). 

\begin{table}[H]
\begin{raggedright}
\caption{Two sample KS tests for marginal selectivity for Experiment 2.}
\par\end{raggedright}
\begin{centering}
\begin{tabular}{ccccccc}
\hline 
\multirow{2}{*}{{\scriptsize{}Exp.}} & \multirow{2}{*}{{\scriptsize{}Subject}} & \multirow{2}{*}{{\scriptsize{}$((0\,\mathrm{px},30\,\mathrm{px}],)$}} & \multirow{2}{*}{{\scriptsize{}$([-30\,\mathrm{px},0\,\mathrm{px}],)$}} & \multirow{2}{*}{{\scriptsize{}$(,(0\,\mathrm{px},30\,\mathrm{px}])$}} & \multirow{2}{*}{{\scriptsize{}$(,[-30\,\mathrm{px},0\,\mathrm{px}])$}} & {\scriptsize{}Marginal }\tabularnewline
 &  &  &  &  &  & {\scriptsize{}selectivity?}\tabularnewline
\hline 
2(a) & S4 & .069(.230) & .049(.653) & .084(.079) & .073(.186) & Yes\tabularnewline
2(a) & S5  & .049(.635) & .034(.952) & .054(.504) & .041(.855) & Yes\tabularnewline
2(a) & S6 & .062(.348) & .084(.089) & .070(.236) & .039(.876) & Yes\tabularnewline
2(b) & S1 & .057(.442) & .073(.191) & .037(.927) & .059(.388) & Yes\tabularnewline
2(b) & S2 & .094(.032) & .107(.008) & .054(.522) & .0620(.351) & No\tabularnewline
2(b) & S3 & .057(.476) & .120(.003) & .063(.336) & .078(.142) & No\tabularnewline
2(c) & S17 & .167(.009) & .211(.000) & .141(.034) & .089(.437) & No\tabularnewline
2(c) & S18 & .134(.046) & .148(.028) & .067(.742) & .120(.112) & Yes\tabularnewline
2(c) & S19 & .091(.394) & .070(.677) & .108(.174) & .081(.548) & Yes\tabularnewline
2(c) & S20 & .128(.070) & .199(.001) & .072(.701) & .115(.116) & No\tabularnewline
2(c) & S21 & .128(.094) & .197(.001) & .151(.018) & .082(.530) & No\tabularnewline
\hline 
\end{tabular}
\par\end{centering}
Note: Each number outside of the brackets is the KS statistic value
and each number in the brackets is the $p$ value.

\label{test marginal selecitivity_shape}
\end{table}

For those who passed the test of marginal selectivity, we investigated
if $(A,B)\looparrowleft(\alpha,\beta)$ secured by conducting LFT.
We created two levels for $A$: \{smaller than or equal to 0 px, larger
than 0 px\}, labeled as $\left\{ a_{1},a_{2}\right\} $, and two levels
for $B$: \{smaller than or equal to 0 px, larger than 0 px\}, labeled
as $\left\{ b_{1},b_{2}\right\} $. The numbers in the cells of Table
\ref{joint probabilities_shape} are the joint probabilities for the
discretized $(A_{ij},\:B_{ij}),i,j\in\left\{ 1,2\right\} $, and the
numbers outside are the marginal probabilities.

\begin{table}[H]
\caption{Joint distributions of the discretized $(A_{ij},\:B_{ij}),i,j\in\left\{ 1,2\right\} $
for S4, S5, S6, S1, S18, and S19 in Experiment 2. S4, S5, and S6 participated
in Experiment 2(a). S1 participated in Experiment 2(b). S18 and S19
participated in Experiment 2(c). }
\begin{centering}
\begin{tabular}{|c|c|c|c|c|c|c|c|}
\hline 
\multicolumn{8}{|c|}{S4}\tabularnewline
\hline 
$\alpha_{1}\beta_{1}$ & $B_{11}=b_{1}$ & $B_{11}=b_{2}$ &  & $\alpha_{1}\beta_{2}$ & $B_{12}=b_{1}$ & $B_{12}=b_{2}$ & \tabularnewline
\cline{1-3} \cline{5-7} 
$A_{11}=a_{1}$ & 0 & .0271 & .0271 & $A_{12}=a_{1}$ & .0505 & 0 & .0505\tabularnewline
\cline{1-3} \cline{5-7} 
$A_{11}=a_{2}$ & .0146 & .9583 & .9729 & $A_{12}=a_{2}$ & .9183 & .0313 & .9496\tabularnewline
\cline{1-3} \cline{5-7} 
\multicolumn{1}{|c}{} & \multicolumn{1}{c}{.0146} & \multicolumn{1}{c}{.9854} & \multicolumn{1}{c}{} & \multicolumn{1}{c}{} & \multicolumn{1}{c}{.9688} & \multicolumn{1}{c}{.0313} & \tabularnewline
\cline{1-3} \cline{5-7} 
$\alpha_{2}\beta_{1}$ & $B_{21}=b_{1}$ & $B_{21}=b_{2}$ &  & $\alpha_{2}\beta_{2}$ & $B_{22}=b_{1}$ & $B_{22}=b_{2}$ & \tabularnewline
\cline{1-3} \cline{5-7} 
$A_{21}=a_{1}$ & .0370 & .9261 & .9631 & $A_{22}=a_{1}$ & .9214 & .0175 & .9389\tabularnewline
\cline{1-3} \cline{5-7} 
$A_{21}=a_{2}$ & 0 & .0370 & .0370 & $A_{22}=a_{2}$ & .0611 & 0 & .0611\tabularnewline
\cline{1-3} \cline{5-7} 
\multicolumn{1}{|c}{} & \multicolumn{1}{c}{.0370} & \multicolumn{1}{c}{.9631} & \multicolumn{1}{c}{} & \multicolumn{1}{c}{} & \multicolumn{1}{c}{.9825} & \multicolumn{1}{c}{.0175} & \tabularnewline
\hline 
\end{tabular}
\par\end{centering}
\begin{centering}
\begin{tabular}{|c|c|c|c|c|c|c|c|}
\hline 
\multicolumn{8}{|c|}{S5}\tabularnewline
\hline 
$\alpha_{1}\beta_{1}$ & $B_{11}=b_{1}$ & $B_{11}=b_{2}$ &  & $\alpha_{1}\beta_{2}$ & $B_{12}=b_{1}$ & $B_{12}=b_{2}$ & \tabularnewline
\cline{1-3} \cline{5-7} 
$A_{11}=a_{1}$ & 0 & .0134 & .0134 & $A_{12}=a_{1}$ & .0312 & 0 & .0312\tabularnewline
\cline{1-3} \cline{5-7} 
$A_{11}=a_{2}$ & .0201 & .9664 & .9865 & $A_{12}=a_{2}$ & .9310 & .0379 & .9689\tabularnewline
\cline{1-3} \cline{5-7} 
\multicolumn{1}{|c}{} & \multicolumn{1}{c}{.0201} & \multicolumn{1}{c}{.9798} & \multicolumn{1}{c}{} & \multicolumn{1}{c}{} & \multicolumn{1}{c}{.9622} & \multicolumn{1}{c}{.0379} & \tabularnewline
\cline{1-3} \cline{5-7} 
$\alpha_{2}\beta_{1}$ & $B_{21}=b_{1}$ & $B_{21}=b_{2}$ &  & $\alpha_{2}\beta_{2}$ & $B_{22}=b_{1}$ & $B_{22}=b_{2}$ & \tabularnewline
\cline{1-3} \cline{5-7} 
$A_{21}=a_{1}$ & .0299 & .9124 & .9423 & $A_{22}=a_{1}$ & .9531 & .0094 & .9625\tabularnewline
\cline{1-3} \cline{5-7} 
$A_{21}=a_{2}$ & 0 & .0577 & .0577 & $A_{22}=a_{2}$ & .0376 & 0 & .0376\tabularnewline
\cline{1-3} \cline{5-7} 
\multicolumn{1}{|c}{} & \multicolumn{1}{c}{.0299} & \multicolumn{1}{c}{.9701} & \multicolumn{1}{c}{} & \multicolumn{1}{c}{} & \multicolumn{1}{c}{.9907} & \multicolumn{1}{c}{.0094} & \tabularnewline
\hline 
\end{tabular}
\par\end{centering}
\begin{centering}
\begin{tabular}{|c|c|c|c|c|c|c|c|}
\hline 
\multicolumn{8}{|c|}{S6}\tabularnewline
\hline 
$\alpha_{1}\beta_{1}$ & $B_{11}=b_{1}$ & $B_{11}=b_{2}$ &  & $\alpha_{1}\beta_{2}$ & $B_{12}=b_{1}$ & $B_{12}=b_{2}$ & \tabularnewline
\cline{1-3} \cline{5-7} 
$A_{11}=a_{1}$ & .0023 & .0465 & .0488 & $A_{12}=a_{1}$ & .0458 & 0 & .0458\tabularnewline
\cline{1-3} \cline{5-7} 
$A_{11}=a_{2}$ & .0256 & .9256 & .9512 & $A_{12}=a_{2}$ & .9259 & .0283 & .9542\tabularnewline
\cline{1-3} \cline{5-7} 
\multicolumn{1}{|c}{} & \multicolumn{1}{c}{.0279} & \multicolumn{1}{c}{.9721} & \multicolumn{1}{c}{} & \multicolumn{1}{c}{} & \multicolumn{1}{c}{.9717} & \multicolumn{1}{c}{.0283} & \tabularnewline
\cline{1-3} \cline{5-7} 
$\alpha_{2}\beta_{1}$ & $B_{21}=b_{1}$ & $B_{21}=b_{2}$ &  & $\alpha_{2}\beta_{2}$ & $B_{22}=b_{1}$ & $B_{22}=b_{2}$ & \tabularnewline
\cline{1-3} \cline{5-7} 
$A_{21}=a_{1}$ & .0393 & .9076 & .9469 & $A_{22}=a_{1}$ & .9438 & .0202 & .9640\tabularnewline
\cline{1-3} \cline{5-7} 
$A_{21}=a_{2}$ & 0 & .0531 & .0531 & $A_{22}=a_{2}$ & .0360 & 0 & .0360\tabularnewline
\cline{1-3} \cline{5-7} 
\multicolumn{1}{|c}{} & \multicolumn{1}{c}{.0393} & \multicolumn{1}{c}{.9607} & \multicolumn{1}{c}{} & \multicolumn{1}{c}{} & \multicolumn{1}{c}{.9798} & \multicolumn{1}{c}{.0202} & \tabularnewline
\hline 
\end{tabular}
\par\end{centering}
\begin{centering}
\begin{tabular}{|c|c|c|c|c|c|c|c|}
\hline 
\multicolumn{8}{|c|}{S1}\tabularnewline
\hline 
$\alpha_{1}\beta_{1}$ & $B_{11}=b_{1}$ & $B_{11}=b_{2}$ &  & $\alpha_{1}\beta_{2}$ & $B_{12}=b_{1}$ & $B_{12}=b_{2}$ & \tabularnewline
\cline{1-3} \cline{5-7} 
$A_{11}=a_{1}$ & 0 & .0343 & .0343 & $A_{12}=a_{1}$ & .0277 & 0 & .0277\tabularnewline
\cline{1-3} \cline{5-7} 
$A_{11}=a_{2}$ & .0114 & .9542 & .9656 & $A_{12}=a_{2}$ & .9574 & .0149 & .9723\tabularnewline
\cline{1-3} \cline{5-7} 
\multicolumn{1}{|c}{} & \multicolumn{1}{c}{.0114} & \multicolumn{1}{c}{.9885} & \multicolumn{1}{c}{} & \multicolumn{1}{c}{} & \multicolumn{1}{c}{.9851} & \multicolumn{1}{c}{.0149} & \tabularnewline
\cline{1-3} \cline{5-7} 
$\alpha_{2}\beta_{1}$ & $B_{21}=b_{1}$ & $B_{21}=b_{2}$ &  & $\alpha_{2}\beta_{2}$ & $B_{22}=b_{1}$ & $B_{22}=b_{2}$ & \tabularnewline
\cline{1-3} \cline{5-7} 
$A_{21}=a_{1}$ & .0046 & .9771 & .9817 & $A_{22}=a_{1}$ & .9452 & .0160 & .9612\tabularnewline
\cline{1-3} \cline{5-7} 
$A_{21}=a_{2}$ & 0 & .0183 & .0183 & $A_{22}=a_{2}$ & .0388 & 0 & .0388\tabularnewline
\cline{1-3} \cline{5-7} 
\multicolumn{1}{|c}{} & \multicolumn{1}{c}{.0046} & \multicolumn{1}{c}{.9954} & \multicolumn{1}{c}{} & \multicolumn{1}{c}{} & \multicolumn{1}{c}{.9840} & \multicolumn{1}{c}{.0160} & \tabularnewline
\hline 
\end{tabular}
\par\end{centering}
\label{joint probabilities_shape}
\end{table}

\begin{table}[H]
\begin{raggedright}
Table 15: Joint distributions of the discretized $(A_{ij},\:B_{ij}),i,j\in\left\{ 1,2\right\} $
for S4, S5, S6, S1, S18, and S19 in Experiment 2 (continued). S4,
S5, and S6 participated in Experiment 2(a). S1 participated in Experiment
2(b). S18 and S19 participated in Experiment 2(c). 
\par\end{raggedright}
\begin{centering}
\begin{tabular}{|c|c|c|c|c|c|c|c|}
\hline 
\multicolumn{8}{|c|}{S18}\tabularnewline
\hline 
$\alpha_{1}\beta_{1}$ & $B_{11}=b_{1}$ & $B_{11}=b_{2}$ &  & $\alpha_{1}\beta_{2}$ & $B_{12}=b_{1}$ & $B_{12}=b_{2}$ & \tabularnewline
\cline{1-3} \cline{5-7} 
$A_{11}=a_{1}$ & 0 & .0381 & .0381 & $A_{12}=a_{1}$ & .0896 & 0 & .0896\tabularnewline
\cline{1-3} \cline{5-7} 
$A_{11}=a_{2}$ & .0095 & .9523 & .9618 & $A_{12}=a_{2}$ & .8806 & .0299 & .9105\tabularnewline
\cline{1-3} \cline{5-7} 
\multicolumn{1}{|c}{} & \multicolumn{1}{c}{.0095} & \multicolumn{1}{c}{.9904} & \multicolumn{1}{c}{} & \multicolumn{1}{c}{} & \multicolumn{1}{c}{.9702} & \multicolumn{1}{c}{.0299} & \tabularnewline
\cline{1-3} \cline{5-7} 
$\alpha_{2}\beta_{1}$ & $B_{21}=b_{1}$ & $B_{21}=b_{2}$ &  & $\alpha_{2}\beta_{2}$ & $B_{22}=b_{1}$ & $B_{22}=b_{2}$ & \tabularnewline
\cline{1-3} \cline{5-7} 
$A_{21}=a_{1}$ & .0105 & .9634 & .9739 & $A_{22}=a_{1}$ & .9521 & .0106 & .9627\tabularnewline
\cline{1-3} \cline{5-7} 
$A_{21}=a_{2}$ & 0 & .0262 & .0262 & $A_{22}=a_{2}$ & .0372 & 0 & .0372\tabularnewline
\cline{1-3} \cline{5-7} 
\multicolumn{1}{|c}{} & \multicolumn{1}{c}{.0105} & \multicolumn{1}{c}{.9896} & \multicolumn{1}{c}{} & \multicolumn{1}{c}{} & \multicolumn{1}{c}{.9893} & \multicolumn{1}{c}{.0106} & \tabularnewline
\hline 
\end{tabular}
\par\end{centering}
\begin{centering}
\begin{tabular}{|c|c|c|c|c|c|c|c|}
\hline 
\multicolumn{8}{|c|}{S19}\tabularnewline
\hline 
$\alpha_{1}\beta_{1}$ & $B_{11}=b_{1}$ & $B_{11}=b_{2}$ &  & $\alpha_{1}\beta_{2}$ & $B_{12}=b_{1}$ & $B_{12}=b_{2}$ & \tabularnewline
\cline{1-3} \cline{5-7} 
$A_{11}=a_{1}$ & 0 & .0204 & .0204 & $A_{12}=a_{1}$ & .0718 & 0 & .0718\tabularnewline
\cline{1-3} \cline{5-7} 
$A_{11}=a_{2}$ & .0102 & .9694 & .9796 & $A_{12}=a_{2}$ & .9116 & .0166 & .9282\tabularnewline
\cline{1-3} \cline{5-7} 
\multicolumn{1}{|c}{} & \multicolumn{1}{c}{.0102} & \multicolumn{1}{c}{.9898} & \multicolumn{1}{c}{} & \multicolumn{1}{c}{} & \multicolumn{1}{c}{.9834} & \multicolumn{1}{c}{.0166} & \tabularnewline
\cline{1-3} \cline{5-7} 
$\alpha_{2}\beta_{1}$ & $B_{21}=b_{1}$ & $B_{21}=b_{2}$ &  & $\alpha_{2}\beta_{2}$ & $B_{22}=b_{1}$ & $B_{22}=b_{2}$ & \tabularnewline
\cline{1-3} \cline{5-7} 
$A_{21}=a_{1}$ & .0474 & .8957 & .9431 & $A_{22}=a_{1}$ & .9646 & .0051 & .9697\tabularnewline
\cline{1-3} \cline{5-7} 
$A_{21}=a_{2}$ & 0 & .0569 & .0569 & $A_{22}=a_{2}$ & .0303 & 0 & .0303\tabularnewline
\cline{1-3} \cline{5-7} 
\multicolumn{1}{|c}{} & \multicolumn{1}{c}{.0474} & \multicolumn{1}{c}{.9526} & \multicolumn{1}{c}{} & \multicolumn{1}{c}{} & \multicolumn{1}{c}{.9949} & \multicolumn{1}{c}{.0051} & \tabularnewline
\hline 
\end{tabular}
\par\end{centering}
\label{joint probabilities_shape-1}
\end{table}

The equations (\ref{marginal selectivity}) did not strictly hold
in Table \ref{joint probabilities_shape}. We modified the values
of the marginal probabilities and joint probabilities for each subject
in the same way as in Experiment 1. We were able to find nonnegative
solutions for LFT (Table \ref{LFT_shape}), indicating selective influence
of $\alpha$ and $\beta$ on $A$ and $B$ was established for S4,
S5, S6, S1, S18, and S19. Of course one can choose values other than
0 px to discretize $A$ or $B$. We found LFT passed for all the other
values that we tried. We then considered $(T_{\alpha},T_{\beta})\looparrowleft(\alpha,\beta)$
was successfully established for these subjects. 

\begin{table}[h]
\caption{Solutions for LFT for S4, S5, S6, S1, S18, and S19 in Experiment 2.
S4, S5, and S6 participated in Experiment 2(a). S1 participated in
Experiment 2(b). S18 and S19 participated in Experiment 2(c). }
\begin{centering}
\begin{tabular}{cc}
\hline 
Subject & $(Q_{a_{1}a_{1}b_{1}b_{1}},\,Q_{a_{1}a_{1}b_{1}b_{2}},\,\ldots,\,Q_{a_{2}a_{2}b_{2}b_{2}})^{T}$\tabularnewline
\hline 
S4 & $(0,0,0,0,0,0,.0388,0,.0014,.0244,.9252,0,0,0,.0102,0)^{T}$\tabularnewline
S5 & $(0,0,0,0,0,0,.0223,0,.0014,.0236,.9274,0,0,0,.0253,0)^{T}$\tabularnewline
S6 & $(.0023,0,.0131,0,0,0,.0319,0,.0197,.0116,.9087,0,0,0,0,.0127)^{T}$\tabularnewline
S1 & $(0,0,.0026,.0033,0,0,.0251,0,0,.0046,.9535,.0075,.0034,0,0,0)^{T}$\tabularnewline
S18 & $(0,0,.0361,0,0,0,.0277,0,0,.01,.916,.0062,0,0,0,.004)^{T}$\tabularnewline
\multirow{1}{*}{S19} & $(0,0,.0026,0,0,0,.0436,0,.018,.0108,.925,0,0,0,0,0)^{T}$\tabularnewline
\hline 
\end{tabular}
\par\end{centering}
\label{LFT_shape}
\end{table}

\subsubsection*{Testing Stochastic Dominance}

We tested the assumption of stochastic dominance (\ref{eq:sd}) for
all the participants in Experiment 2. For each participant, we considered
any trial with RT outside of 5 standard deviations of the set of RTs
an outlier and it was not included in the further analysis. 

The left column of Figure \ref{shape} presents the survival functions
of RT for the subjects who passed the test for $(A,B)\looparrowleft(\alpha,\beta)$.
Two one tail KS tests were performed on each of the four paired variables
(\ref{eq:sd}). The statistical results (Table \ref{test stochastic dominance_shape})
support the assumption of stochastic dominance for these subjects
with the assignment (\ref{level_shape}) as for each subject the $p$
values in the bottom row were larger than the critical value $\mathrm{alpha}=.05/4$.
Note for S1, the $p$ value for $S_{\alpha_{1}\beta_{1}}<S_{\alpha_{2}\beta_{1}}$
was not larger than the critical value. We loosely considered S1 passed
the test of stochastic dominance as $p=.011$ was close to the critical
value and other paired comparisons of this person passed the statistical
criterion. 

Subject S2, S3, S17, S20, and S21 did not pass the test of selective
influence. We found that the ordering of RT was tortured for S2, S3,
S17, and S21. For S3 and S17, the RT for the stimuli with opposite
signs of amplitudes consumed more time to make responses than the
stimuli with the same sign of amplitudes. For S2, the RT for stimulus
$\alpha_{1}\beta_{2}$ was the shortest. For S20, we found $S_{\alpha_{2}\beta_{1}}>S_{\alpha_{1}\beta_{1}}>S_{\alpha_{2}\beta_{2}}>S_{\alpha_{1}\beta_{2}}$.
S21 passed the test of stochastic dominance (Table \ref{test stochastic dominance_failed subjects-1})
and the ordering of survival functions was plotted in the left column
of Figure \ref{SIC_MIC_failed subjects-2}.

\begin{table}[H]
\caption{One tail KS tests for stochastic dominance for S4, S5, S6, S1, S18,
and S19 in Experiment 2. S4, S5, and S6 participated in Experiment
2(a). S1 participated in Experiment 2(b). S18 and S19 participated
in Experiment 2(c). }
\begin{centering}
\begin{tabular}{cccc}
\hline 
\multicolumn{4}{c}{S4}\tabularnewline
\hline 
$S_{\alpha_{1}\beta_{1}}>S_{\alpha_{1}\beta_{2}}$ & $S_{\alpha_{1}\beta_{1}}>S_{\alpha_{2}\beta_{1}}$ & $S_{\alpha_{1}\beta_{2}}>S_{\alpha_{2}\beta_{2}}$ & $S_{\alpha_{2}\beta_{1}}>S_{\alpha_{2}\beta_{2}}$\tabularnewline
.035(.580) & .040(.489) & .141(.000) & .151(.000)\tabularnewline
$S_{\alpha_{1}\beta_{1}}<S_{\alpha_{1}\beta_{2}}$ & $S_{\alpha_{1}\beta_{1}}<S_{\alpha_{2}\beta_{1}}$ & $S_{\alpha_{1}\beta_{2}}<S_{\alpha_{2}\beta_{2}}$ & $S_{\alpha_{2}\beta_{1}}<S_{\alpha_{2}\beta_{2}}$\tabularnewline
.053(.285) & .081(.050) & .000(1.0) & .002(.997)\tabularnewline
\hline 
\multicolumn{4}{c}{S5}\tabularnewline
\hline 
$S_{\alpha_{1}\beta_{1}}>S_{\alpha_{1}\beta_{2}}$ & $S_{\alpha_{1}\beta_{1}}>S_{\alpha_{2}\beta_{1}}$ & $S_{\alpha_{1}\beta_{2}}>S_{\alpha_{2}\beta_{2}}$ & $S_{\alpha_{2}\beta_{1}}>S_{\alpha_{2}\beta_{2}}$\tabularnewline
.109(.005) & .043(.432) & .093(.023) & .161(.000)\tabularnewline
$S_{\alpha_{1}\beta_{1}}<S_{\alpha_{1}\beta_{2}}$ & $S_{\alpha_{1}\beta_{1}}<S_{\alpha_{2}\beta_{1}}$ & $S_{\alpha_{1}\beta_{2}}<S_{\alpha_{2}\beta_{2}}$ & $S_{\alpha_{2}\beta_{1}}<S_{\alpha_{2}\beta_{2}}$\tabularnewline
.009(.961) & .018(.858) & .005(.990) & .002(.996)\tabularnewline
\hline 
\multicolumn{4}{c}{S6}\tabularnewline
\hline 
$S_{\alpha_{1}\beta_{1}}>S_{\alpha_{1}\beta_{2}}$ & $S_{\alpha_{1}\beta_{1}}>S_{\alpha_{2}\beta_{1}}$ & $S_{\alpha_{1}\beta_{2}}>S_{\alpha_{2}\beta_{2}}$ & $S_{\alpha_{2}\beta_{1}}>S_{\alpha_{2}\beta_{2}}$\tabularnewline
.238(.000) & .081(.063) & .095(.017) & .306(.000)\tabularnewline
$S_{\alpha_{1}\beta_{1}}<S_{\alpha_{1}\beta_{2}}$ & $S_{\alpha_{1}\beta_{1}}<S_{\alpha_{2}\beta_{1}}$ & $S_{\alpha_{1}\beta_{2}}<S_{\alpha_{2}\beta_{2}}$ & $S_{\alpha_{2}\beta_{1}}<S_{\alpha_{2}\beta_{2}}$\tabularnewline
.020(.833) & .065(.168) & .004(.991) & .007(.978)\tabularnewline
\hline 
\multicolumn{4}{c}{S1}\tabularnewline
\hline 
$S_{\alpha_{1}\beta_{1}}>S_{\alpha_{1}\beta_{2}}$ & $S_{\alpha_{1}\beta_{1}}>S_{\alpha_{2}\beta_{1}}$ & $S_{\alpha_{1}\beta_{2}}>S_{\alpha_{2}\beta_{2}}$ & $S_{\alpha_{2}\beta_{1}}>S_{\alpha_{2}\beta_{2}}$\tabularnewline
.053(.284) & .027(.721) & .144(.000) & .220(.000)\tabularnewline
$S_{\alpha_{1}\beta_{1}}<S_{\alpha_{1}\beta_{2}}$ & $S_{\alpha_{1}\beta_{1}}<S_{\alpha_{2}\beta_{1}}$ & $S_{\alpha_{1}\beta_{2}}<S_{\alpha_{2}\beta_{2}}$ & $S_{\alpha_{2}\beta_{1}}<S_{\alpha_{2}\beta_{2}}$\tabularnewline
.043(.429) & .102(.011) & .006(.986) & .007(.980)\tabularnewline
\hline 
\multicolumn{4}{c}{S18}\tabularnewline
\hline 
$S_{\alpha_{1}\beta_{1}}>S_{\alpha_{1}\beta_{2}}$ & $S_{\alpha_{1}\beta_{1}}>S_{\alpha_{2}\beta_{1}}$ & $S_{\alpha_{1}\beta_{2}}>S_{\alpha_{2}\beta_{2}}$ & $S_{\alpha_{2}\beta_{1}}>S_{\alpha_{2}\beta_{2}}$\tabularnewline
.098(.142) & .084(.245) & .144(.019) & .169(.005)\tabularnewline
$S_{\alpha_{1}\beta_{1}}<S_{\alpha_{1}\beta_{2}}$ & $S_{\alpha_{1}\beta_{1}}<S_{\alpha_{2}\beta_{1}}$ & $S_{\alpha_{1}\beta_{2}}<S_{\alpha_{2}\beta_{2}}$ & $S_{\alpha_{2}\beta_{1}}<S_{\alpha_{2}\beta_{2}}$\tabularnewline
-2.082(1.0) & .023(.898) & 1.306(1.0) & .000(1.0)\tabularnewline
\hline 
\multicolumn{4}{c}{S19}\tabularnewline
\hline 
$S_{\alpha_{1}\beta_{1}}>S_{\alpha_{1}\beta_{2}}$ & $S_{\alpha_{1}\beta_{1}}>S_{\alpha_{2}\beta_{1}}$ & $S_{\alpha_{1}\beta_{2}}>S_{\alpha_{2}\beta_{2}}$ & $S_{\alpha_{2}\beta_{1}}>S_{\alpha_{2}\beta_{2}}$\tabularnewline
.155(.011) & .076(.312) & .111(.099) & .308(.000)\tabularnewline
$S_{\alpha_{1}\beta_{1}}<S_{\alpha_{1}\beta_{2}}$ & $S_{\alpha_{1}\beta_{1}}<S_{\alpha_{2}\beta_{1}}$ & $S_{\alpha_{1}\beta_{2}}<S_{\alpha_{2}\beta_{2}}$ & $S_{\alpha_{2}\beta_{1}}<S_{\alpha_{2}\beta_{2}}$\tabularnewline
.017(.948) & .147(.013) & .041(.728) & .011(.977)\tabularnewline
\hline 
\end{tabular}
\par\end{centering}
Note: Each number outside of the brackets is the KS statistic value
and each number in the brackets is the $p$ value.

\label{test stochastic dominance_shape}
\end{table}

\begin{table}[H]
\caption{One tail KS tests for stochastic dominance for S21 in Experiment 2(c).}
\begin{centering}
\begin{tabular}{cccc}
\hline 
\multicolumn{4}{c}{S21}\tabularnewline
\hline 
$S_{\alpha_{1}\beta_{1}}>S_{\alpha_{1}\beta_{2}}$ & $S_{\alpha_{1}\beta_{1}}>S_{\alpha_{2}\beta_{1}}$ & $S_{\alpha_{1}\beta_{2}}>S_{\alpha_{2}\beta_{2}}$ & $S_{\alpha_{2}\beta_{1}}>S_{\alpha_{2}\beta_{2}}$\tabularnewline
.085(.265) & .029(.842) & .102(.142) & .178(.001)\tabularnewline
$S_{\alpha_{1}\beta_{1}}<S_{\alpha_{1}\beta_{2}}$ & $S_{\alpha_{1}\beta_{1}}<S_{\alpha_{2}\beta_{1}}$ & $S_{\alpha_{1}\beta_{2}}<S_{\alpha_{2}\beta_{2}}$ & $S_{\alpha_{2}\beta_{1}}<S_{\alpha_{2}\beta_{2}}$\tabularnewline
.023(.906) & .064(.445) & .089(.225) & .014(.958)\tabularnewline
\hline 
\end{tabular}
\par\end{centering}
Note: Each number outside of the brackets is the KS statistic value
and each number in the brackets is the $p$ value.

\label{test stochastic dominance_failed subjects-1}
\end{table}

\subsubsection*{Diagnosing Architectures According to SIC and MIC in the Presence
of $(A,B)\looparrowleft(\alpha,\beta)$ and Stochastic Dominance}

\begin{figure}[H]
\includegraphics[scale=0.4]{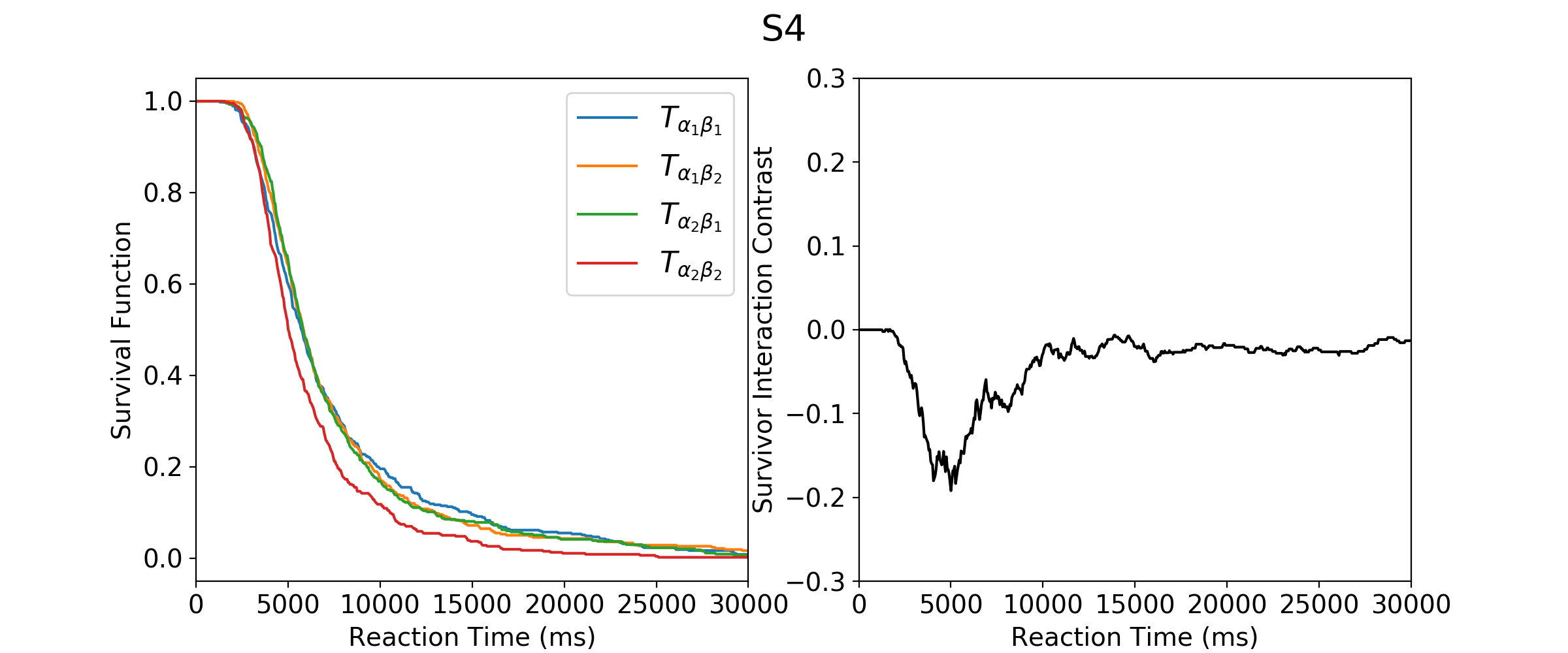}

\includegraphics[scale=0.4]{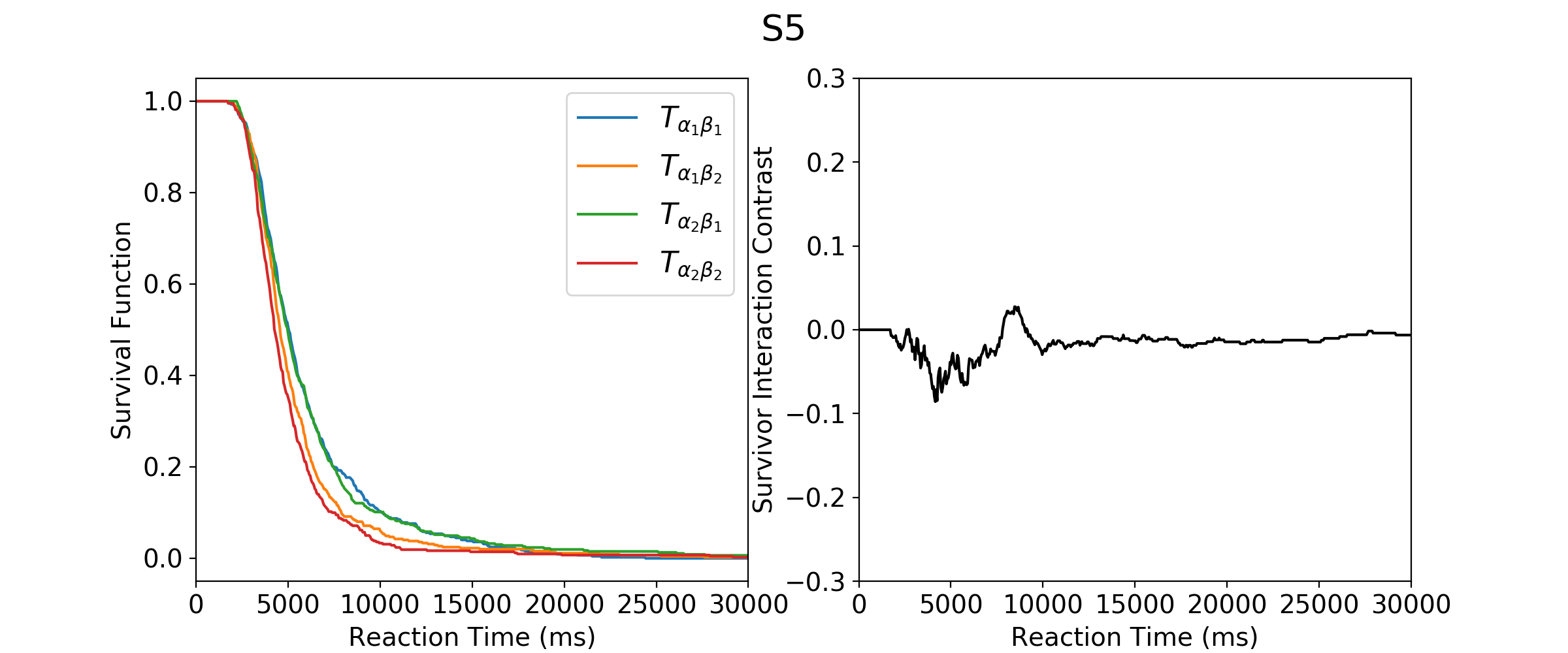}

\includegraphics[scale=0.4]{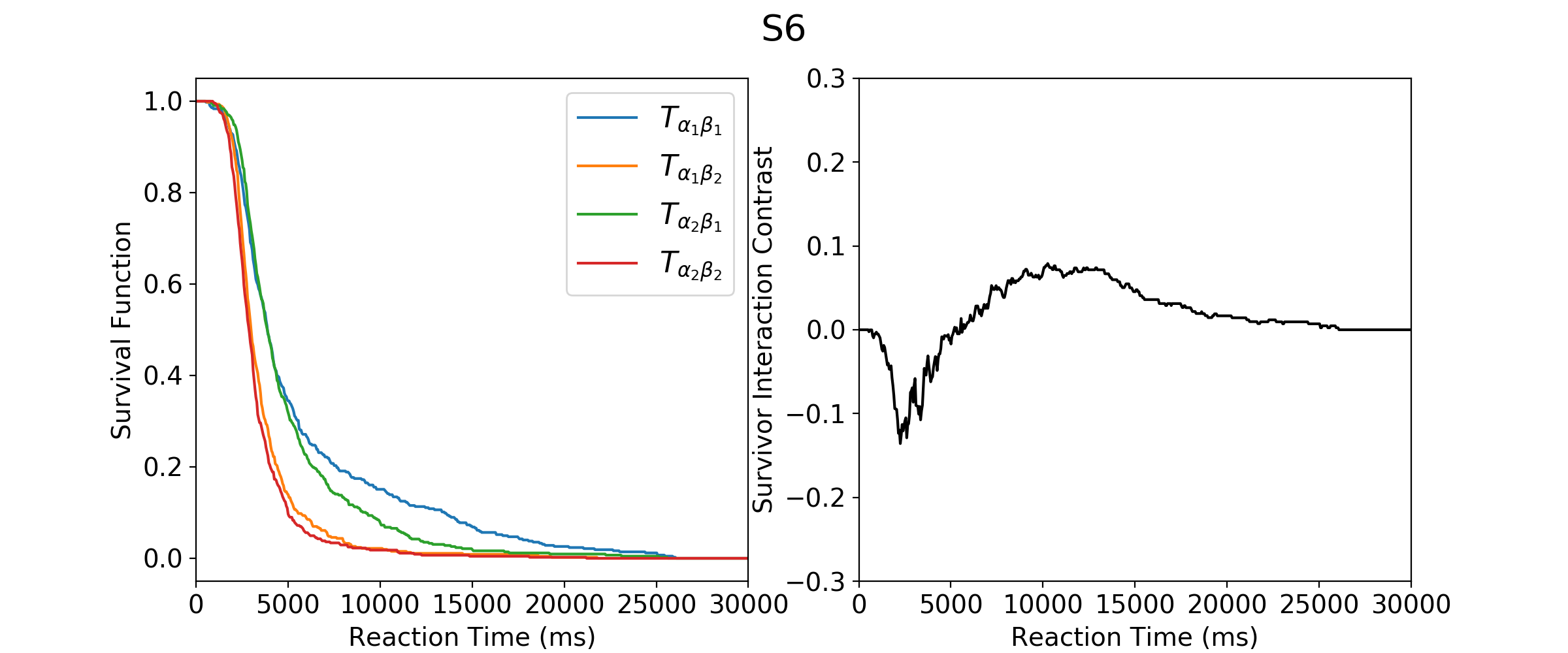}

\includegraphics[scale=0.4]{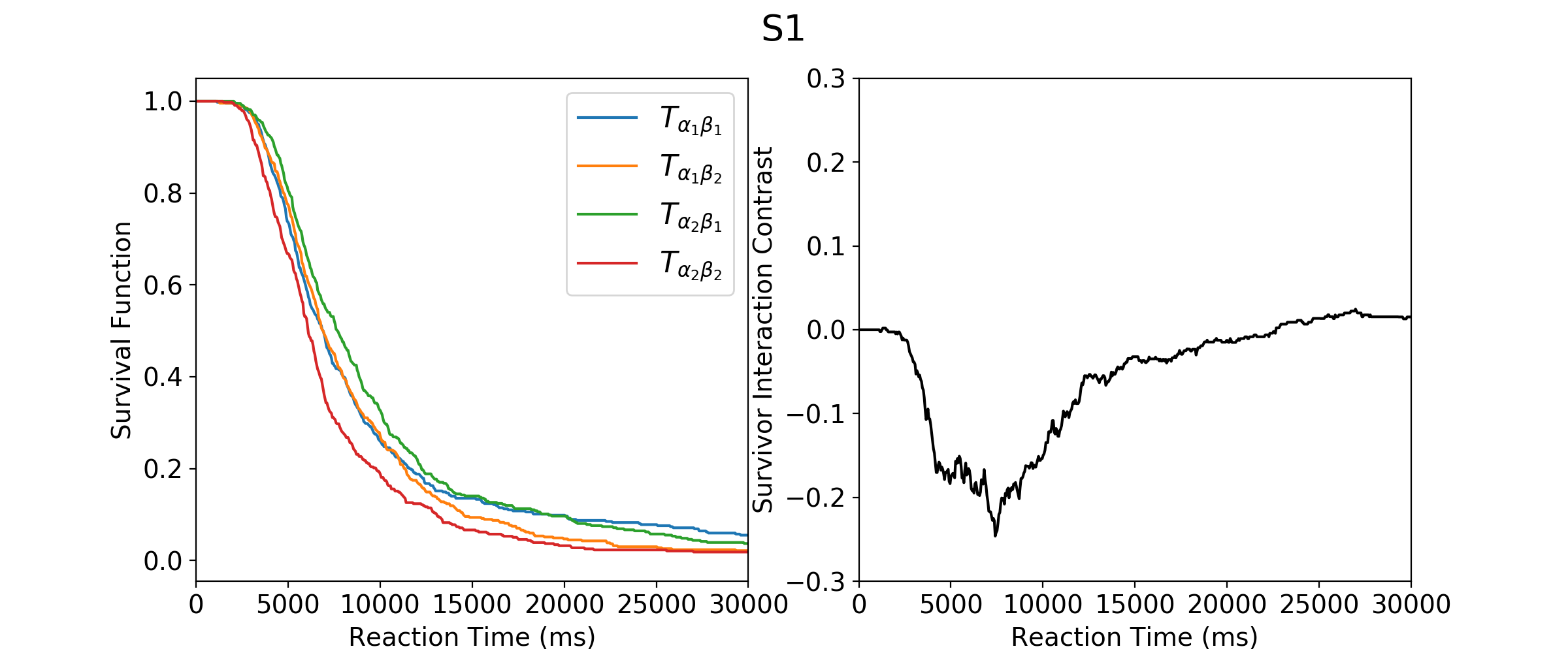}

\caption{Survival functions of RT and SIC for S4, S5, S6, S1, S18, and S19
in Experiment 2. S4, S5, and S6 participated in Experiment 2(a). S1
participated in Experiment 2(b). S18 and S19 participated in Experiment
2(c). }

\label{shape}
\end{figure}

With the confirmation of the three assumptions for S4, S5, S6, S1,
S18, and S19, we then diagnosed how the amplitudes of the test shape
were adjusted by these subjects by investigating the behavior of SIC
and MIC. The SIC curves are displayed in the right column of Figure
\ref{shape}. Table 19 includes the statistics for SIC and MIC and
the inferred architectures from SFT. S4, S5, S1, and S18 were diagnosed
to implement the parallel AND manner to adjust $A$ and $B$ as $D^{+}$
was not significant, $D^{-}$ was significant, and MIC was significantly
less than zero ($\mathrm{alpha}=.33$). S6 had significant $D^{+}$
and $D^{-}$, which agreed with the signature of serial AND or coactive.
The MIC of this person was positive but not significant, which seemed
to support the model of serial AND. Or one can suspect it a lack of
statistical power for a coactive model. Hence we conclude the architecture
for S6 uncertain. S19 was coactive since $D^{+}$ and $D^{-}$ was
significant and MIC was significantly greater than zero. Overall according
to SFT, the strategies the subjects implemented in the floral shape
reproduction task were either parallel AND or coactive, which were
not contradicted with the researchers' expectation and the trajectory
of the trackball move. 

\begin{figure}[H]
\includegraphics[scale=0.4]{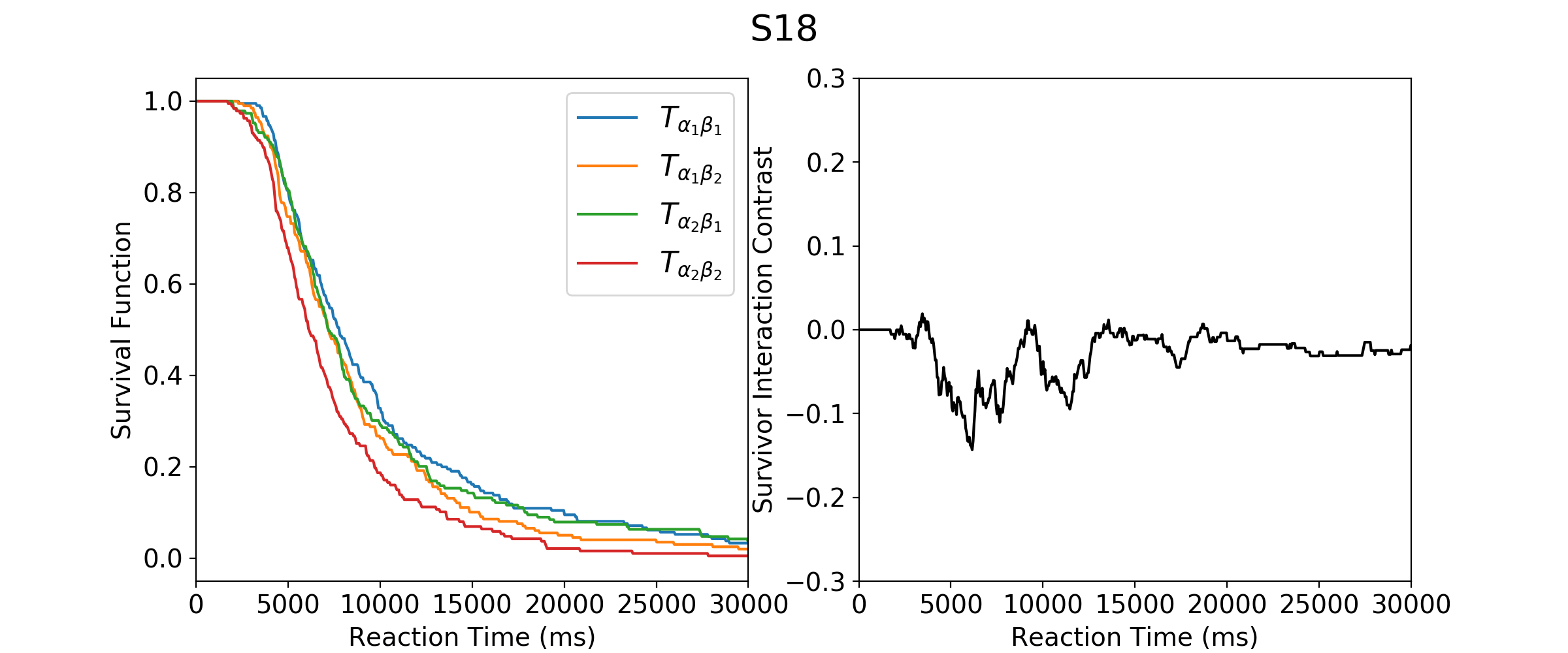}

\includegraphics[scale=0.4]{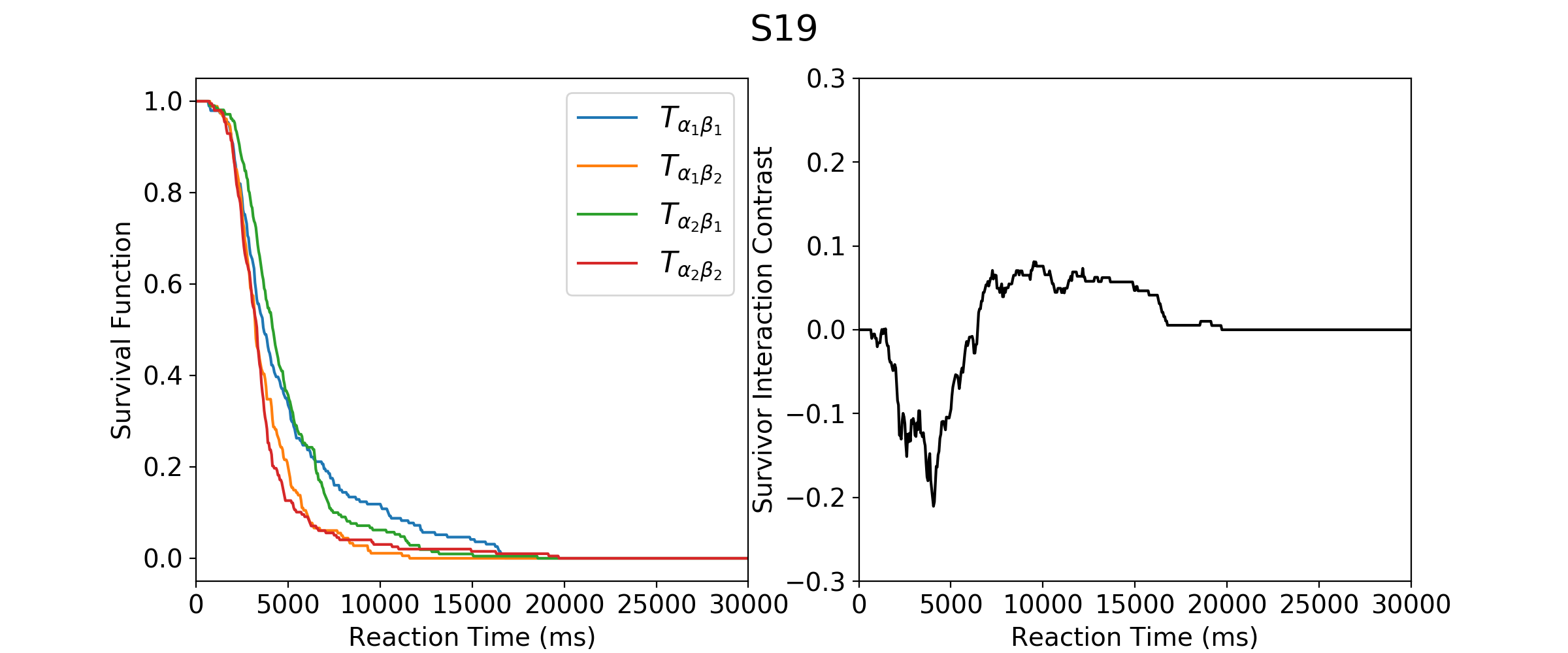}
\begin{raggedright}
Figure 8: Survival functions of RT and SIC for S4, S5, S6, S1, S18,
and S19 in Experiment 2 (continued). S4, S5, and S6 participated in
Experiment 2(a). S1 participated in Experiment 2(b). S18 and S19 participated
in Experiment 2(c). 
\par\end{raggedright}
\label{shape-1}
\end{figure}

\begin{table}[h]
\begin{raggedright}
\caption{The statistics of SIC and MIC and the inferred architectures for Experiment
2. S4, S5, and S6 participated in Experiment 2(a). S1 participated
in Experiment 2(b). S18, S19, and S21 participated in Experiment 2(c). }
\par\end{raggedright}
\begin{centering}
\begin{tabular}{ccccc}
\hline 
Subject & $D^{+}$($p$ value) & $D^{-}$($p$ value) & MIC($p$ value) & Architecture\tabularnewline
\hline 
S4 & .002(.999) & .191(.000) & -1189.2(.002) & Parallel AND\tabularnewline
S5  & .026(.864) & .095(.136) & -520.27(.234) & Parallel AND\tabularnewline
S6 & .080(.242) & .134(.019) & 580.1(.436) & Uncertain\tabularnewline
S1 & .019(.921) & .253(.000) & -1636.6(.000) & Parallel AND\tabularnewline
S18 & .018(.968) & .155(.095) & -2178.9(.242) & Parallel AND\tabularnewline
S19 & .086(.487) & .208(.015) & 202.41(.122) & Coactive\tabularnewline
S21 & .090(.457) & .147(.124) & -1163(.278) & Parallel AND\tabularnewline
\hline 
\end{tabular}
\par\end{centering}
\label{SIC MIC shape}
\end{table}

\subsubsection*{The Consequence of Absence of $(A,B)\looparrowleft(\alpha,\beta)$}

\begin{figure}[H]
\includegraphics[scale=0.4]{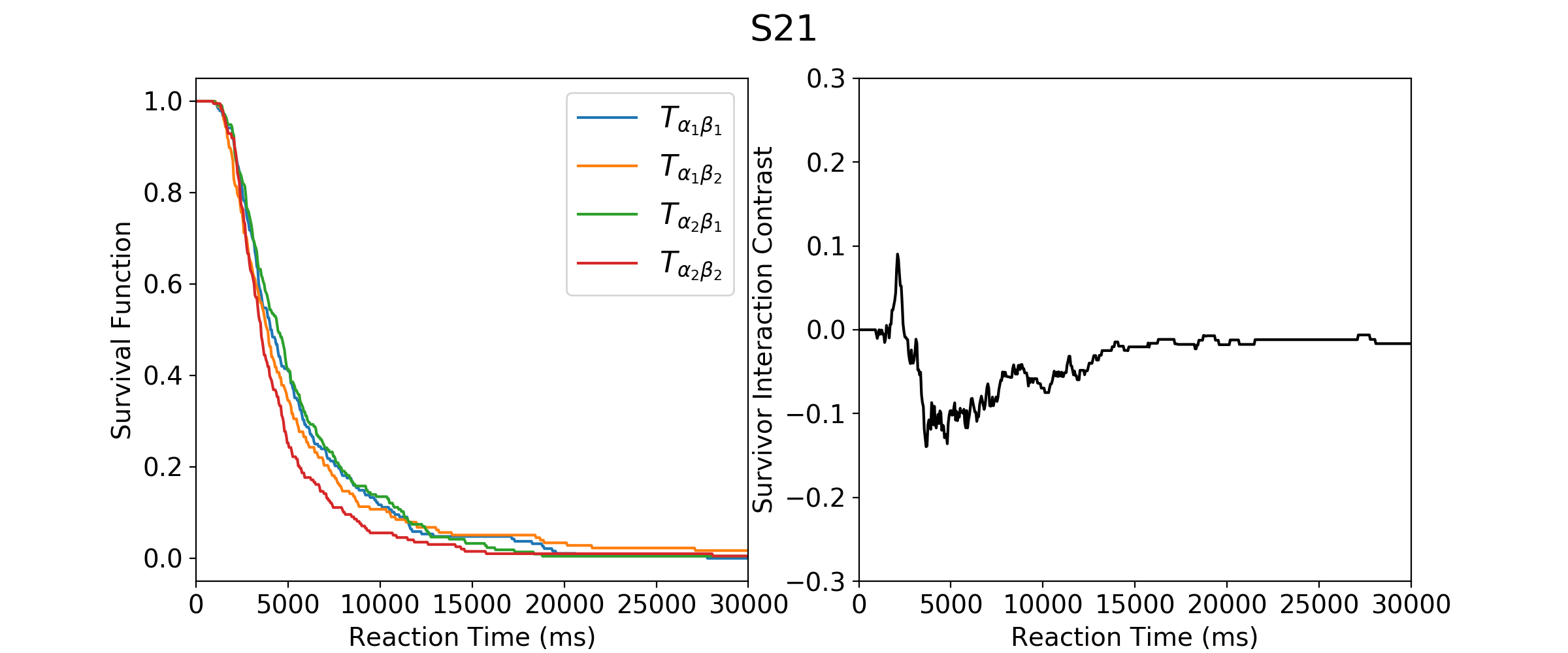}

\caption{Survival functions of RT and SIC for S21 in Experiment 2(c). }

\label{SIC_MIC_failed subjects-2}
\end{figure}

Subject S2, S3, S17, S20, and S21 did not pass the test for $(A,B)\looparrowleft(\alpha,\beta)$.
We considered it a failure for the establishment of $(T_{\alpha},T_{\beta})\looparrowleft(\alpha,\beta)$.
S2, S3, S17, and S20 tortured the ordering of RT distributions, so
it was impossible to investigate their SIC and MIC. Only S21 passed
the test of stochastic dominance (Table \ref{test stochastic dominance_failed subjects-1}).
We investigated the architecture for this person even without the
security of selective influence. The SIC curve of S21 is displayed
in the right column of Figure \ref{SIC_MIC_failed subjects-2}. The
statistics for SIC and MIC and the inferred architecture are in Table
19. The diagnosis was parallel AND because of nonsignificant $D^{+}$,
significant $D^{-}$, and significant negative MIC ($\mathrm{alpha}=.33$).

\subsubsection*{Discussions}

In this experiment, when the subjects passed the test for $(A,B)\looparrowleft(\alpha,\beta)$,
the RTs were all ordered right and the diagnosed architectures were
as anticipated: Some were parallel AND and some were coactive. By
contrast if they failed the test for $(A,B)\looparrowleft(\alpha,\beta)$,
four out of the five subjects' ordering of RTs was tortured. The results
in this experiment seem to imply a success of $(A,B)\looparrowleft(\alpha,\beta)$
was associated with a success of $(T_{\alpha},T_{\beta})\looparrowleft(\alpha,\beta)$;
a failure of $(A,B)\looparrowleft(\alpha,\beta)$ was associated with
a failure of $(T_{\alpha},T_{\beta})\looparrowleft(\alpha,\beta)$.

In Experiment 2, the chance to obtain $(A,B)\looparrowleft(\alpha,\beta)$
was 6 out of 11, higher than 5 out of 13 in Experiment 1, which met
the earlier prediction. It indicates by removing the straightforward
correspondence between hand move and the change of the test stimulus
increases the chance of $(A,B)\looparrowleft(\alpha,\beta)$.

Experiment 2(c) replicated the results obtained from Experiment 2(b):
There was no systematic difference between the two experiments in
terms of the absence or presence of $(A,B)\looparrowleft(\alpha,\beta)$,
the way the RTs were ordered, and the behavior of SIC and MIC.

\section*{Summary}

We investigated how people moved the trackball in the trackball movement
tasks through two lines of approach. One was the trajectory of the
trackball movements and the other was SFT. According to the trajectory
of trackball movements, we observed the subjects implemented parallel
AND or coactive strategy (The trajectory cannot distinguish them)
to adjust the location of the dot or modify the floral shape. SFT
can distinguish the two strategies: In the shape reproduction task,
some subjects implemented the parallel AND manner and some were coactive.
The conclusions from the two lines of approach agreed with each other
for both tasks. 

We proposed a paradigm that can test the assumptions of SFT that are
usually unobservable. In our paradigm, we recorded the physical parameters
labeled as $A$ and $B$ in response to the stimulus features $\alpha$
and $\beta$ and the trajectory of $A$ and $B$ in addition to the
reaction time for each trial. We showed that selective influence of
$T_{\alpha}$ and $T_{\beta}$ can be established through testing
selective influences of $A$ and $B$. The results indicate it is
a valid approach since when $(A,B)\looparrowleft(\alpha,\beta)$ was
established, the behavior of SIC and MIC was as anticipated; when
the $(A,B)\looparrowleft(\alpha,\beta)$ was violated, the ordering
of RT may break or the diagnosis about the architecture may be misleading. 

Conventionally researchers consider stochastic dominance a successful
establishment of selective influence for $T_{\alpha}$ and $T_{\beta}$.
We agree stochastic dominance is correlated with selective influence
$T_{\alpha}$ and $T_{\beta}$ to some extent, but stochastic dominance
is neither a sufficient nor a necessary condition for $(T_{\alpha},T_{\beta})\looparrowleft(\alpha,\beta)$.
Relying on stochastic dominance as an evidence of $(T_{\alpha},T_{\beta})\looparrowleft(\alpha,\beta)$
is risky as it can lead to an incorrect diagnosis for the architecture.
However we understand due to the empirical limit, stochastic dominance
can be the only way to exam $(T_{\alpha},T_{\beta})\looparrowleft(\alpha,\beta)$
as not all the experimental paradigms can afford recording $A$ and
$B$ in response to $\alpha$ and $\beta$. Under this condition,
we agree stochastic dominance is a useful test for $(T_{\alpha},T_{\beta})\looparrowleft(\alpha,\beta)$,
but one should keep in mind these two properties are not equal to
each other. 

Another fundamental assumption for SFT is the subject uses only one
architecture to make responses from trial to trial. This assumption
is impossible to be tested in most studies. In our paradigm, we show,
for the first time, the subjects indeed were stable with their strategies
to respond to the stimuli as reflected by the trajectory over trials.
It demonstrates at least in this particular paradigm, this assumption
is valid. 

One may suggest the exact values of $T_{\alpha}$ and $T_{\beta}$
can be estimated according to the moment the corresponding changing
coordinates or amplitudes terminates. $(T_{\alpha},T_{\beta})\looparrowleft(\alpha,\beta)$
can then be tested using the values of $T_{\alpha}$ and $T_{\beta}$
rather than through testing $(A,B)\looparrowleft(\alpha,\beta)$.
However this approach does not work as we frequently observed in a
particular trial the subjects modified $A$ and $B$ simultaneously
for a while then proceeded only $A$ or $B$ for a while and so on
(e.g. Figure \ref{trackball move dot}). It is practically impossible
to estimate $T_{\alpha}$ and $T_{\beta}$ when they are broken in
parts. So the values of $A$ and $B$ are a better map for the values
of $T_{\alpha}$ and $T_{\beta}$. 

We believe the experimental paradigm we have developed can be applied
in other similar type of tasks, for instance eye-tracking. In the
eye-tracking study, the physical parameters $A$ and $B$ in response
to the stimulus features $\alpha$ and $\beta$ and the trajectory
of $A$ and $B$ can be recorded in the similar way. Then one can
inspect if $(T_{\alpha},T_{\beta})\looparrowleft(\alpha,\beta)$ holds
by examining if $(A,B)\looparrowleft(\alpha,\beta)$ is present. With
the establishment of $(T_{\alpha},T_{\beta})\looparrowleft(\alpha,\beta)$,
SFT can be applied to diagnose the architecture and estimate capacity.
SFT can provide deeper information about the architecture than observing
the trajectory only: One cannot differentiate parallel AND from coactive
according to the trajectory but SFT can tell the two models apart. 

Moreover, most existing studies on mental architectures focus on the
tasks with short response time. Subjects in those studies make a response
within one second. In the current study, it usually took several seconds
to make a response. Our work extends the application of SFT to a broader
field. 

The capacity coefficient can be estimated in Experiment 1. Experiment
2 did not allow the computation for capacity because of the absence
of single-channel trials. In Experiment 2(a), 2(b), and 2(c), the
amplitudes of the reference shapes were selected from the interval
{[}-30 px, 30 px{]} and the amplitudes of the test shape were initialized
randomly in the range {[}-35 px, 35 px{]}. In order to create the
single-channel trials for this task, we introduced the reference shapes
with amplitudes generated from {[}-30 px, 30 px{]} $\times$0 px or
0 px $\times${[}-30 px, 30 px{]} and the test shapes were initialized
with the amplitudes 0 px $\times$0 px. We observed the context effect
regardless of if the single-channel trials and double-channel trials
were displayed to the subjects in a mixed way or the single-channel
trials came after all the double-channel trials were seen. The context
effect we observed were the RT for the reference shapes with opposite-sign
amplitudes were longer than those with same-sign amplitudes. Since
the RTs were not ordered in an expected manner, the capacity was not
computable. 

\section*{Acknowledgements }

This study was supported by NSF grant SES-1155956, AFOSR grant FA9550-14-1-0318,
and MOST (Taiwan) grant 105-2410-H-006-020-MY2.

We thank Dr. Ehtibar Dzhafarov for the suggestions about the experimental
design and the discussions about the concept of selective influence,
ordering of RTs, and stochastic dominance.

\end{document}